\documentclass[aps,footinbib, notitlepage,superscriptaddress, longbibliography,eqsecnum]{revtex4-1}
\usepackage{graphicx}
 \usepackage{amssymb}
 \usepackage{amsmath}
 \usepackage{amsfonts}
\usepackage{overpic}
\usepackage{enumerate}
\usepackage{footmisc}
\usepackage{color}
\usepackage[dvipsnames]{xcolor}
\usepackage{xspace}
\usepackage[normalsize]{subfigure}
\usepackage{hyperref}
\usepackage{bm}
\usepackage{empheq}
\usepackage[capitalize]{cleveref}

\crefname{section}{Sec.}{Secs.}

\DeclareFontFamily{OT1}{pzc}{}
\DeclareFontShape{OT1}{pzc}{m}{it}{<-> s * [1.10] pzcmi7t}{}
\DeclareMathAlphabet{\mathpzc}{OT1}{pzc}{m}{it}

\providecommand{\st}[1]{_{\text{#1}}}
\providecommand{\sfrac}[2]{#1/#2}

\providecommand{\pfrac}[2]{\left(\frac{#1}{#2}\right)}

\def\onehalf{\frac{1}{2}}

\def\eq{\st{eq}}

\def\const{\mathrm{const}}

\def\pd{\partial}

\def\im{\mathrm{i}}

\def\rv{\bv{r}}

\def\b0{\bv{0}}

\def\Fcal{\mathcal{F}}

\def\Hcal{\mathcal{H}}
\def\Hc2{\Hcal^{(2)}}
\def\Ccal{\mathcal{C}}

\def\Kcal{\mathcal{K}}

\def\Mcal{\mathcal{M}}

\def\Ocal{\mathcal{O}}

\def\Tcal{\mathcal{T}}
\def\Ucal{\mathcal{U}}

\def\hyp13{{_1 F_3}}

\def\op{\mathpzc{m}}

\def\bcs{boundary conditions\xspace}

\def\izero{^{(0)}}

\def\tred{\tau}
\def\tscal{x}
\def\time{\theta}

\def\amplPhit{\phi_t\izero}

\def\amplXip{\xi_+\izero}

\def\amplHsurf{l_{h_1}\izero}
\def\amplTrelax{\mathpzc{t}_{+}\izero}
\def\d{\mathrm{d}}

\def\trelax{t_R}
\def\tLG{r}
\def\zdyn{\mathpzc{z}}

\newcommand{\bitem}{\begin{itemize}}
\newcommand{\eitem}{\end{itemize}}
\newcommand{\benum}{\begin{enumerate}}
\newcommand{\eenum}{\end{enumerate}}
\newcommand{\btab}[1]{\begin{tabular}{#1}}
\newcommand{\etab}{\end{tabular}}
\newcommand{\beq}{\begin{equation}}
\newcommand{\eeq}{\end{equation}}
\newcommand{\beqn}{\begin{equation*}}
\newcommand{\eeqn}{\end{equation*}}

\newcommand{\bv}[1]{\mathbf{#1}}

\begin{document}
\title{Surface-induced non-equilibrium dynamics and critical Casimir forces\\for model B in film geometry}
\author{Markus Gross}
\email{gross@is.mpg.de}
\affiliation{Max-Planck-Institut f\"{u}r Intelligente Systeme, Heisenbergstra{\ss}e 3, 70569 Stuttgart, Germany}
\affiliation{IV.\ Institut f\"{u}r Theoretische Physik, Universit\"{a}t Stuttgart, Pfaffenwaldring 57, 70569 Stuttgart, Germany}
 \author{Andrea Gambassi}
 \affiliation{SISSA -- International School for Advanced Studies and INFN, via Bonomea 265, 34136 Trieste, Italy}
 \author{S. Dietrich}
 \affiliation{Max-Planck-Institut f\"{u}r Intelligente Systeme, Heisenbergstra{\ss}e 3, 70569 Stuttgart, Germany}
 \affiliation{IV.\ Institut f\"{u}r Theoretische Physik, Universit\"{a}t Stuttgart, Pfaffenwaldring 57, 70569 Stuttgart, Germany}
\date{\today}

\begin{abstract}
Using analytic and numerical approaches, we study the spatio-temporal evolution of a conserved order parameter of a fluid in film geometry, following an instantaneous quench to the critical temperature $T_c$ as well as to supercritical temperatures.
The order parameter dynamics is chosen to be governed by \emph{model B} within mean field theory and is subject to no-flux \bcs as well as to symmetric surface fields at the confining walls.
The latter give rise to critical adsorption of the order parameter at both walls and provide the driving force for the non-trivial time evolution of the order parameter.
During the dynamics, the order parameter is locally and globally conserved; thus, at thermal equilibrium, the system represents the canonical ensemble.
We furthermore consider the dynamics of the nonequilibrium critical Casimir force, which we obtain based on the generalized force exerted by the order parameter field on the confining walls.
We identify various asymptotic regimes concerning the time evolution of the order parameter and the critical Casimir force and we provide, within our approach, exact expressions of the corresponding dynamic scaling functions.
\end{abstract}


\maketitle

\section{Introduction}

A fluid at its critical point exhibits scale-invariant long-ranged fluctuations and a drastic slowing-down of its dynamics.
The critical behavior is characterized by universal features which are determined by general properties of the fluid, such as the dimensionality of the order parameter (OP), conservation laws, and possibly secondary fields coupled to the OP \cite{hohenberg_theory_1977}. 
We recall that, in a one-component fluid, the OP $\phi$ is proportional to the deviation of the actual number density $n$ from its critical value $n_c$, i.e., $\phi\propto n-n_c$, while for a binary liquid mixture, $\phi$ is proportional to the deviation of the concentration $C_A$ of species A from its critical value $C_{A,c}$, i.e., $\phi\propto C_A- C_{A,c}$.

Dynamic critical phenomena have been extensively studied in bulk fluids (see, e.g., Refs.\ \cite{folk_critical_2006, tauber_critical_2014} for reviews) and in a semi-infinite geometry \cite{dietrich_effects_1983, diehl_boundary_1992, diehl_universality_1994, wichmann_dynamic_1995, ritschel_universal_1995, majumdar_nonequilibrium_1996, pleimling_aging_2004}.
However, in the case of more strongly confined systems, such as \emph{films}, theoretical results on dynamic criticality are more scarce.
In fact, previous studies \cite{diehl_dynamical_1993, ritschel_long-time_1995, ritschel_dynamical_1996, gambassi_critical_2006, diehl_dynamic_2009, gambassi_relaxation_2008} mostly addressed purely relaxational dynamics, i.e., \emph{model A} in the nomenclature of Ref.\ \cite{hohenberg_theory_1977}, which captures the dynamics of a single non-conserved density, e.g., the magnetization in the case of a uniaxial ferromagnet near its Curie point.
The critical dynamics of a fluid, instead, is described by \emph{model H}, which, in its simplest realization, encompasses an advection-diffusion equation for the conserved OP, coupled to a diffusive transport equation for the transverse fluid momentum \cite{hohenberg_theory_1977}.
In passing, we mention that a significant number of studies of (partly) confined fluids exists addressing specific sub-critical phenomena, such as surface-induced phase separation \cite{ ball_spinodal_1990, binder_dynamics_1991, puri_dynamics_1993, lee_filler-induced_1999, onuki_phase_2002}.

Introducing confinement in a near-critical fluid gives rise to the so-called \emph{critical Casimir force} (CCF) acting on the confining boundaries \cite{fisher_wall_1978,krech_free_1992}. Generally, the CCF can result from a confinement-induced long-wavelength cutoff of the fluctuation spectrum as well as from the appearance of slowly decaying OP profiles in the film (see, e.g., Refs.\ \cite{krech_casimir_1994, brankov_theory_2000, gambassi_casimir_2009} for reviews).
In the latter case, the CCF lends itself to a description within \emph{mean field theory} (MFT), which entails neglecting effects of thermal noise.
Similarly to critical dynamics, studies on the dynamics of thermal Casimir-like forces in films focused so far mostly on model A \cite{gambassi_critical_2006, brito_generalized_2007, gambassi_relaxation_2008, rodriguez-lopez_dynamical_2011, dean_out--equilibrium_2010, dean_out--equilibrium_2012, dean_relaxation_2014, hanke_non-equilibrium_2013}, with the exception of Ref.\ \cite{rohwer_transient_2017}, where model B-like dynamics in a quench far from criticality has been investigated.

Here, we consider a film after an instantaneous quench from a quasi-infinite temperature right to the critical point as well as to supercritical temperatures. 
Thus in the initial state the mean OP profile across the film vanishes \cite{gross_critical_2016}.  
However, at finite temperatures, the presence of effective surface fields at the confining walls gives rise to the build-up of a spatially varying adsorption profile across the film at late times ($t\to\infty$).
Accordingly, in the present MFT case the dynamics of the OP and of the resulting CCF is induced solely by the action of surface fields.
In order to facilitate an analytical study, we approximate the actual critical fluid dynamics in terms of the mean-field limit of \emph{model B}, which describes the diffusive dynamics of a single conserved OP but neglects fluctuations.  
Accordingly, we also assume heat diffusion to be sufficiently fast to provide an effective isothermal environment directly after the quench.

In Ref.\ \cite{gross_critical_2016} it has been shown that in a film, which is close to criticality and confined by the walls of the container, the behavior of the \emph{equilibrium} CCF depends crucially on whether the OP of the confined fluid is conserved or not.
In particular, for so-called $(++)$ \bcs, i.e., if the confined fluid is adsorbed with equal strength at both walls, the CCF is attractive in the grand canonical ensemble (globally non-conserved OP), while it is repulsive in the canonical ensemble (globally conserved OP).
For model B dynamics in a film with no-flux \bcs, the total OP is conserved at all times, i.e.,
\beq \Phi(t) \equiv \int_V \d^d r\, \phi(t,\rv) = \const,
\label{eq_mass_t}\eeq 
where the integral runs over the $d$-dimensional volume $V$ of the film.
In Ref.\ \cite{gross_statistical_2017}, ensemble-induced differences for the CCF have been discussed within a field theoretical treatment and for further \bcs.
In the present study, we investigate how the equilibrium CCF in the canonical ensemble considered in Ref.\ \cite{gross_critical_2016} emerges dynamically within model B after a temperature quench. 

\section{General scaling considerations}

Here, we formulate the general dynamic scaling behavior expected for the OP and the CCF for a confined fluid (see, e.g., Refs.\ \cite{diehl_field-theoretical_1986, gambassi_critical_2006, gross_critical_2016}).
We consider systems which are translationally invariant along the lateral film directions, such that, as a consequence of the mean-field approximation, only the transverse coordinate $z$ enters the description.
We consider symmetric [$(++)$] \bcs, such that the influence of the confining walls, placed at $z=0$ and $z=L$, is accounted for by a single parameter $h_1$, describing the strength of both surface fields.
In a near-critical film of thickness $L$, the OP $\phi$ fulfills the general homogeneity relation 
\beq \phi(t,z,\tred, h_1, L) = b^{-\beta/\nu} \phi(t b^{-\zdyn}, z/b, \tred b^{1/\nu}, h_1 b^{\Delta_1/\nu}, L/b)
\label{eq_OP_gen_scal}\eeq 
where $b$ is a scaling factor,
\beq \tred \equiv \frac{T-T_c}{T_c}
\label{eq_tred}\eeq 
is the reduced temperature, $\beta$ and $\nu$ are standard bulk critical exponents, $\Delta_1$ is a surface critical exponent \cite{gross_critical_2016,diehl_field-theoretical_1986}, and $\zdyn$ (not to be confused with the spatial coordinate $z$) is the dynamic bulk critical exponent.
In model B, one has $\zdyn=4-\eta$, where $\eta$ is a standard static critical exponent \cite{tauber_critical_2014,hohenberg_theory_1977}.
Within MFT, one has $\eta=0$ and thus
\beq \zdyn = 4.
\label{eq_zdyn}\eeq 
Upon setting $b=L$ in \cref{eq_OP_gen_scal} and by introducing appropriate length and time scales, one obtains the following finite-size scaling relation \cite{gross_critical_2016,gambassi_critical_2006}:
\beq \phi(t,z,\tred, h_1, L) =  \amplPhit \left(\frac{L}{\amplXip}\right)^{-\beta/\nu} \op( \time, \zeta, \tscal, H_1)
\label{eq_FSS_op}\eeq
where $\op$ is a universal scaling function. We have introduced the following scaling variables:
\begin{subequations}\label{eq_FSS_vars}\begin{align}
\zeta &\equiv z/L, \label{eq_FSS_z}\\
\tscal &\equiv \pfrac{L}{\amplXip}^{1/\nu} \tred \, = \left( \frac{L}{\xi}\right)^{1/\nu}, \label{eq_FSS_temp}\\
H_1 &\equiv \pfrac{L}{\amplHsurf}^{\Delta_1/\nu} h_1, \label{eq_FSS_H1}\\
\time &\equiv  \pfrac{\amplXip}{L}^\zdyn \frac{t}{\amplTrelax} = \left(\frac{\xi}{L}\right)^\zdyn \tau^{\nu \zdyn} \frac{t}{\amplTrelax}\, = \left(\frac{\tau}{\tscal}\right)^{\nu\zdyn} \frac{t}{\amplTrelax}  . \label{eq_FSS_time}
\end{align}\end{subequations}
The non-universal amplitudes $\amplXip$ and $\amplPhit$ are defined in terms of the critical behavior of the bulk correlation length $\xi$ above $T_c$, i.e., $\xi(\tau\to 0) = \amplXip \tred^{-\nu}$, and of the bulk OP $\phi_b$ below $T_c$, i.e., $\phi_b = \amplPhit(-\tred)^\beta$. 
The non-universal amplitude $\amplHsurf$ relates the characteristic length scale $l_{h_1}$ for the OP decay close to the wall (i.e., the so-called ``extrapolation length'') to the strength $h_1$ of the effective surface field via $l_{h_1} = \amplHsurf |h_1|^{-\nu/\Delta_1}$ (see Ref.\ \cite{gross_critical_2016} for further details).
The non-universal amplitude $\amplTrelax$ is defined via the critical divergence of the relaxation time $\trelax$ above $T_c$, i.e., $\trelax = \amplTrelax \tred^{-\nu \zdyn}$.
The bulk correlation length and the relaxation time can be inferred from the exponential decay in space and time of the dynamical bulk correlation function.

In thermal equilibrium, the CCF $\Kcal$ can be defined as the difference of the thermodynamic pressure $p_f=-\d \Fcal_f / \d L$ of the film and the pressure $p_b$ of the surrounding bulk medium:
\beq \Kcal\eq = p_f - p_b.
\label{eq_CCF_eq_def}\eeq 
Here, $\Fcal_f$ denotes the total free energy of the film (per area $A$ of a single wall and $k_B T$), while the bulk pressure is obtained as $p_b = \lim_{L\to\infty} p_f$. Note that the limit is to be performed by keeping the relevant thermodynamic control parameter fixed, which is the external bulk field $\mu$ in the grand canonical ensemble and the mean mass density $\Phi/(AL)$ [see \cref{eq_mass_t}] in the canonical ensemble. In the presence of an OP constraint, the definition of the CCF is, in fact, subtle and we refer to Ref.\ \cite{gross_critical_2016} for further discussion.
In \cref{sec_CCF}, we shall extend the definition of the CCF to \emph{non-equilibrium} situations and we shall show that this definition reduces to \cref{eq_CCF_eq_def} in the equilibrium limit.

Since near criticality the correlation length $\xi$ and the relaxation time $\trelax$ represent the dominant length and time scales, one expects a scaling form analogous to \cref{eq_FSS_op} to apply also for the general non-equilibrium CCF $\Kcal$, i.e.,
\beq \Kcal(t,\tred,h_1) = L^{-d} \Xi( \time, \tscal, H_1),
\label{eq_FSS_CCF}\eeq 
where $\Xi$ is a scaling function and the scaling variables $\time$, $\tscal$, and $H_1$ are given in \cref{eq_FSS_vars}.

\section{Model}
\label{sec_model}

In this section we introduce the dynamic model to be analyzed below.
In the following all extensive quantities are understood to be divided by the transverse area $A$.
Within our approach the \emph{static} properties of the OP field $\phi(t,z)$ follow from the standard  Landau-Ginzburg free energy functional (defined per area and per $k_B T$):
\beq \Fcal[\phi] = \int_0^L \d z\, \left[ (\pd_z \phi)^2 + \Ucal_b(\phi) + \Ucal_s(\phi) \right],
\label{eq_freeEn}\eeq 
where 
\beq \Ucal_b(\phi) \equiv \frac{\tLG}{2}\phi^2 + \frac{g}{4!}\phi^4
\label{eq_freeEn_pot}\eeq
and
\beq \Ucal_s(\phi) \equiv \left[ \onehalf c \phi^2 - h_1 \phi \right] \left[ \delta(z) + \delta(z-L) \right]
\label{eq_freeEn_potSurf}\eeq 
denote the bulk and the surface contribution, respectively.
In \cref{eq_freeEn}, the integral runs from $z=0^-$ up to $z=L^+$.
Within MFT, the coupling constants $\tLG= (\amplXip)^{-2} \tred$ and $g=6(\amplPhit \amplXip)^{-2}$ are given in terms of the non-universal amplitudes $\amplXip$ and $\amplPhit$ introduced in \cref{eq_FSS_op,eq_FSS_vars} \cite{pelissetto_critical_2002}. 
The parameters $c$ and $h_1$ in \cref{eq_freeEn_potSurf} represent the surface enhancement and the surface adsorption strength for the OP, respectively. Here, we focus on the case $c=0$ and $h_1>0$, corresponding to the $(++)$ surface universality class, which describes critical adsorption as observed generically for a confined fluid \cite{gambassi_critical_2009}.
Minimization of $\Fcal$ leads to the well-known equilibrium \bcs \cite{gross_critical_2016}
\beq \pd_z\phi(t,z=0) = -h_1\qquad \text{and}\qquad \pd_z\phi(t,z=L) = h_1,
\label{eq_bcs_CA}\eeq 
which we will impose also during the non-equilibrium time evolution.

The \emph{dynamics} of $\phi$ is governed by model B, which, within MFT, is given by \cite{hohenberg_theory_1977}
\beq \pd_t \phi = -\pd_z J = D\pd_z^{\,2} \mu = D\left\{ -\pd_z^{\,4} \phi + \pd_z^{\,2} \left[ \Ucal_b'(\phi)\right] \right\}.
\label{eq_modelB}\eeq 
Here, the diffusivity $D$ is a kinetic coefficient, $J\equiv -D\pd_z \mu$ is the flux, and 
\beq \mu \equiv \frac{\delta \Fcal_b}{\delta \phi} = -\pd_z^2\phi + \tLG\phi + \frac{g}{3!} \phi^3
\label{eq_chempot}\eeq 
is the bulk chemical potential, defined in terms of the corresponding bulk free energy functional $\Fcal_b \equiv \Fcal-\int_0^L \d z\, \Ucal_s(\phi)$.
In order to ensure global mass conservation, no-flux \bcs are imposed, i.e., $J(t,z=0) = 0 = J(t,z=L)$, or, equivalently,
\beq 0 = -\pd_z\mu = \pd_z^3\phi(t,z)- \left[\tLG + \frac{g}{2}\phi^2(t,z)\right] \pd_z \phi(t,z), \quad \text{for}\quad  z=0,L.
\label{eq_bcs_noflx}\eeq 
The \bcs in \cref{eq_bcs_CA,eq_bcs_noflx} are imposed at all times $t>0$. They have also been derived in Ref.\ \cite{diehl_boundary_1992} within a full field theoretical treatment of model B in a half-space.

We focus on the dynamics induced by \cref{eq_modelB} after an instantaneous quench from a high temperature ($\tred\to \infty$) to a temperature close to $T_c$.
In the limit $\tred\to\infty$ the equilibrium OP profile resulting from \cref{eq_freeEn} vanishes so that, accordingly, the initial condition is 
\beq \phi(t=0,z) = 0.
\label{eq_prof_init}\eeq 
In the course of time, the OP profile attains its equilibrium shape characteristic for critical adsorption under $(++)$ \bcs \cite{gross_critical_2016}.
In the present case, owing to no-flux \bcs [\cref{eq_bcs_noflx}], the OP is globally conserved. \Cref{eq_prof_init} therefore implies
\beq \int_0^L \d z\, \phi(t,z) =0,
\label{eq_mass}\eeq 
i.e., the so-called ``mass'' $\Phi(t)$ [per area $A$, see \cref{eq_mass_t}] vanishes at all times $t$.

Within MFT, the dynamical critical exponent is $\zdyn=4$ [\cref{eq_zdyn}] and the finite-size scaling variables in \cref{eq_FSS_op,eq_FSS_vars} take the forms
\begin{subequations}\label{eq_scalvar_MFT}\begin{align}
\zeta &= z/L,\\
\tscal &= L^2\tLG, \label{eq_tau_resc}\\
H_1 &= \sqrt{\frac{g}{6}} L^2 h_1,\\
\time &= \frac{D}{L^4}t, \\
\op(\time,\zeta) &= \sqrt{\frac{g}{6}} L\phi\left(L^4\time/D , \zeta L \right).
\end{align}\end{subequations}
According to \cref{eq_FSS_time}, the kinetic coefficient $D$ can be expressed in terms of non-universal amplitudes as $D=(\amplXip)^4 / \amplTrelax$. 
Using \cref{eq_scalvar_MFT}, the dynamic equation of model B [\cref{eq_modelB}] assumes the dimensionless form
\beq \pd_\time \op = \pd_\zeta^2 \left(-\pd_\zeta^2 \op +  \tscal\, \op + \op^3 \right)
\label{eq_modelB_resc}\eeq 
and the \bcs in \cref{eq_bcs_CA,eq_bcs_noflx} become
\begin{subequations}\begin{align}
\pd_\zeta\op(\time,\zeta=0) = -H_1, \qquad \pd_\zeta\op(\time,\zeta=1) &= H_1, \label{eq_bcs_CA_red}\\
\pd_\zeta^3\op(\time,\zeta) - \left[\tscal + 3 \op^2(\time,\zeta)\right] \pd_\zeta\op(\time,\zeta)  &= 0, \quad \text{for}\quad \zeta=0,1. \label{eq_bcs_noflx_red}
\end{align}\label{eq_bcs_red}\end{subequations}
In the following we proceed with an analysis of this set of equations.

\section{Linear quench dynamics}
\label{sec_lin_dyn}

In order to facilitate an analytical study of \cref{eq_modelB_resc}, in the free energy functional [\cref{eq_freeEn_pot}] we disregard the term $\propto \phi^4$.
This amounts to studying the linearized (or Gaussian) model B,
\beq \pd_\time \op = -\pd_\zeta^4 \op + \tscal \pd_\zeta^2 \op ,
\label{eq_modelB_resc_lin}\eeq 
together with \cref{eq_bcs_CA} and the linearized no-flux boundary condition in \cref{eq_bcs_noflx_red}, i.e., $\pd_\zeta^3\op - x\pd_\zeta\op=0$.
Upon introducing the Laplace transform
\beq \hat\op(s,\zeta) = \int_0^\infty \d \time\, \op(\time,\zeta) e^{-s \time},
\eeq 
\cref{eq_modelB_resc_lin} turns into
\beq s \hat\op(s,\zeta) = \op(\theta=0,\zeta) - \pd_\zeta^4 \hat \op(s,\zeta) + \tscal \pd_\zeta^2 \hat \op(s,\zeta).
\label{eq_linB_lapl}\eeq
Assuming as initial condition a flat profile $\op(\time=0, \zeta)=0$ [see \cref{eq_prof_init}], \cref{eq_linB_lapl} reduces to a homogeneous fourth-order differential equation:
\beq s \hat\op(s,\zeta) = - \pd_\zeta^4 \hat \op(s,\zeta) + \tscal\pd_\zeta^2\hat\op(s,\zeta).
\label{eq_linB_lapl_flat}\eeq
This equation is solved by the ansatz $\op(s,\zeta) \propto \exp(\Omega \zeta)$, with the parameter $\Omega$ taking one of the four possible values $\{\lambda_\pm, -\lambda_\pm\}$ where 
\beq \lambda_\pm = \frac{1}{\sqrt{2}} \sqrt{\tscal\pm \sqrt{\tscal^2 -4s}}.
\label{eq_lambda_pm}\eeq 
Imposing the (linearized version of the) \bcs in \cref{eq_bcs_red} yields \footnote{Note that in Laplace space \cref{eq_bcs_CA_red}, being time-independent, reduces to $\pd_\zeta\hat{\op}(s,\zeta\in \{ 0,1 \})=\mp H_1/s$. These \bcs also fix the prefactor of the solution in \cref{eq_linB_lapl_sol}.
}
\begin{equation} \hat\op(s,\zeta)/H_1 = \frac{\tscal - \lambda_+^2}{s \left( e^{\lambda_-}-1 \right) \lambda_- (\lambda_-^2 - \lambda_+^2)} \left[ e^{\zeta\lambda_-} + e^{(1-\zeta)\lambda_-} \right]  +
\frac{\tscal - \lambda_-^2}{s \left( e^{\lambda_+}-1 \right) \lambda_+ (\lambda_+^2 - \lambda_-^2)} \left[ e^{\zeta\lambda_+} + e^{(1-\zeta)\lambda_+} \right]
\label{eq_linB_lapl_sol}\end{equation}
as the solution of \cref{eq_linB_lapl_flat}. 
Since \cref{eq_linB_lapl_sol} has only a simple pole in $s$, at $s=0$, the long-time limit follows from \cref{eq_linB_lapl_sol} as \cite{poularikas_transforms_2010}
\beq \op\eq(\zeta) \equiv \op(\time\to\infty,\zeta) 
= \lim_{s\to 0^+} s\hat\op(s,\zeta) 
= \frac{H_1}{\tscal}  \left(\sqrt{\tscal} \frac{ \cosh\left[\left(1/2-\zeta\right)\sqrt{\tscal}\,\right]}{\sinh\left(\sqrt{\tscal}/2\right)} -2 \right)
\label{eq_eqprof_lin}\eeq 
corresponding, in unrescaled quantities, to
\beq \phi\eq(z) 
= \frac{h_1}{L\tLG}  \left(L\sqrt{\tLG} \frac{\cosh\left[\left(L/2-z\right)\sqrt{\tLG}\,\right]}{\sinh\left(L\sqrt{\tLG}/2\right)} -2 \right).
\label{eq_eqprof_lin_bare}\eeq 
These expressions agree with the ones obtained by minimizing the equilibrium free energy in \cref{eq_freeEn} under the constraint of vanishing total mass, $\int_0^L\d z\,\phi\eq(\zeta)=0$ \cite{gross_critical_2016}.
For later use, we determine $\op\eq$ in a few particular limits.  
At criticality (i.e., $\tscal=0$), \cref{eq_eqprof_lin} reduces to
\beq \op\eq(\zeta)\big|_{\tscal = 0} =  H_1 \left(\frac{1}{6} - \zeta + \zeta^2\right),
\label{eq_eqprof_lin_Tc}\eeq 
and, for general $\tscal\geq 0$ and at $\zeta=0$, to
\beq \op\eq(\zeta=0)\big|_{\tscal\geq 0} = H_1 \left[ \frac{\coth(\sqrt{\tscal}/2)}{\sqrt{\tscal}} - \frac{2}{\tscal}\right].
\label{eq_eqprof_lin_wall}\eeq
Asymptotically for $\tscal\to \infty$, instead, the OP profile behaves as
\beq \op\eq(\zeta)\big|_{\tscal\gg 1} \simeq \frac{H_1}{\sqrt{\tscal}} \left( e^{-\zeta\sqrt{\tscal}} + e^{-(1-\zeta)\sqrt{\tscal}}\right).
\label{eq_eqprof_largeT}
\eeq
We remark that, within linear MFT, the strength $H_1$ of the surface field appears as an overall prefactor in the expression for the OP profile. Within nonlinear MFT, the limit $H_1\to\infty$, corresponding to the fixed-point of the critical adsorption universality class \cite{diehl_field-theoretical_1986}, is ill-defined in the presence of the global OP conservation (see Ref.\ \cite{gross_critical_2016}). In fact, studying the limit $H_1\to\infty$ requires to include thermal fluctuations, which is beyond the scope of the present study.

Using \cref{eq_eqprof_lin}, the solution for the OP profile can be conveniently written as
\beq \hat\op(s,\zeta) = \frac{\op\eq(\zeta)}{s} + \hat\psi(s, \zeta),
\label{eq_linB_lapl_sol_split}\eeq 
where the first term describes the equilibrium profile and $\hat\psi$ encodes the relaxation towards it.
The function $\hat\psi(s,\zeta)$ has a regular expansion around $s=0$ in terms of positive integer powers of $s$ and, therefore, it is defined over the whole complex plane without a branch cut. 
The time dependence of $\op$ follows from the inverse Laplace transform of $\hat\op$, i.e.,
\beq \op(\time,\zeta) = \frac{1}{2\pi \im} \int_\Ccal \d s\, \hat\op(s,\zeta) e^{s \time},
\label{eq_inv_lapl}\eeq 
where the contour $\Ccal$ runs parallel to the imaginary axis and to the right of the pole of $\hat\op$ at $s=0$.
Inserting \cref{eq_linB_lapl_sol_split} into \cref{eq_inv_lapl}, the contribution from the pole at $s=0$ renders $\op\eq(\zeta)$ as the residue.
The contribution involving $\hat\psi$ reduces to an inverse Fourier transform since the contour $\Ccal$ can be shifted onto the imaginary axis.
One accordingly obtains
\beq \op(\time,\zeta) = \op\eq(\zeta) + \frac{1}{2\pi} \int_{-\infty}^\infty \d v\, \hat\psi(\im v,\zeta) e^{\im v \time}
\label{eq_inv_lapl_four}\eeq 
where the second term on the r.h.s.\ must vanish as $\time\to\infty$.

Closed analytical expressions for the Laplace inversion in \cref{eq_inv_lapl} can be obtained in certain asymptotic limits, which are discussed in the following sections.
In the general case, the Laplace inversion has to be computed numerically. 
An efficient and robust approach, which we use here, is provided by the Talbot method \cite{abate_multi-precision_2004}. 

\subsection{Quench to the critical point ($\tscal=0$)}
\label{sec_quench_crit}

\begin{figure}[t!]\centering
    \subfigure[]{\includegraphics[width=0.41\linewidth]{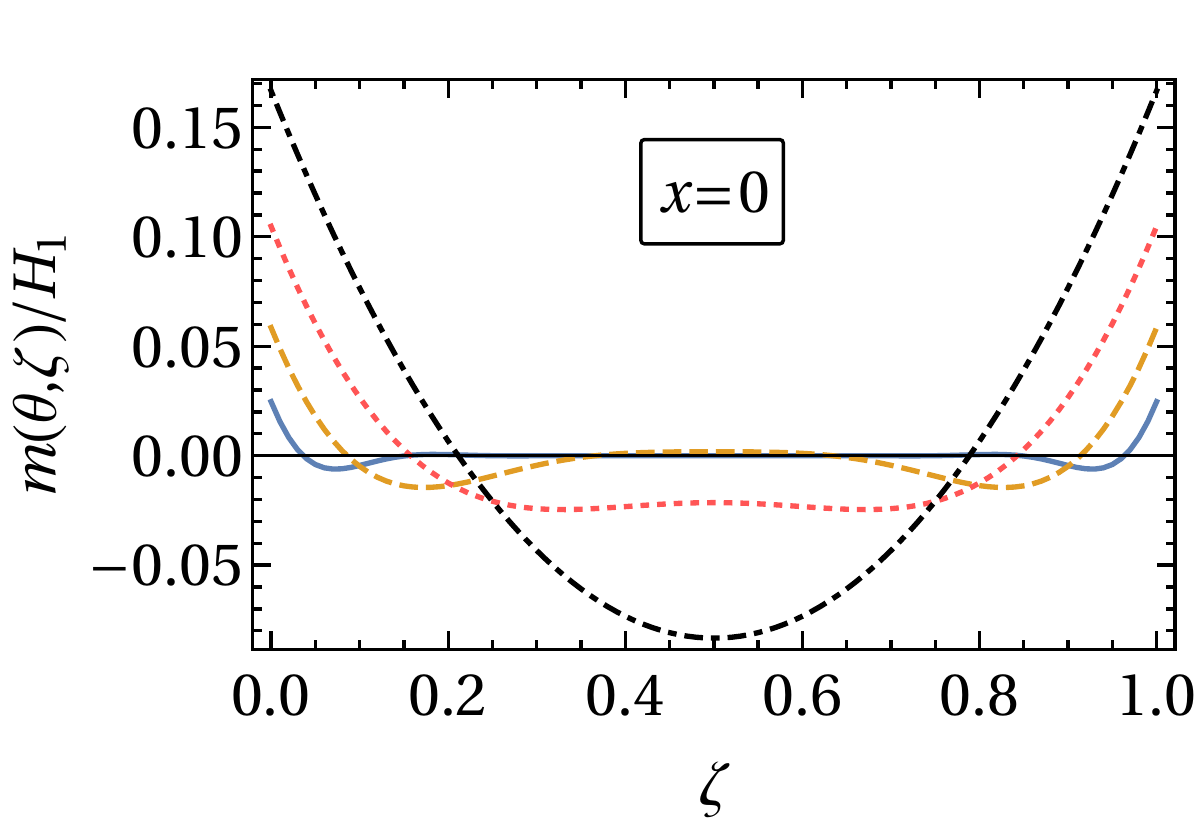} \label{fig_prof_evol_quench_Tc} }\qquad
    \subfigure[]{\includegraphics[width=0.41\linewidth]{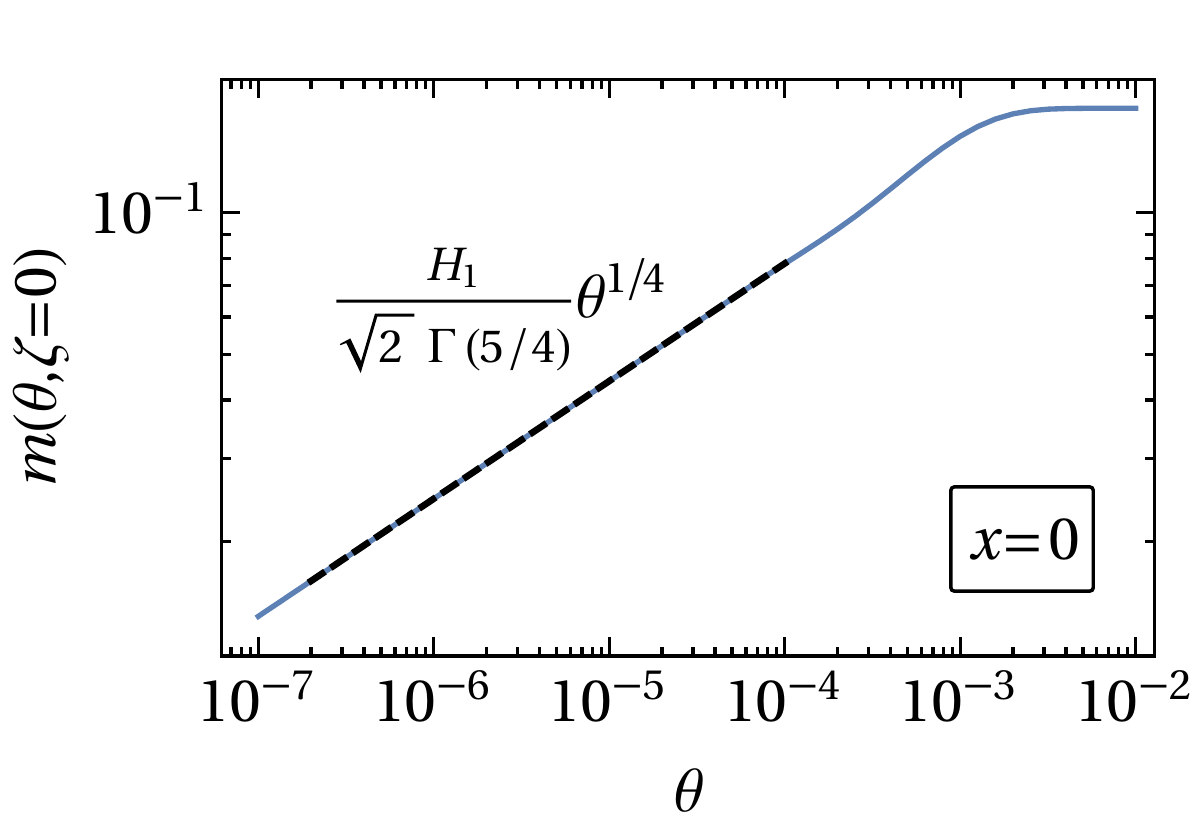} \label{fig_OPwall_small_t_Tc}} 
    \subfigure[]{\includegraphics[width=0.41\linewidth]{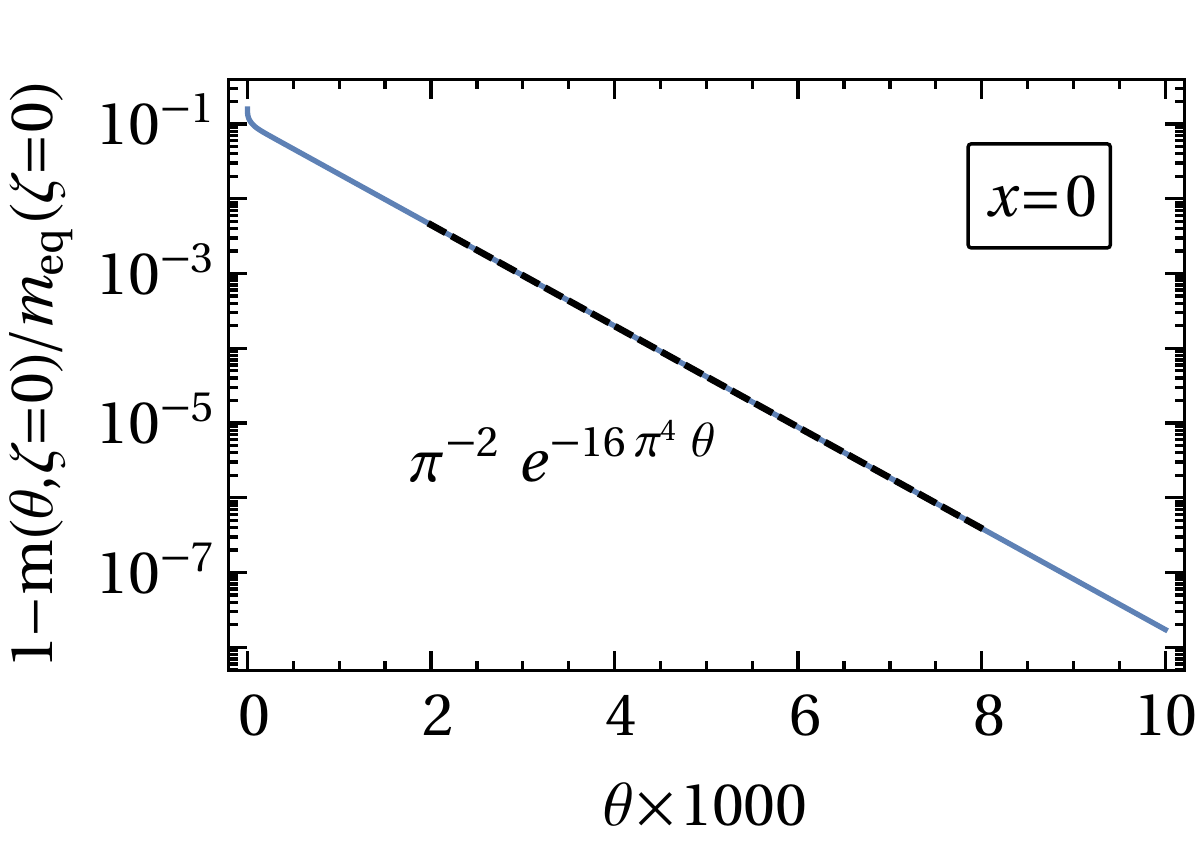} \label{fig_OPwall_large_t_Tc}} \qquad
    \subfigure[]{\includegraphics[width=0.415\linewidth]{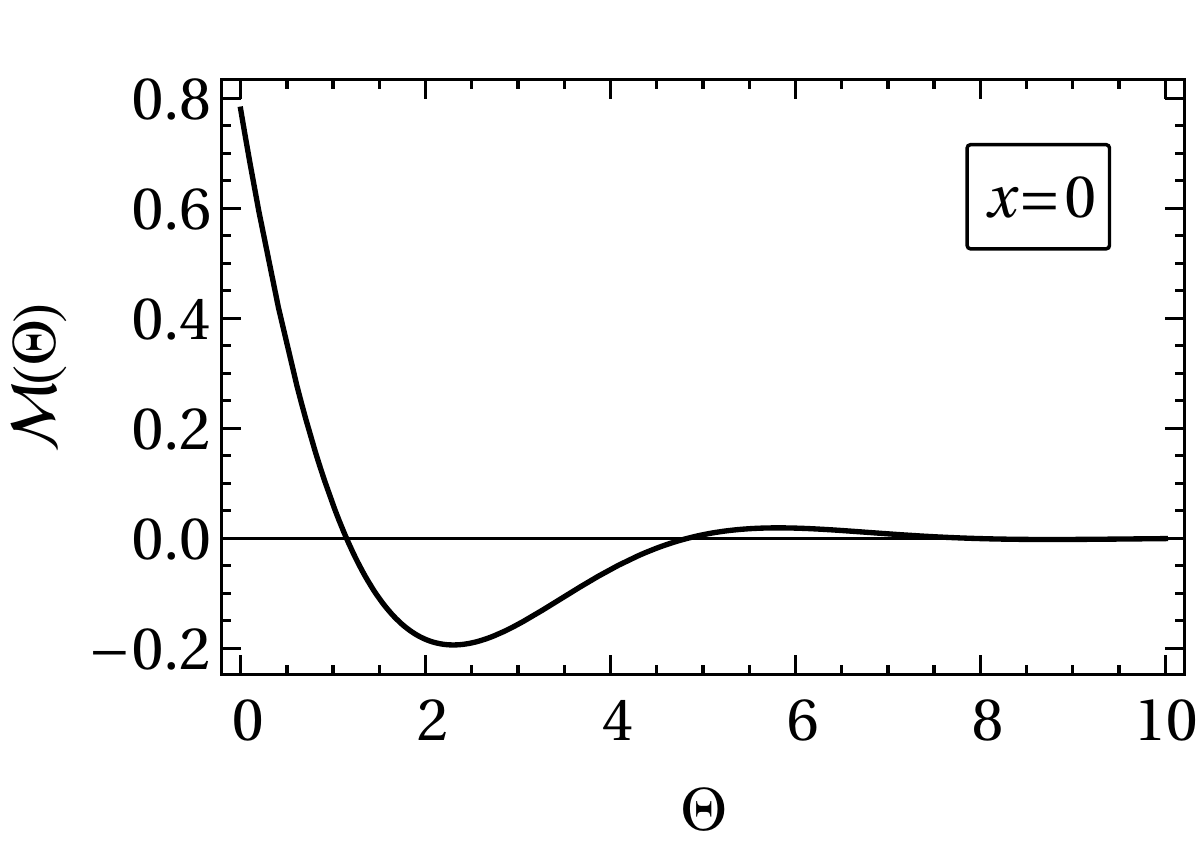} \label{fig_prof_scalf_shortt_Tc} }
    \subfigure[]{\includegraphics[width=0.4\linewidth]{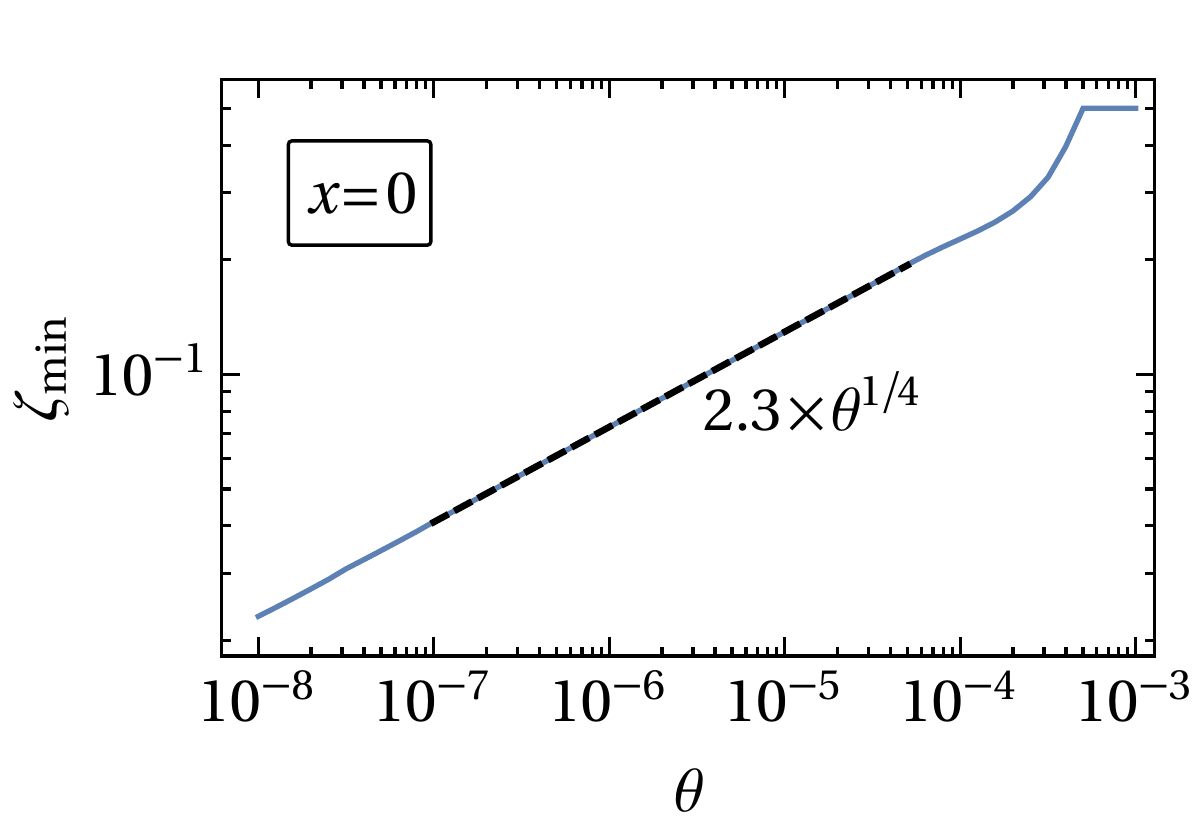} \label{fig_minpos_evol_Tc}}
    \caption{Time-evolution of the OP profile $m(\time,\zeta)$, subject to the linearized model B dynamics according to \cref{eq_modelB_resc_lin,eq_bcs_CA_red,eq_bcs_noflx_red} and to the initial condition \cref{eq_prof_init}, at criticality $\tscal=0$ [see \cref{eq_tau_resc}]. (a) Profile shape for times $\time=10^{-6}, 3\times 10^{-5}, 3\times 10^{-4}, \infty$ corresponding to the solid (blue), dashed (orange), dotted (red), and dashed-dotted (black) curves, respectively. For $\time\gtrsim 10^{-2}$, the OP has essentially reached equilibrium. (b) OP at the wall at early times $\time\lesssim \time^*_e\simeq 10^{-4}$ [see \cref{eq_time_early_crit}]. The dashed (black) line indicates the asymptotic result in \cref{eq_linBTc_earlytime_wall}. (c) OP at the wall at late times $\time\gtrsim \time^*_e$. The dashed (black) line indicates the asymptotic result in \cref{eq_linBTc_latetime_wall}. (d) Short-time scaling function of the profile, obtained asymptotically for $\zeta\to 0$ and for short times  $\time\to 0$, as function of the scaling variable $\Theta=\zeta/\time^{1/4}$ [\cref{eq_linB_profTc_scalf}]. (e) Position $\zeta\st{min}$ of the global minimum of the OP profile for $0\leq \zeta< 1/2$ as a function of the rescaled time $\time$, as determined numerically via the Laplace inversion of \cref{eq_linBTc_lapl_sol}. The dashed (black) line represents the prediction given in \cref{eq_linB_wall_earlyScal}.}
    \label{fig_quench_Tc}
\end{figure}

For $\tscal=0$, \cref{eq_lambda_pm} reduces to
\beq \lambda_\pm = \frac{1}{\sqrt{2}} (\pm 1 + \im) s^{1/4}
\label{eq_lambda_pm_crit}\eeq 
and the expression for the profile in Laplace space given in \cref{eq_linB_lapl_sol} becomes
\beq \hat\op(s,\zeta) = \frac{H_1}{2\omega^{5}} \big[ \cos(\omega \zeta)\cot(\omega/2) - \cosh(\omega \zeta)\coth(\omega/2) + \sin(\omega \zeta) + \sinh(\omega \zeta) \big]
\label{eq_linBTc_lapl_sol}\eeq 
where we introduced 
\beq \omega\equiv (-s)^{1/4} = \lambda_+ =  s^{1/4}\exp(\im\pi/4).
\label{eq_omega}\eeq 
For definiteness, we consider $(-1)^{1/4} = e^{\im \pi/4}$ as the principal branch of the complex root.
Since all physical quantities considered here are real-valued, they are not affected by this particular choice. 
\Cref{fig_prof_evol_quench_Tc} illustrates the time evolution of the scaled profile $\op(\time,\zeta)$ for the linearized model B and for $\tscal=0$ as determined by the numerical Laplace inversion of \cref{eq_linBTc_lapl_sol}.
The profile starts from a flat configuration [\cref{eq_prof_init}, not shown] and evolves towards the equilibrium solution given in \cref{eq_eqprof_lin_Tc} (dashed-dotted curve).
We proceed with an analysis of the characteristic spatial and temporal scaling behavior of the profile.

\subsubsection{Order parameter at the boundary: $\op(\time,\zeta=0)$}
\label{sec_quench_crit_bound}

In order to determine the time evolution of the profile \emph{at} the boundary, we evaluate \cref{eq_linBTc_lapl_sol} at $\zeta=0$:
\beq \hat\op(s,\zeta=0) = \frac{H_1}{2\omega^5} \left[ \cot(\omega/2) - \coth(\omega/2) \right].
\label{eq_laplsol_wall}\eeq
In order to determine the Laplace inversion of this expression, we expand the functions $\cot$ and $\coth$ in terms of simple fractions (see \S 1.42 in Ref.\ \cite{gradshteyn_table_2014}), resulting in
\beq \hat\op(s,\zeta=0) = \frac{H_1}{\omega^4} \sum_{k=1}^\infty \frac{16\pi^2 k^2}{\omega^4 - 16\pi^4 k^4}
\eeq 
and finally 
\beq \op(\time,\zeta=0) = H_1 \left[ \frac{1}{6} - \sum_{k=1}^\infty \frac{\exp(-16 \pi^4 k^4 \time)}{\pi^2 k^2} \right],
\label{eq_linBTc_lapl_inv}\eeq 
where we used $\sum_{k=1}^\infty k^{-2} = \pi^2/6$.
The first term on the r.h.s.\ can be identified with $\op\eq(\zeta=0)$ [see \cref{eq_eqprof_lin_Tc}].
We note that the expression in \cref{eq_linBTc_lapl_inv} occurs in similar form in various contexts and its asymptotic behavior has been analyzed previously (see Ref.\ \cite{gross_first-passage_2018} and references therein).
For times $\time\gg \time^*_e$, with 
\beq \time^*_e \simeq \frac{1}{16\pi^4}\sim \Ocal(10^{-4}),
\label{eq_time_early_crit}\eeq 
the term for $k=1$ in \cref{eq_linBTc_lapl_inv} dominates the sum, implying that
\beq \op(\time\gg 1,\zeta=0) \simeq \op\eq(0) - \frac{1}{6\pi^2} H_1 e^{-16\pi^4 \time}
\label{eq_linBTc_latetime_wall}\eeq 
provides the late-time asymptotic behavior of the OP.
The asymptotic behavior of \cref{eq_linBTc_lapl_inv} for $\time\to 0$ can be obtained by replacing the sum by an integral, i.e.,
\beq \op(\time\to 0,\zeta=0) \simeq H_1\int_0^\infty \d k \frac{1-\exp(-16\pi^4 k^4\time)}{\pi^2 k^2} = \time^{1/4} \frac{H_1}{2\pi}\int_{0}^\infty \d p \frac{1-e^{-p}}{p^{5/4}}. 
\label{eq_linBTc_wall_asympt_int}\eeq 
Numerical analysis shows that this approximation is reliable up to times $\time\simeq \time^*_e$, such that the asymptotic early-time behavior of the OP at the wall results, upon evaluating the integral in  \cref{eq_linBTc_wall_asympt_int}, as
\beq \op(\time,\zeta=0) \simeq \frac{H_1}{\sqrt{2}\, \Gamma(5/4)} \time^{1/4},\qquad \time\lesssim \time^*_e .
\label{eq_linBTc_earlytime_wall}\eeq 
The early-time and late-time behavior of the OP at the wall for $\tscal=0$ are illustrated in \cref{fig_quench_Tc}(b) and (c), respectively, where we find excellent agreement between the numerical Laplace inversion of \cref{eq_laplsol_wall} (solid curves) and the asymptotic results in \cref{eq_linBTc_latetime_wall,eq_linBTc_earlytime_wall} (dashed lines).

\subsubsection{Short-time scaling function of the profile}

Next we consider the full spatio-temporal evolution of the profile [\cref{eq_linBTc_lapl_sol}] at \emph{short} times near one wall, i.e., for $\time\to 0$ and $0<\zeta\ll 1/2$. 
Inserting \cref{eq_linBTc_lapl_sol} into \cref{eq_inv_lapl} and changing the integration variable to $\sigma = s \time$ yields 
\beq \op(\time,\zeta) = \time^{1/4} \frac{H_1}{4\pi \im} \int_\Ccal \d\sigma\, (-\sigma)^{-5/4} \chi_0\left((-\sigma/\time)^{1/4}, \zeta\right) e^{\sigma}
\label{eq_laplsol_wall_inv0}\eeq 
with $\chi_0(\kappa,\zeta) \equiv \cos(\kappa \zeta)\cot(\kappa/2)- \cosh(\kappa \zeta)\coth(\kappa/2) + \sin(\kappa \zeta) + \sinh(\kappa \zeta)$ and $\kappa\equiv (-\sigma/\time)^{1/4}$.
The integration path $\Ccal$ in the above integral is parametrized as $\sigma=\epsilon + \im \sigma''$ with $\sigma''\in \mathbb{R}$ and $\epsilon\in\mathbb{R}^+$ fixed in order to avoid the pole at $\sigma=0$ (with the final result being independent of the choice of $\epsilon$).
Decomposing $\kappa$ into its real and imaginary parts, $\kappa=\kappa'+\im \kappa''$, for $\time\to 0$ one has $|\kappa'|,|\kappa''|\to\infty$ and, correspondingly, in this limit, keeping the dominant terms of $\chi_0$, leads to \footnote{Upon performing the asymptotic expansion of the expression for $\chi_0$ given below \cref{eq_laplsol_wall_inv0}, one has to take into account that $\kappa\zeta\ll \kappa$ because we assume $\zeta\ll 1/2$.} 
\beq \chi_0(\kappa,\zeta) \simeq  - \exp(-\kappa \zeta) - \im \exp(\im \kappa \zeta).
\label{eq_chi0_asmpt}\eeq 
Accordingly, \cref{eq_laplsol_wall_inv0} takes the scaling form
\beq \op(\time \ll \time^*_e,\zeta\ll 1/2) = H_1 \time^{1/4} \Mcal(\zeta/\time^{1/4} =\Theta),
\label{eq_linB_wall_earlyScal}\eeq
with the scaling function
\beq \Mcal(\Theta) = \frac{1}{4\pi\im} \int_\Ccal \d\sigma\, (-\sigma)^{-5/4} \chi\left((-\sigma)^{1/4}\Theta \right) e^{\sigma}
\label{eq_linB_wall_scalf}\eeq 
where $\chi(\kappa) \equiv \chi_0(\kappa,1)$. 
A numerical analysis of the involved approximations reveals that \cref{eq_linB_wall_earlyScal} holds reliably if $(|\sigma| \zeta^4 / \time)^{1/4} \gtrsim \Ocal(10)$. Since $\zeta \ll 1/2$ and contributions from the integral in \cref{eq_linB_wall_scalf} are negligible for large $\sigma''$ \footnote{This is formally a consequence of the lemma of Riemann-Lebesgue \cite{bender_advanced_1999}}, it follows that \cref{eq_linB_wall_earlyScal} applies for times $\time\lesssim 10^{-4}\ll \time^*_e$, as indicated.
The inverse Laplace transform in \cref{eq_linB_wall_scalf} can be calculated analytically \footnote{The calculation is facilitated by using known Laplace transforms of hypergeometric functions (see \S 3.38.1 in Ref.\ \cite{prudnikov_direct_1992})}, yielding
\beq \Mcal(\Theta) =  \frac{2}{\pi} \Gamma(\sfrac{3}{4})\, {_1 F_3} \left(-\frac{1}{4}; \frac{1}{4}, \frac{1}{2}, \frac{3}{4}; \pfrac{\Theta}{4}^4\right) + \frac{1}{\pi} \Theta^2  \Gamma(\sfrac{5}{4})\, {_1 F_3}\left(\frac{1}{4}; \frac{3}{4}, \frac{5}{4}, \frac{3}{2}; \pfrac{\Theta}{4}^4\right) - \Theta
\label{eq_linB_profTc_scalf}\eeq 
where ${_1 F_3}$ is a standard hypergeometric function \cite{olver_nist_2010}.
For $\zeta=0$ and by using \cref{eq_linB_profTc_scalf}, \cref{eq_linB_wall_earlyScal} reduces to \cref{eq_linBTc_earlytime_wall}.
$\Mcal(\Theta)$, which is displayed in \cref{fig_prof_scalf_shortt_Tc}, represents the exact asymptotic short-time scaling function of the film profile in model B for $\zeta\ll 1/2$ and coincides with the corresponding expression obtained for a half-space (see \cref{app_halfspace}).

Since, according to \cref{eq_linB_wall_earlyScal}, $\op(\time,\zeta)/(H_1 \time^{1/4})$ is solely a function of the scaling variable $\Theta= \zeta/\time^{1/4}$, in the asymptotic short-time regime any spatial feature of the profile scales subdiffusively $\propto \time^{1/4}$ with time $\time$. This applies, in particular, to the position $\zeta\st{min}$ of the global minimum of $\op(\time,\zeta<1/2)$, for which one finds
\beq \zeta\st{min} \simeq 2.3\times \time^{1/4},\qquad \time\ll \time^*_e.
\label{eq_linB_wall_zmin}\eeq 
The prefactor follows from the numerically determined minimum of the scaling function $\Mcal$ in \cref{eq_linB_profTc_scalf}.
In terms of dimensional quantities, \cref{eq_linB_wall_zmin} corresponds to $z\st{min}\simeq 2.3\times (D t)^{1/4}$.
As shown in \cref{fig_quench_Tc}(e), the time evolution of the position of the global OP minimum, as obtained from the numerical Laplace inversion of \cref{eq_linBTc_lapl_sol}, is accurately captured by \cref{eq_linB_wall_zmin}.

\subsection{Quench to a supercritical temperature ($\tscal\gg 1$)}
\label{sec_quench_super}

\begin{figure}[t!]\centering
    \subfigure[]{\includegraphics[width=0.41\linewidth]{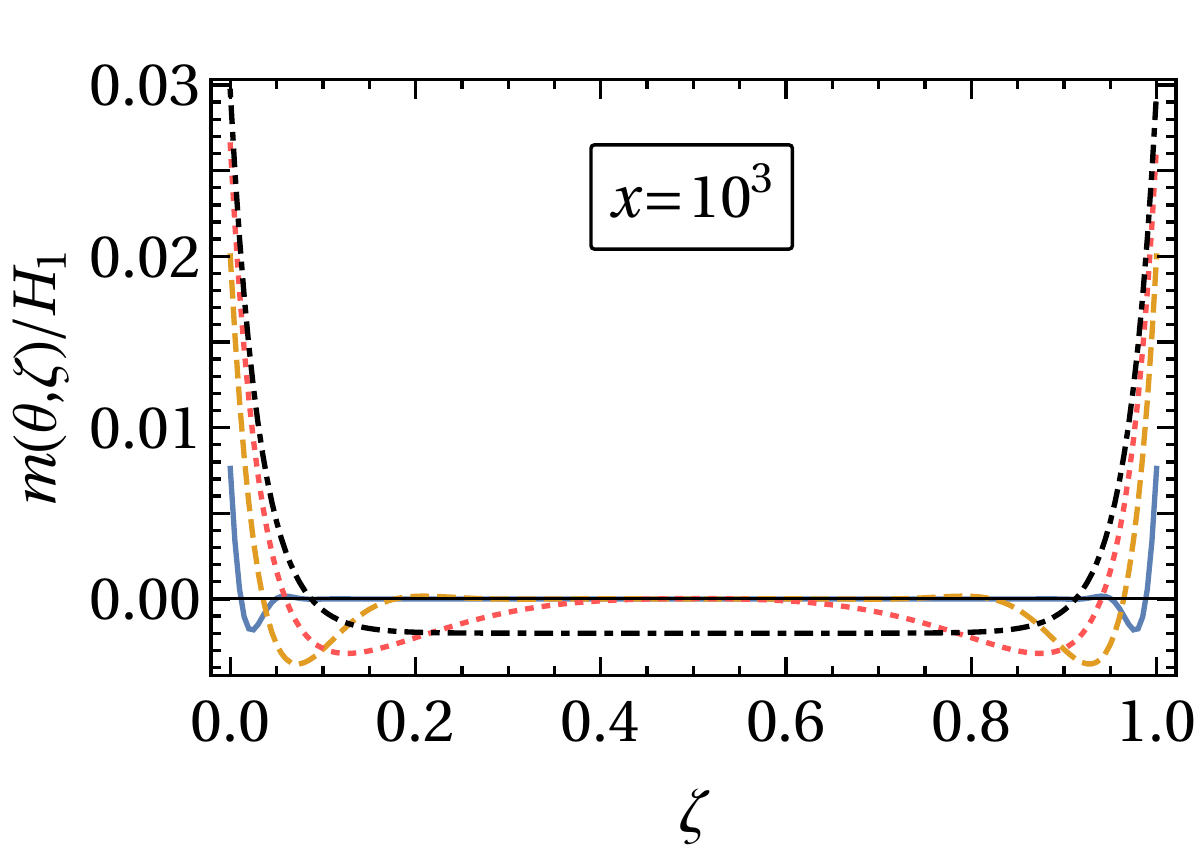} }\qquad
    \subfigure[]{\includegraphics[width=0.415\linewidth]{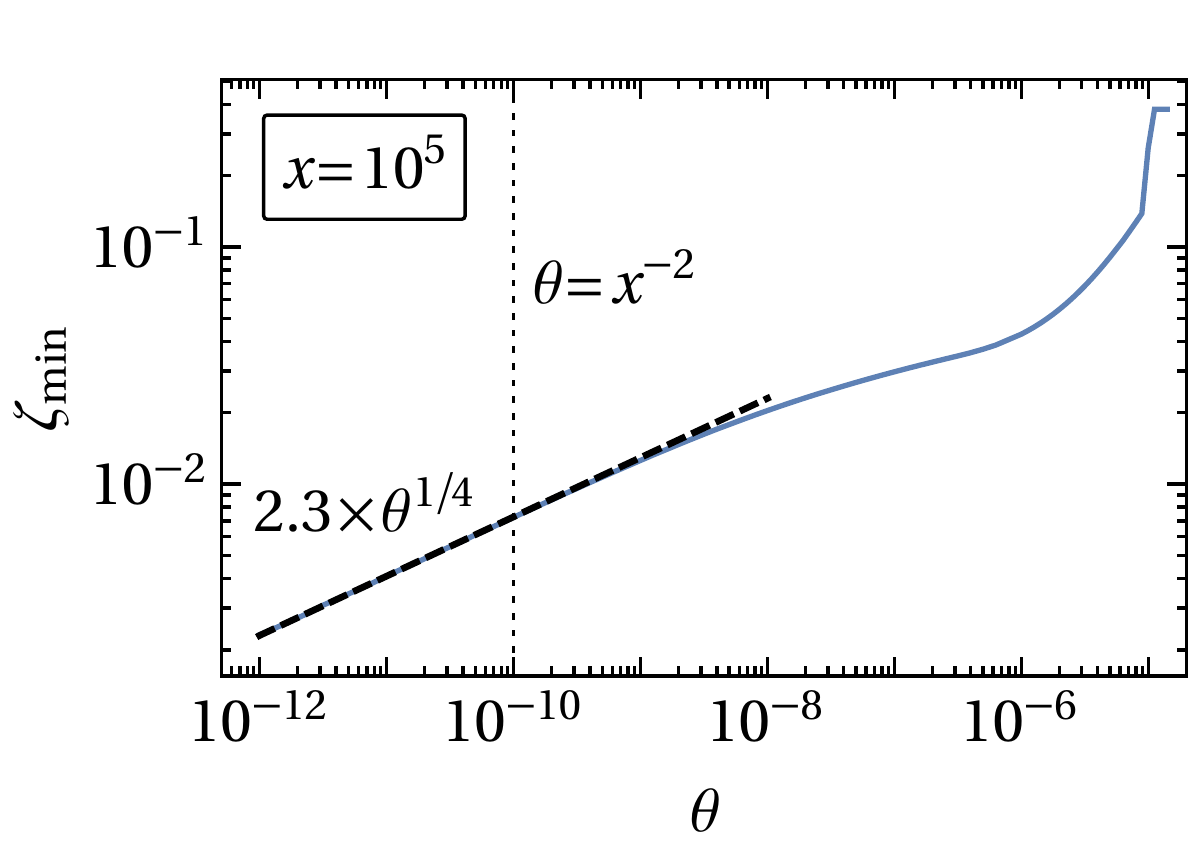} \label{fig_minpos_evol_abvTc}}
    \subfigure[]{\includegraphics[width=0.405\linewidth]{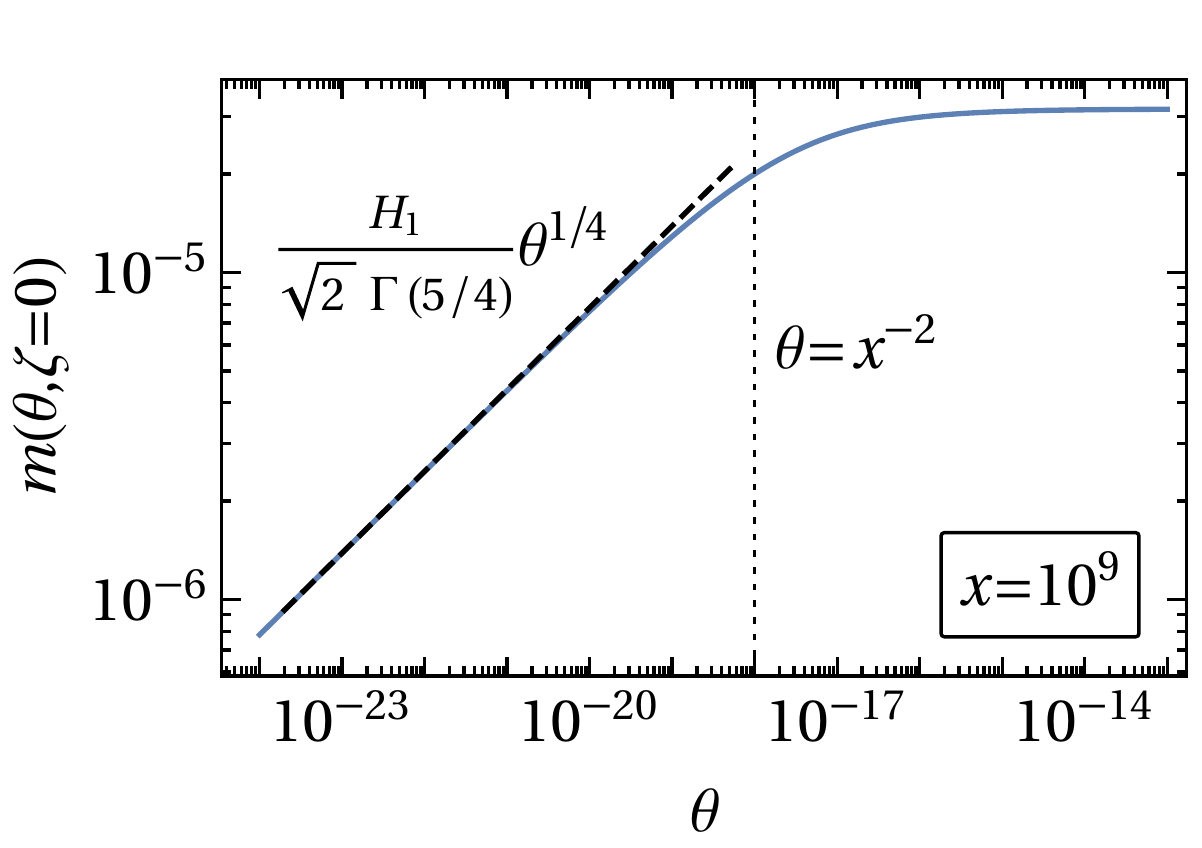} \label{fig_OPwall_small_t_abvTc} } \qquad
    \subfigure[]{\includegraphics[width=0.42\linewidth]{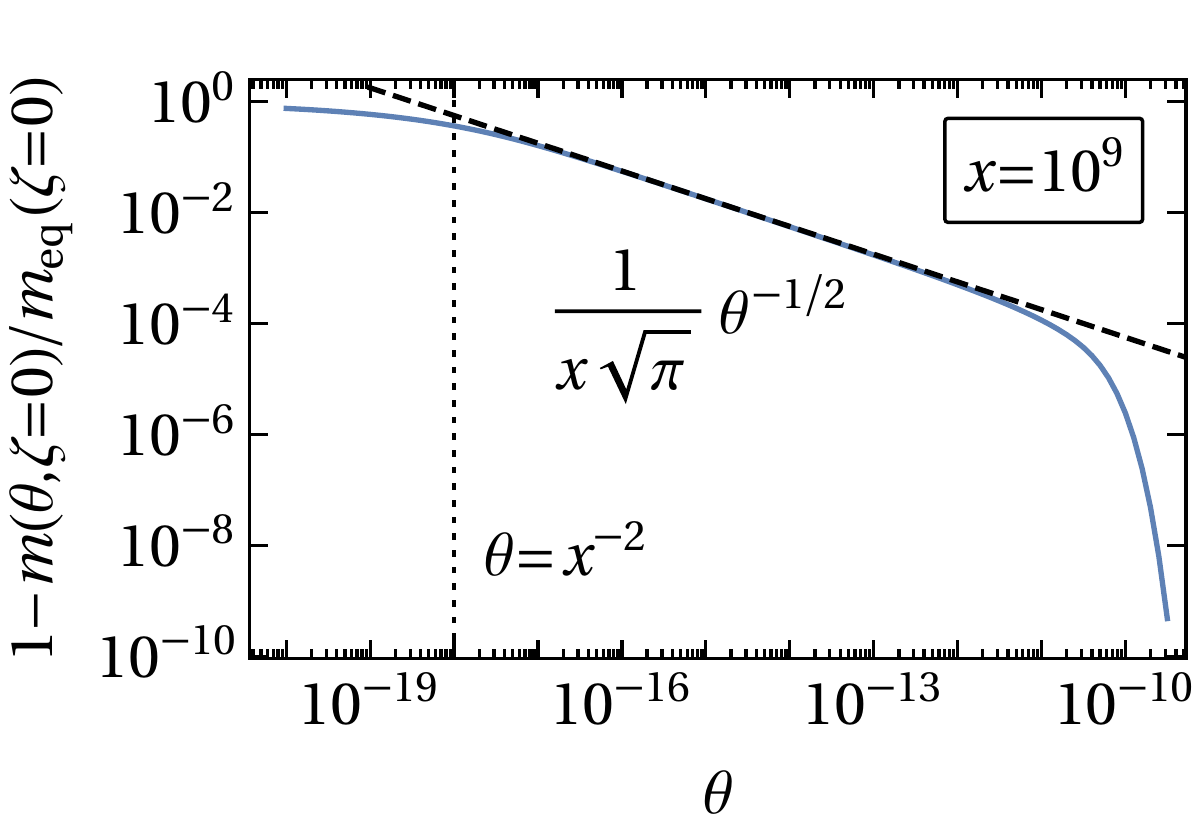} \label{fig_OPwall_interm_t_abvTc} } 
    \subfigure[]{\includegraphics[width=0.42\linewidth]{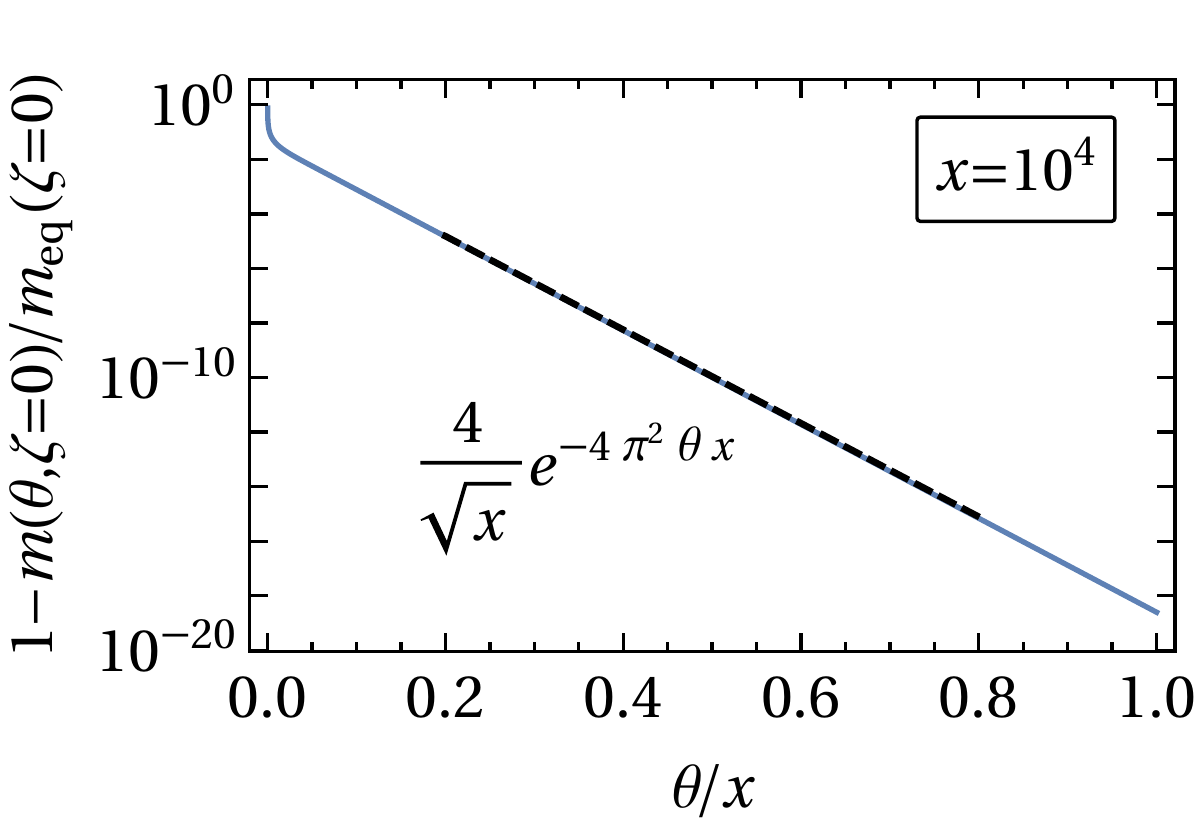} \label{fig_OPwall_large_t_abvTc}} 
    \caption{Time-evolution of an initially vanishing [\cref{eq_prof_init}] OP profile $\op(\time,\zeta)$, subject to the linearized model B dynamics [\cref{eq_modelB_resc_lin,eq_bcs_red}] in the supercritical regime $\tscal\gg 1$ [see \cref{eq_tau_resc}]. (a) $\op(\time,\zeta)$ as a function of $\zeta$ for rescaled times $\time=10^{-8}, 10^{-6}, 10^{-5}$, and $\infty$ corresponding to the solid, dashed, dotted, and dashed-dotted curves, respectively. (b) Position $\zeta\st{min}$ of the global minimum of the profile (within the left half $0\leq \zeta <1/2$ of the film) as a function of the rescaled time $\time$. The dashed line indicates the prediction of \cref{eq_linB_wall_earlyScal}, where the proportionality factor resulting from a fit is $\approx 2.3$ [\cref{eq_linB_wall_zmin}]. (c) OP $\op(\time,\zeta=0)$ at the wall at early times $\time\ll \tscal^{-2}$. The dashed line indicates the asymptotic prediction given in \cref{eq_linBTc_earlytime_wall}. (d) OP $\op(\time,\zeta=0)$ at the wall at intermediate times, i.e., for $\tscal^{-2}\ll \time \ll \tscal^{-1}$. The dashed line represents the intermediate asymptotic law in \cref{eq_linB_wall_evol_interm}. (e) At late times $\time\gtrsim \tscal^{-1}$ the OP $\op(\time,\zeta=0)$ saturates exponentially. The dashed line indicates the asymptotic prediction given in \cref{eq_linB_wall_evol_late}. The dotted vertical lines in panels (b), (c), and (d) mark the approximate boundary between the early- and intermediate asymptotic regimes [see the discussion in \cref{sec_quench_super} as well as \cref{eq_linB_wall_evol} below]. The specific values of $\tscal$ in the various panels are chosen for illustrative purposes, as being representative of the various behaviors of the system.}
    \label{fig_quench_abvTc}
\end{figure}

Here, we focus on the case of large reduced temperatures, $\tscal\gg 1$, and study the associated asymptotic behavior of the OP dynamics, which is expected to differ from the one discussed in the preceding section.
In \cref{fig_quench_abvTc}(a), the time evolution of the profile for large $\tscal$ is illustrated, based on the numerical Laplace inversion of \cref{eq_linB_lapl_sol}.
In order to proceed with the asymptotic analysis, we note that the time dependence of $\op(\time,\zeta)$ is essentially encoded in the dependence of $\lambda_\pm$ on $s$ [\cref{eq_lambda_pm}] and that, according to \cref{eq_inv_lapl}, the dominant contribution to $\op(\time,\zeta)$ stems from values of $|s|\sim 1/\time$. 
For $\tscal \gg 1$, one thus infers from \cref{eq_lambda_pm} the occurrence of three characteristic regimes: (i) $|s|\gg \tscal^2$, (ii) $\tscal\ll |s| \ll \tscal^2$, and (iii) $|s|\ll \tscal$, which translate to an early-, intermediate-, and late-time asymptotic regime defined by (i) $\time\ll \tscal^{-2}$, (ii) $\tscal^{-2}\ll \time\ll \tscal^{-1}$, and (iii) $\time\gg \tscal^{-1}$, respectively.

\subsubsection{Early-time asymptotic regime}
The behavior of $\op(\time,\zeta=0)$ for $\time\to 0$ can be inferred in Laplace space from studying the limit $s\to\infty$ \footnote{The asymptotic behaviors of a function in real space and in Laplace space are inter-related by means of so-called Tauberian theorems (see \S XIII.5 in Ref.\ \cite{feller_introduction_1971})}. 
In this limit, \cref{eq_lambda_pm,eq_linB_lapl_sol} reduce to the expressions in \cref{eq_lambda_pm_crit,eq_linBTc_lapl_sol}, respectively.
This implies that the short-time properties of the profile for any large but finite $\tscal$ in fact obey the critical scaling discussed in \cref{sec_quench_crit}.
Accordingly, at early times, the OP $\op(\time,\zeta=0)$ at the wall increases as in \cref{eq_linBTc_earlytime_wall}, and the spatial behavior of the OP near the wall is described by \cref{eq_linB_wall_earlyScal,eq_linB_profTc_scalf}. 
This is confirmed by the plots in Figs.~\ref{fig_quench_abvTc}(b) and (c), where the position $\zeta\st{min}$ of the global minimum of the OP within the range $0\leq \zeta< 1/2$ and the time evolution of $\op(\time,\zeta=0)$, respectively, is illustrated for $\tscal\gg 1$. 
As discussed above, for $\tscal\gg 1$, the early-time regime (i) crosses over to the intermediate regime (ii) approximately at a time $\time \sim \tscal^{-2}$, which is indicated in \cref{fig_quench_abvTc} by dotted vertical lines.

\subsubsection{Intermediate asymptotic regime}
\label{sec_interm_asympt}

For large $\tscal\gg 1$, an intermediate asymptotic temporal regime is expected to arise for times $\time$ such that $\tscal^{-2}\ll \time \ll \tscal^{-1}$.
In order to determine the behavior of $\op(\time,\zeta=0)$ within this regime, we expand the inner square root in \cref{eq_lambda_pm} around $s=0$ up to leading order in $s/\tscal^2$, i.e., $\sqrt{\tscal^2-4s} \simeq \tscal - 2s/\tscal + \Ocal(s^2/\tscal^3)$, which gives
\beq \lambda_+ \simeq \sqrt{\tscal - s/\tscal} \simeq \sqrt{\tscal} \left(1 - \frac{s}{2\tscal^{2}} \right) ,\qquad \lambda_- \simeq \sqrt{s/\tscal} .
\label{eq_lambda_IA_exp}\eeq 
Inserting \cref{eq_lambda_IA_exp} into \cref{eq_linB_lapl_sol} and keeping only the dominant terms for $\tscal\gg 1$ with $\tscal\ll |s|\ll \tscal^2$ yields, after Laplace inversion, the OP profile in the intermediate asymptotic regime:
\begin{equation} \op(\time,\zeta)\big|_{\tscal\gg 1} \simeq \frac{H_1}{\sqrt{\tscal}} \Bigg[ \exp(-\zeta \sqrt{\tscal}) - \frac{1}{\tscal\sqrt{\pi \time}}\exp\left(-\frac{\zeta^2}{4\time\tscal}\right)  + \exp(-(1-\zeta)\sqrt{\tscal}) - \frac{1}{\tscal \sqrt{\pi\time}} \exp\left(-\frac{(\zeta-1)^2}{4\time\tscal}\right) \Bigg] .
\label{eq_linB_prof_intAs}\end{equation}
The first and the third term on the r.h.s.\ render together the asymptotic equilibrium profile reported in \cref{eq_eqprof_largeT}.
Accordingly, the approach of the OP at the wall toward its long-time value $\op\eq(\zeta=0)$ is described by 
\beq \op\eq(\zeta=0) - \op(\time,\zeta=0) \simeq \frac{H_1}{\sqrt{\pi} \tscal^{3/2}} \time^{-1/2},
\label{eq_linB_wall_evol_interm}\eeq
which, as shown in \cref{fig_quench_abvTc}(d), accurately describes the numerical Laplace inversion of \cref{eq_linB_lapl_sol} within the intermediate asymptotic regime.
One recognizes the expression in \cref{eq_linB_prof_intAs} as the superposition of the two corresponding asymptotic profiles obtained in a half-space [see \cref{eqHS_linB_solSupc} in \cref{app_halfspace}].
Within the intermediate asymptotic regime, the position $\zeta\st{min}$ of the global minimum of the profile in, e.g., the left half of the film ($0\leq \zeta<1/2$), effectively follows, as function of time, a logarithmic behavior [see \cref{eqHS_linB_minScal}]:
\beq 
\zeta\st{min}\simeq \tscal^{-1/2}[\ln(\tscal^2\time)-1],\qquad \text{for}\qquad  \tscal^{-2}\ll \time \ll \tscal^{-1}.
\eeq 

\subsubsection{Late-time asymptotic regime}

The late-time asymptotic regime pertaining to the case $\tscal\gg 1$, arises for times $\time\gg \tscal^{-1}$, corresponding to $|s|\ll \tscal$ in Laplace space.
In order to determine the corresponding behavior of the OP, we proceed as in the preceding subsection and insert in \cref{eq_linB_lapl_sol} the expansion given in \cref{eq_lambda_IA_exp}, keeping only the most relevant terms for $|s|\ll \tscal$ and $\tscal\gg 1$.
In Laplace space, this way one obtains the asymptotic profile
\begin{multline} \hat\op(s\ll \tscal,\zeta)\big|_{\tscal \gg 1} \simeq \frac{2 }{s\sqrt{\tscal}} \exp\left(\frac{s}{4\tscal^{3/2}}-\frac{\sqrt{\tscal}}{2}\right) \cosh\left[\left(\onehalf - \zeta\right)\sqrt{\tscal}\right] \\ - \frac{1}{\sqrt{s}\tscal^{3/2}} \exp\left(\frac{s}{4\tscal^{3/2}}\right) \cosh\left[\left(\onehalf-\zeta\right)\sqrt{\frac{s}{\tscal}}\right]\big/ \sinh\left(\onehalf \sqrt{\frac{s}{\tscal}}\right).
\label{eq_late_prof_abvTc}\end{multline}
Note that, except from a pole at $s=0$, this expression has a regular Laurent expansion in terms of $s$, with no branch cut.
At the wall ($\zeta=0$), \cref{eq_late_prof_abvTc} reduces to $\hat\op(s,\zeta=0) = \exp(s/(4\tscal^{3/2})) \tscal^{-1/2} \left[1/s - \coth(\sqrt{s/\tscal}/2)/(\sqrt{s}\tscal) \right] $.
Using the series representation of $\coth$ in terms of simple fractions (see, e.g., \S 1.421 in Ref.\ \cite{gradshteyn_table_2014}), the Laplace inversion is obtained as
\beq \begin{split} \op(\time\gg 1/\tscal,\zeta=0)/H_1 &\simeq \frac{1}{\sqrt{\tscal}} -\frac{2}{\tscal} - \frac{4}{\tscal} \sum_{k=1}^\infty \exp\left[-4\pi^2 k^2 \left(\time \tscal + \frac{1}{4 \sqrt{\tscal}}\right)\right] \\
&= \frac{1}{\sqrt{\tscal}} - \frac{4}{\tscal}  \vartheta_3\left(0,\, \exp\left[-4\pi^2 \left(\time\tscal + \frac{1}{4 \sqrt{\tscal}}\right)\right]\right) ,
\end{split}\eeq 
where $\vartheta_3$ denotes the elliptic Jacobi theta function \cite{olver_nist_2010}.
In the limit $\time\to\infty$, the leading contribution to the sum is given by the term with $k=1$, such that
\beq \op(\time\to\infty,\zeta=0) \simeq \frac{H_1}{\sqrt{\tscal}} - \frac{4 H_1}{\tscal} \exp\left(-4\pi^2 \time\tscal\right),\qquad \text{for}\qquad  \tscal\gg 1.
\label{eq_linB_wall_evol_late}\eeq 
Accordingly, the equilibrium OP at the wall, $\op\eq(\zeta=0) \simeq H_1/\sqrt{\tscal}$ [see \cref{eq_eqprof_lin_wall}], is approached exponentially at late times.
As shown in \cref{fig_quench_abvTc}(e), \cref{eq_linB_wall_evol_late} accurately matches the behavior of the OP determined numerically from the exact expression in \cref{eq_linB_lapl_sol}.
A numerical analysis reveals that \cref{eq_linB_wall_evol_late} is, in fact, reliable for $\tscal\gtrsim 10^4$.

\subsection{Behavior of the second derivative of the profile}

\subsubsection{Critical quench ($\tscal=0$)}

For the purpose of analyzing the CCF (see \cref{sec_CCF} below), it is useful to determine also the behavior of the second derivative of the OP profile $\op(\time,\zeta)$ at the boundary $\zeta=0$ (or, equivalently, at $\zeta=1$).
Focusing first on a critical quench ($\tscal=0$), \cref{eq_linBTc_lapl_sol} yields 
\beq \pd_\zeta^2 \hat m(s,\zeta)\big|_{\zeta=0} = -\frac{H_1}{2\omega^3} \left[ \coth(\omega/2) + \cot(\omega/2) \right]
\eeq 
for the Laplace transform of $\op$.
Proceeding as in \cref{sec_quench_crit_bound}, one finds 
\beq \pd_\zeta^2 m(\time,\zeta)\big|_{\zeta=0} \simeq \frac{H_1}{\sqrt{2}\ \Gamma(3/4)}\time^{-1/4},\qquad \text{for}\qquad \time\lesssim \time^*_e,
\label{eq_linBTc_earlytime_wall_D2z}\eeq 
as the short-time asymptotic behavior \footnote{The limit $\time\to 0$ is singular because the initial condition [\cref{eq_prof_init}] is incompatible with the \bcs [\cref{eq_bcs_CA_red}].}. 
At late times $\time\gtrsim \time^*_e$, instead, $\pd_\zeta^2\op$ approaches the equilibrium value  [see \cref{eq_eqprof_lin_Tc}]
\beq \pd_\zeta^2\op\eq(0)\big|_{\tscal=0} = 2 H_1
\eeq 
exponentially.

\subsubsection{Off-critical quench ($\tscal\gg 1$)}

From \cref{eq_eqprof_lin} one obtains the late-time limit of $\pd_\zeta^2\op$ for $\tscal\gg 1$ as
\beq \pd_\zeta^2 \op\eq(0) \simeq H_1 \sqrt{\tscal} .
\eeq 
Performing an analysis analogous to the one in \cref{sec_quench_super} yields the following asymptotic behaviors of $\pd^2_\zeta\op$ for $\tscal\gg 1$ [49]:
\begin{subequations}
\label{eq_linB_wall_evol_Dz2}
  \begin{empheq}[left ={ \pd^2_\zeta\op(\time,0) \simeq \empheqlbrace }]{align}
    & \frac{H_1}{\sqrt{2}\ \Gamma(3/4)}\time^{-1/4},\qquad & \time\ll \tscal^{-2}, \label{eq_linB_wall_Dz2_early}\\
    & \pd^2_\zeta\op\eq(0)\left(1 + \frac{1}{2\sqrt{\pi} \tscal^3} \time^{-3/2}\right), & \tscal^{-2}\ll \time \ll \tscal^{-1}, \label{eq_linB_wall_Dz2_interm}\\
    & \pd^2_\zeta\op\eq(0)\left(1 + \frac{16\pi^2}{\tscal^{3/2}} e^{-4\pi^2 \tscal \time } \right), & \time\gg \tscal^{-1}, \label{eq_linB_wall_Dz2_late}
   \end{empheq}
\end{subequations}
where we note that $\pd^2_\zeta\op\eq(0)|_{\tscal\to\infty} \simeq H_1 \sqrt{\tscal}$.

\subsection{Summary}

We have shown that, for $\tscal \gtrsim 10^2$, the behavior of the OP \emph{at} the wall, $\op(\time,\zeta=0)$, exhibits three characteristic regimes (see \cref{sec_quench_super}):
\begin{subequations}
\label{eq_linB_wall_evol}
  \begin{empheq}[left ={ \op(\time,\zeta=0) \simeq \empheqlbrace }]{align}
    & \frac{H_1}{\sqrt{2}\, \Gamma(5/4)} \time^{1/4},\qquad & \time\ll \tscal^{-2}, \label{eq_linB_wall_sum_early}\\
    & \op\eq(0)\left(1 - \frac{1}{\sqrt{\pi} \tscal} \time^{-1/2}\right), & \tscal^{-2}\ll \time \ll \tscal^{-1}, \label{eq_linB_wall_sum_interm}\\
    & \op\eq(0)\left(1 - \frac{4}{\sqrt{\tscal}} e^{-4\pi^2 \tscal \time}\right), & \time\gg \tscal^{-1}, \label{eq_linB_wall_sum_late}
   \end{empheq}
\end{subequations}
where we have used the relationship $\op\eq(0)\simeq H_1/\sqrt{x}$ for $\tscal\to \infty$  [see \cref{eq_eqprof_lin_wall}].
For $\tscal\lesssim 10^2$, the intermediate asymptotic regime in \cref{eq_linB_wall_sum_interm} effectively disappears, leaving only the early- and late-time regimes, which are characteristic for a critical quench (see \cref{sec_quench_crit_bound}):
\begin{subequations}
\label{eq_linB_wall_evol_0x}
  \begin{empheq}[left ={ \op(\time, \zeta=0)\big|_{\tscal=0} \simeq \empheqlbrace }]{align}
    & \frac{H_1}{\sqrt{2}\, \Gamma(5/4)} \time^{1/4},\qquad & \time\ll \time^*_e, \label{eq_linB_wall_sum_early_0x}\\
    & \op\eq(0)\left(1 - \frac{1}{\pi^2} e^{-16\pi^4 \time}\right), & \time\gg \time^*_e, \label{eq_linB_wall_sum_late_0x}
   \end{empheq}
\end{subequations}
where $\time^*_e\simeq 10^{-4}$ [see \cref{eq_time_early_crit}].

At the time $\time^* \in \{ \time^*_e, \tscal^{-2} \}$, at which the crossover from the early-time growth [\cref{eq_linB_wall_sum_early}] to the saturation regime [\cref{eq_linB_wall_sum_interm,eq_linB_wall_sum_late}] occurs, one typically has $\op(\time^*,0)\simeq \op\eq(0)$, i.e., the OP is almost fully equilibrated.
Indeed, for $\tscal\gg 1$,  \cref{eq_eqprof_lin_wall} gives $\op\eq(0)\simeq H_1/\sqrt{x}$. Approximately at $\time\sim \tscal^{-2}$, i.e., at the end of the early-time regime, this value is reached by $\op(\time,\zeta=0)$ evolving according to \cref{eq_linB_wall_sum_early}. 
For $\tscal\ll 1$, correspondingly, the equilibrium value $\op\eq(0)\simeq H_1/6$ predicted by \cref{eq_eqprof_lin_Tc} is reached at a time $\time\simeq (\sqrt{2}\Gamma(5/4)/6)^{4}\simeq 2\times 10^{-3}$, which follows by equating $\op\eq(0)$ with $\op(\time,0)$ in \cref{eq_linB_wall_sum_early_0x} and which is consistent with the estimate of $\time^*_e$ in \cref{eq_time_early_crit}.

Notably, for large $\tscal>0$, at the wall the OP $\op$ attains equilibrium much \emph{earlier} than far from the wall. 
Indeed, at the end of the early-time regime, i.e., at a time $\time\simeq \time^*\simeq x^{-2}$, the OP minimum has propagated only a distance $\zeta\st{min}(\time^*)\sim \tscal^{-1/2}\ll 1/2$ into the bulk [see \cref{eq_linB_wall_zmin,fig_minpos_evol_abvTc}].
A numerical analysis of the full solution in \cref{eq_linB_lapl_sol} reveals that the two OP minima in the film (see \cref{fig_minpos_evol_abvTc}) merge in its middle around a characteristic time  
\beq \time\st{min}^* \sim \min(\tscal^{-1},\time^*_e).
\label{eq_min_merge_time}\eeq 
This time scale is close to that of the onset of the exponential saturation regime of the profile [see \cref{eq_linB_wall_sum_late,eq_linB_wall_sum_late_0x}].
This regime is characteristic for a film, whereas the profile in a half-space only shows the algebraic growth and saturation behaviors reported in \cref{eq_linB_wall_sum_early,eq_linB_wall_sum_interm} (see \cref{app_halfspace} as well as Ref.\ \cite{diehl_boundary_1992}).
Accordingly, within the linear model B and for $(++)$ \bcs, the time $\time^*\st{min}$ in \cref{eq_min_merge_time} can be understood as the typical time scale for the onset of the interaction between the two walls after a quench.

\section{Nonlinear quench dynamics}
\label{sec_nonlin_dyn}

\begin{figure}[t!]\centering
    \subfigure[]{\includegraphics[width=0.42\linewidth]{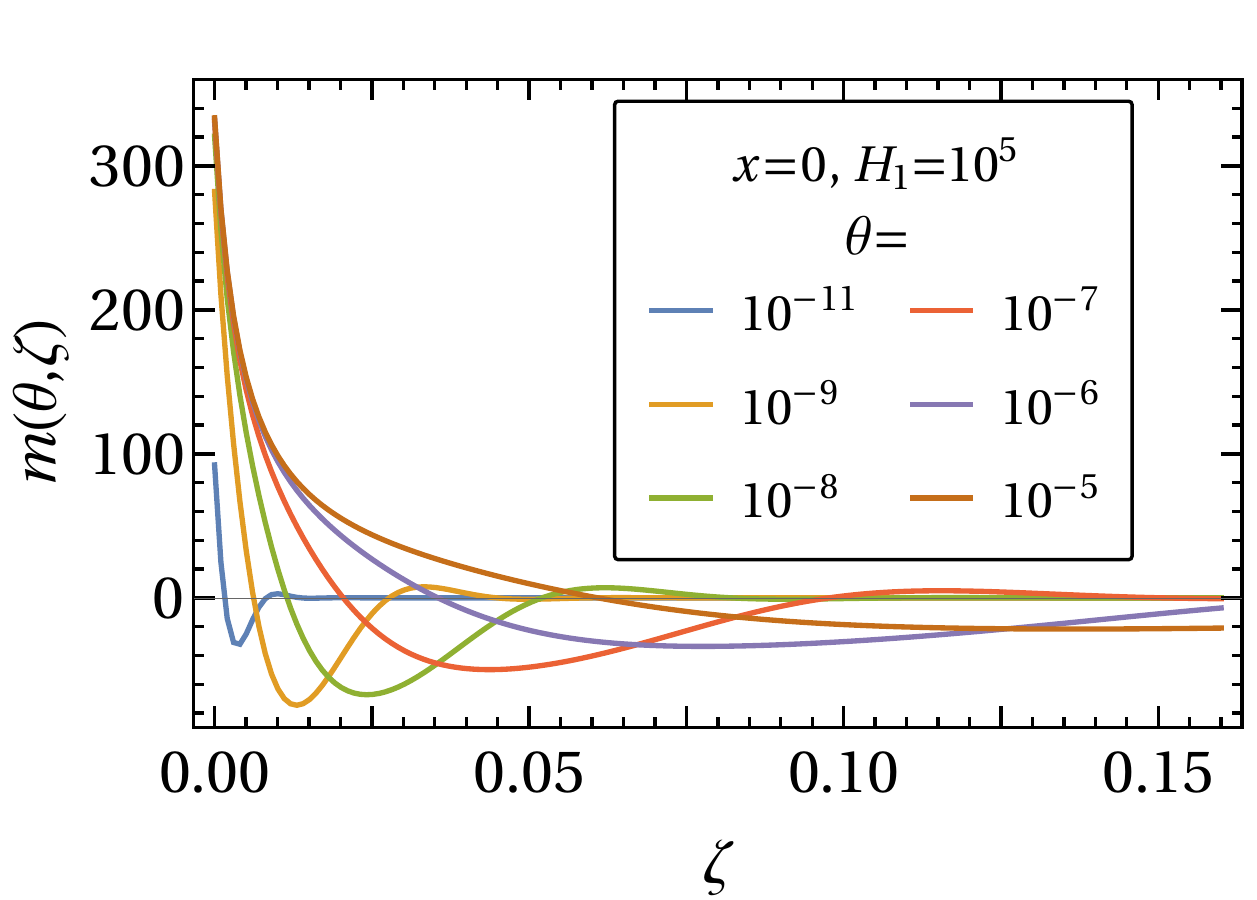} } \qquad
    \subfigure[]{\includegraphics[width=0.44\linewidth]{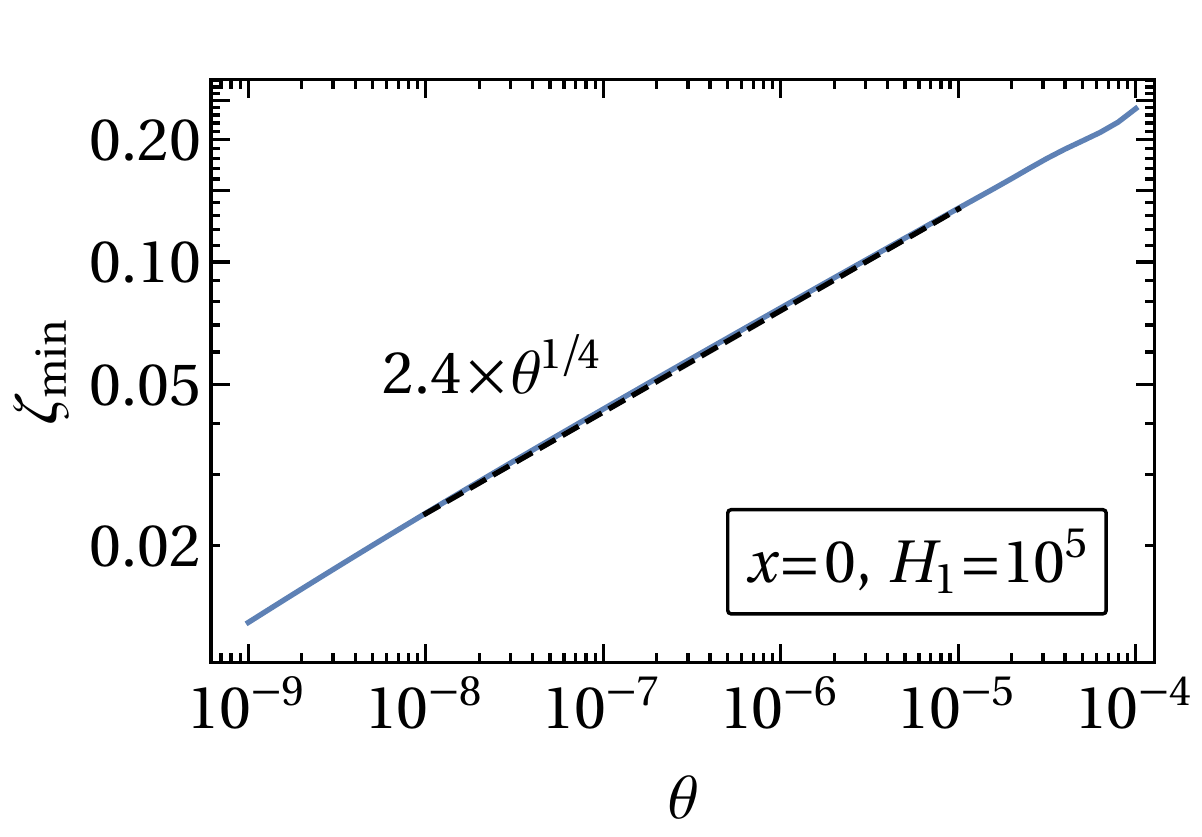} \label{fig_nlin_minPos_Tc} } 
    \subfigure[]{\includegraphics[width=0.44\linewidth]{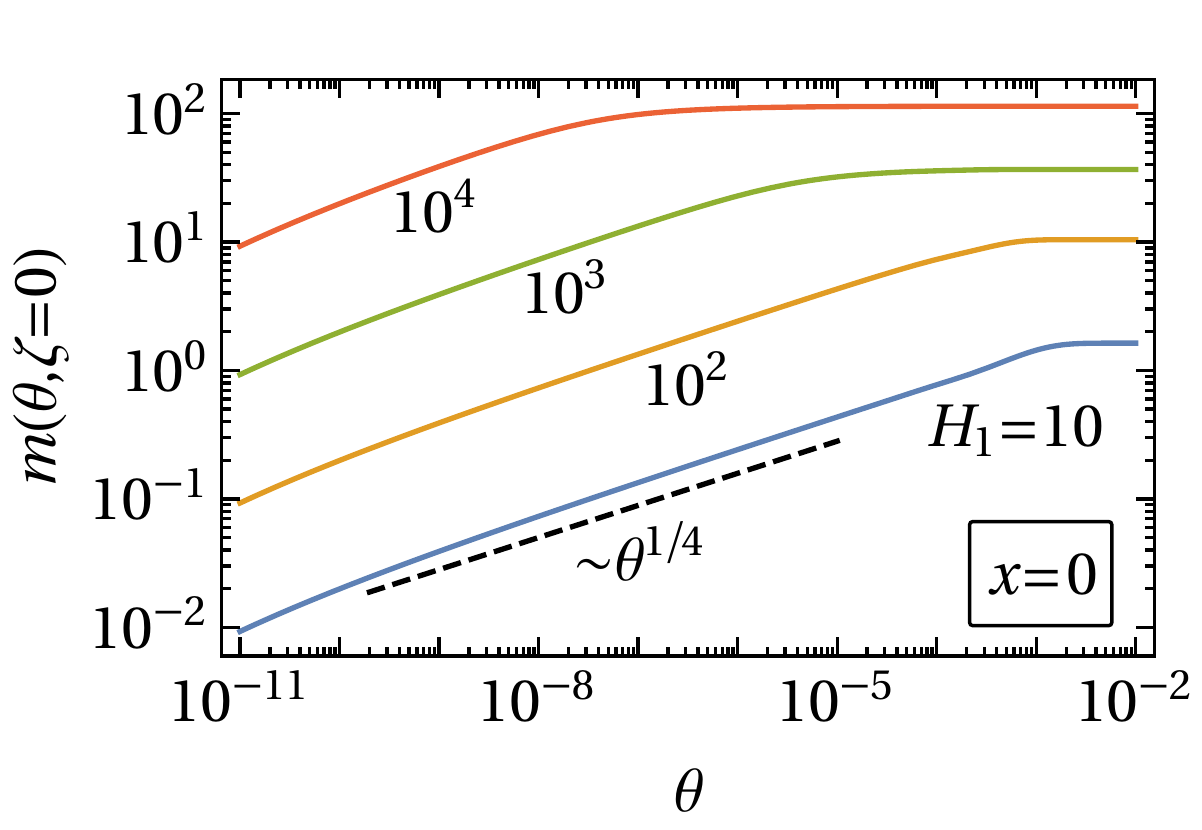} \label{fig_nlin_OPwall_small_t_Tc} } \qquad
    \subfigure[]{\includegraphics[width=0.43\linewidth]{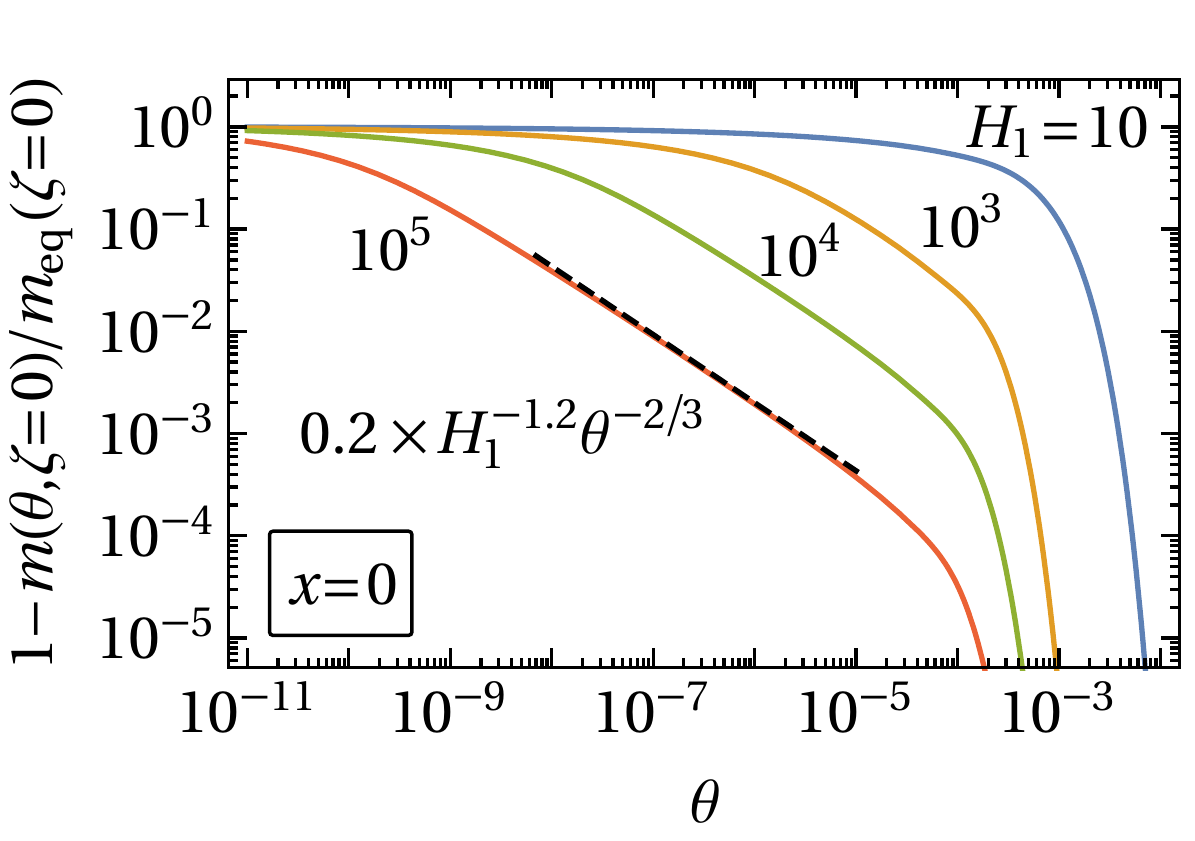} \label{fig_nlin_OPwall_interm_t_Tc} }
    \subfigure[]{\includegraphics[width=0.42\linewidth]{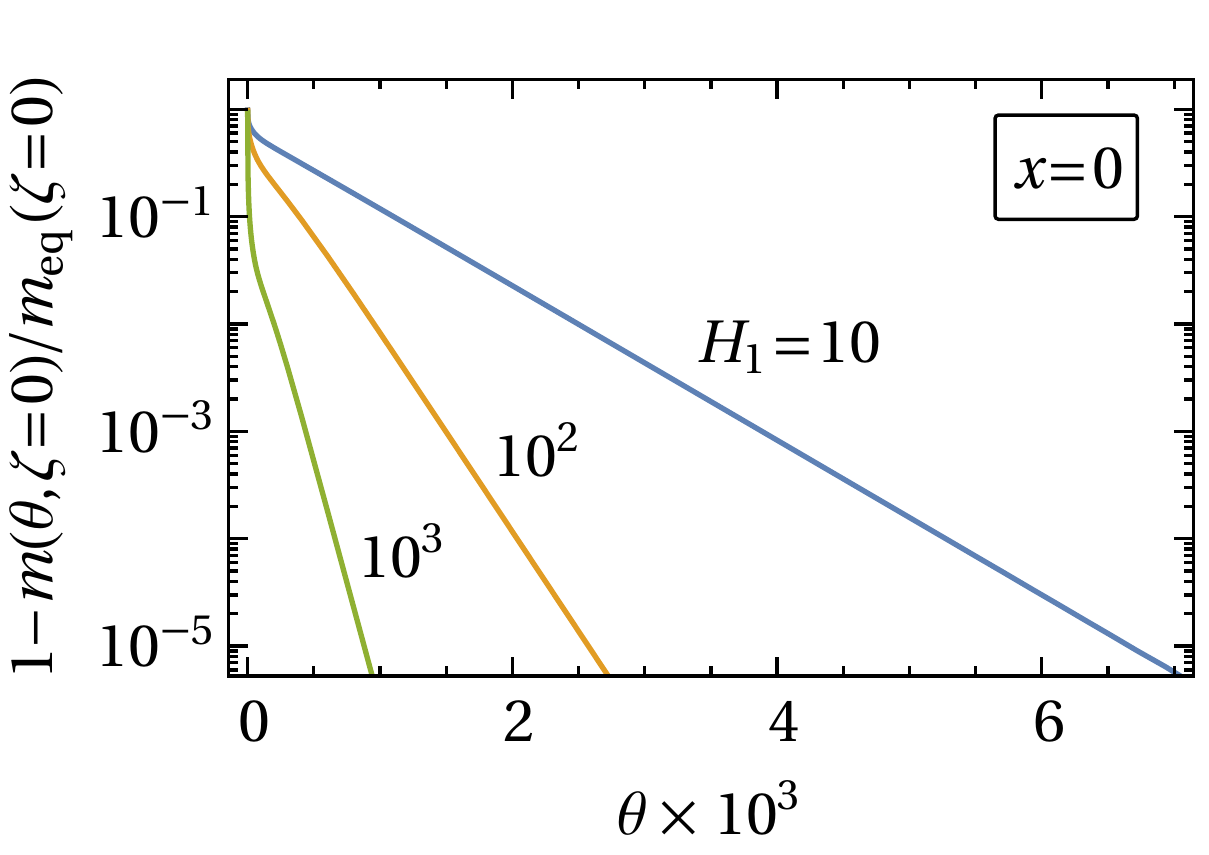} \label{fig_nlin_OPwall_large_t_Tc} } 
    \caption{Time evolution, within the nonlinear model B [\cref{eq_modelB_resc}], of the OP profile at criticality ($\tscal=0$) and for $H_1\gg 1$, starting from a flat, vanishing initial configuration [\cref{eq_prof_init}]. (a) Profile $\op(\time,\zeta)$ near one wall for various rescaled times $\time$. (b) Position of the global minimum of $\op(\time,\zeta)$ as a function of time $\time$; for all values of $H_1\gg 1$, $\zeta\st{min}$ exhibits essentially the same behavior. (c) Increase of the OP at the wall $\op(\time,\zeta=0)$ for various values of $H_1$. The dashed line indicates the scaling behavior implied by \cref{eq_linB_wall_sum_early_0x}. (d) Relaxation of the OP at the wall to its equilibrium value $\op\eq(\zeta=0)$. For large $H_1\gg 1$, an intermediate asymptotic law [\cref{eq_nlinB_wall_interm}] emerges. (e) For times $\time\gtrsim \time^*_e\simeq 10^{-4}$, the relaxation of the OP proceeds exponentially [see \cref{eq_nlinB_wall_late}].}
    \label{fig_quench_Tc_nlin}
\end{figure}

\begin{figure}[t]\centering
    \subfigure[]{\includegraphics[width=0.5\linewidth]{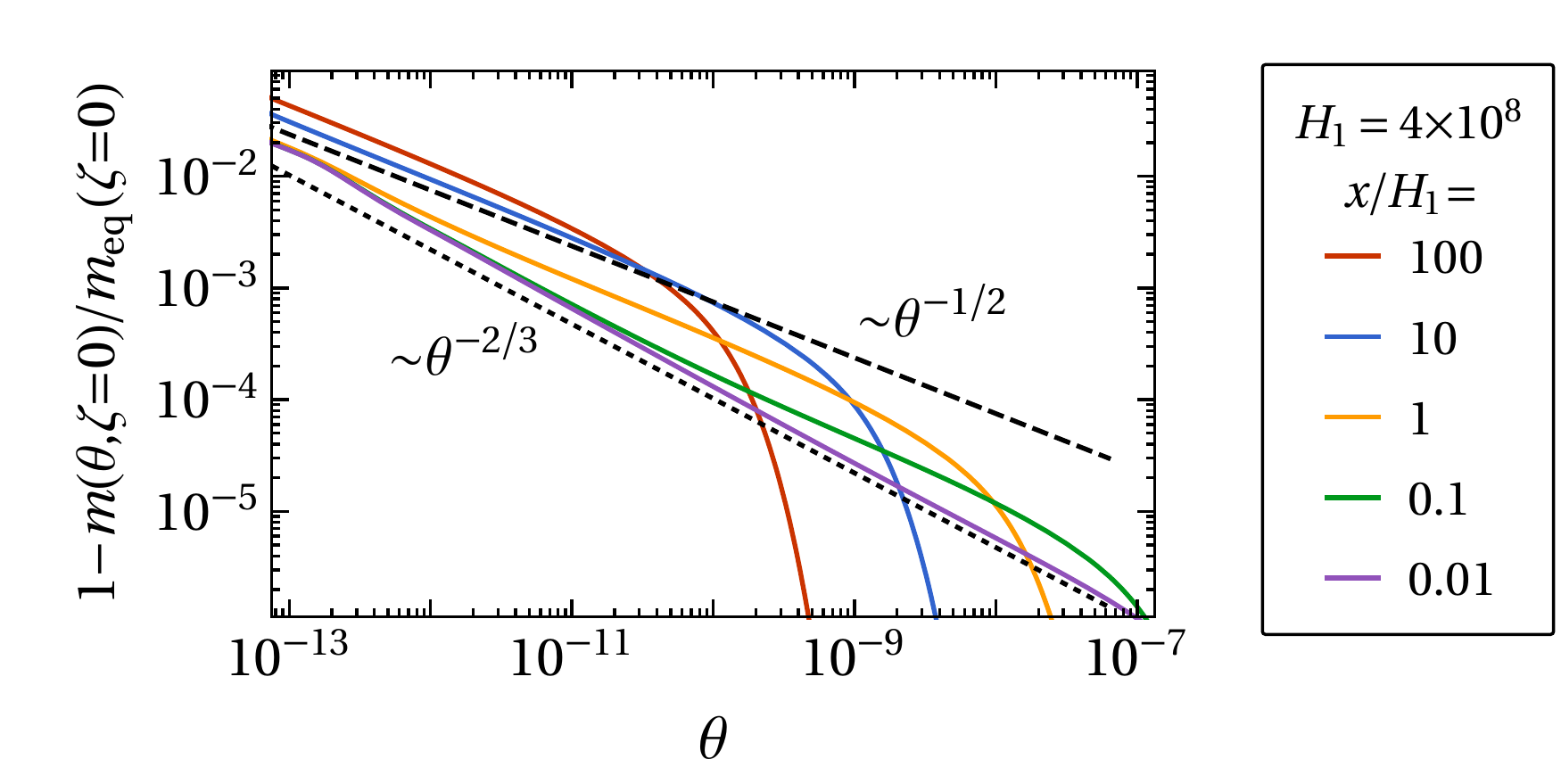} \label{fig_nlin_OPwall_interm_t_abvTc}} \hfill
    \subfigure[]{\includegraphics[width=0.48\linewidth]{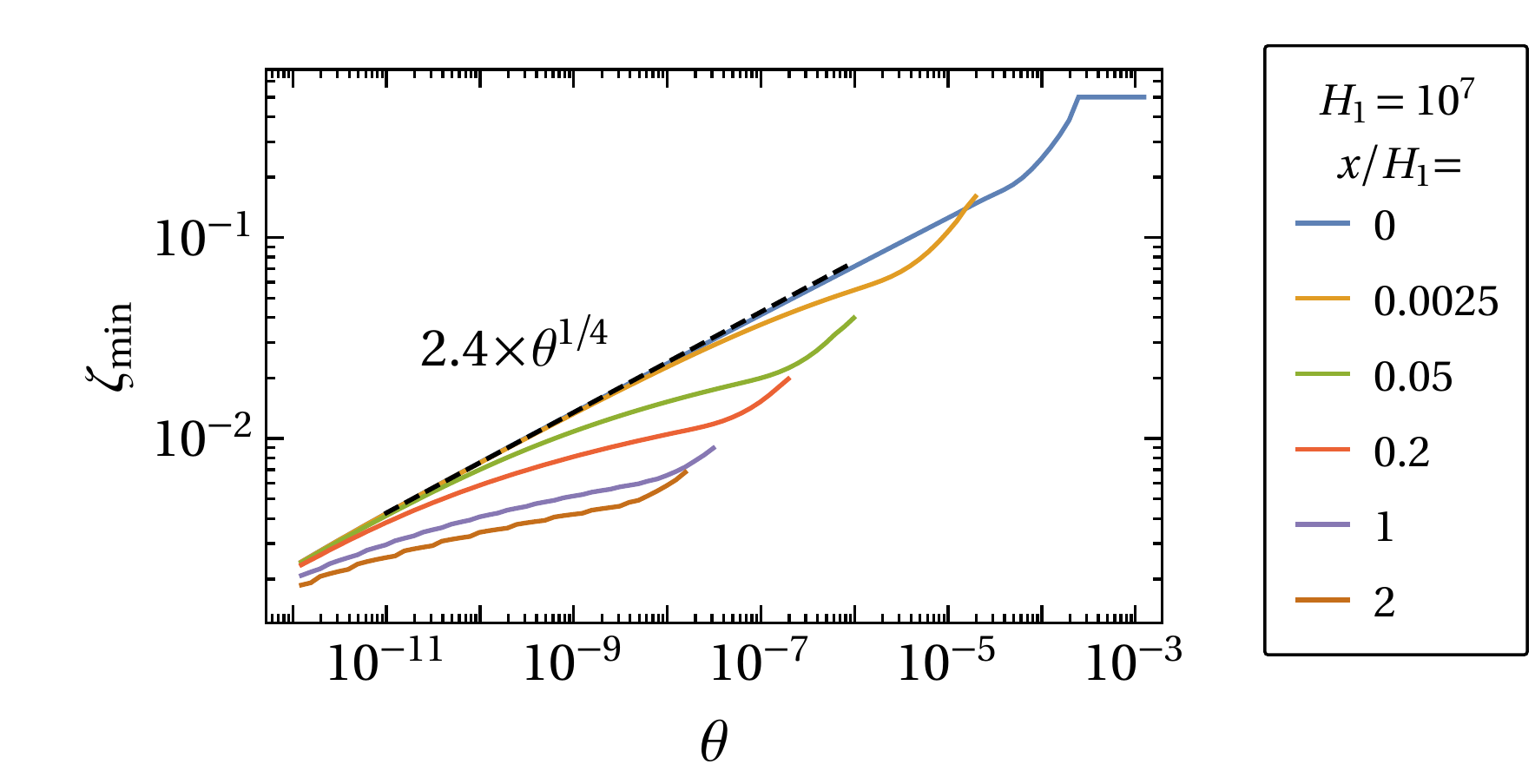} } 
    \caption{Evolution of the rescaled OP profile $\op(\time,\zeta)$ in a film within the nonlinear model B. (a) Temporal crossover of the OP at the boundary $\op(\time,\zeta=0)$ between the linear and the nonlinear intermediate asymptotic behavior, as characterized by the value of $\tscal/H_1$. The dashed and the dotted lines represent the intermediate asymptotic laws given in \cref{eq_linB_wall_sum_interm,eq_nlinB_wall_interm}, respectively. (b) Position of the global minimum of $\op(\time,\zeta)$ as a function of time $\time$ for large $H_1\gg 1$ and various values of $\tscal/H_1$. The dashed line represents the asymptotic law in \cref{eq_nlinB_OPmin}, identified from the numerical data. Due to the limited numerical resolution, the evolution of $\zeta\st{min}$ could not be followed up to $\zeta=1/2$. However, a visual inspection of the profiles reveals that at the times at which the plotted curves end, the two minima of the film profile (corresponding to $\zeta\st{min}$ and $1-\zeta\st{min}$) have in fact merged.}
    \label{fig_quench_abvTc_nlin}
\end{figure}

\begin{figure}[t]\centering
    \subfigure[]{\includegraphics[width=0.49\linewidth]{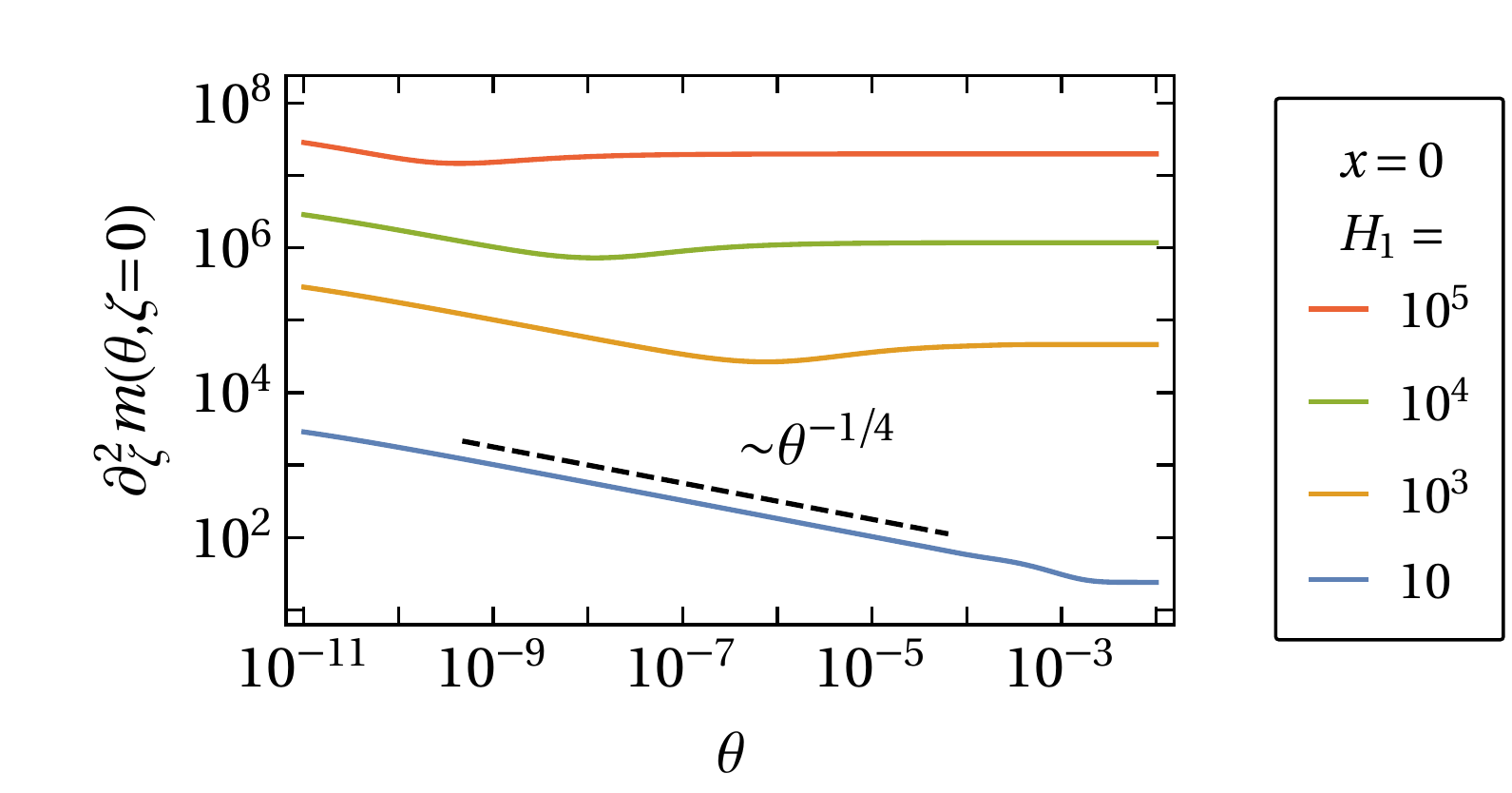} \label{}} \hfill
    \subfigure[]{\includegraphics[width=0.49\linewidth]{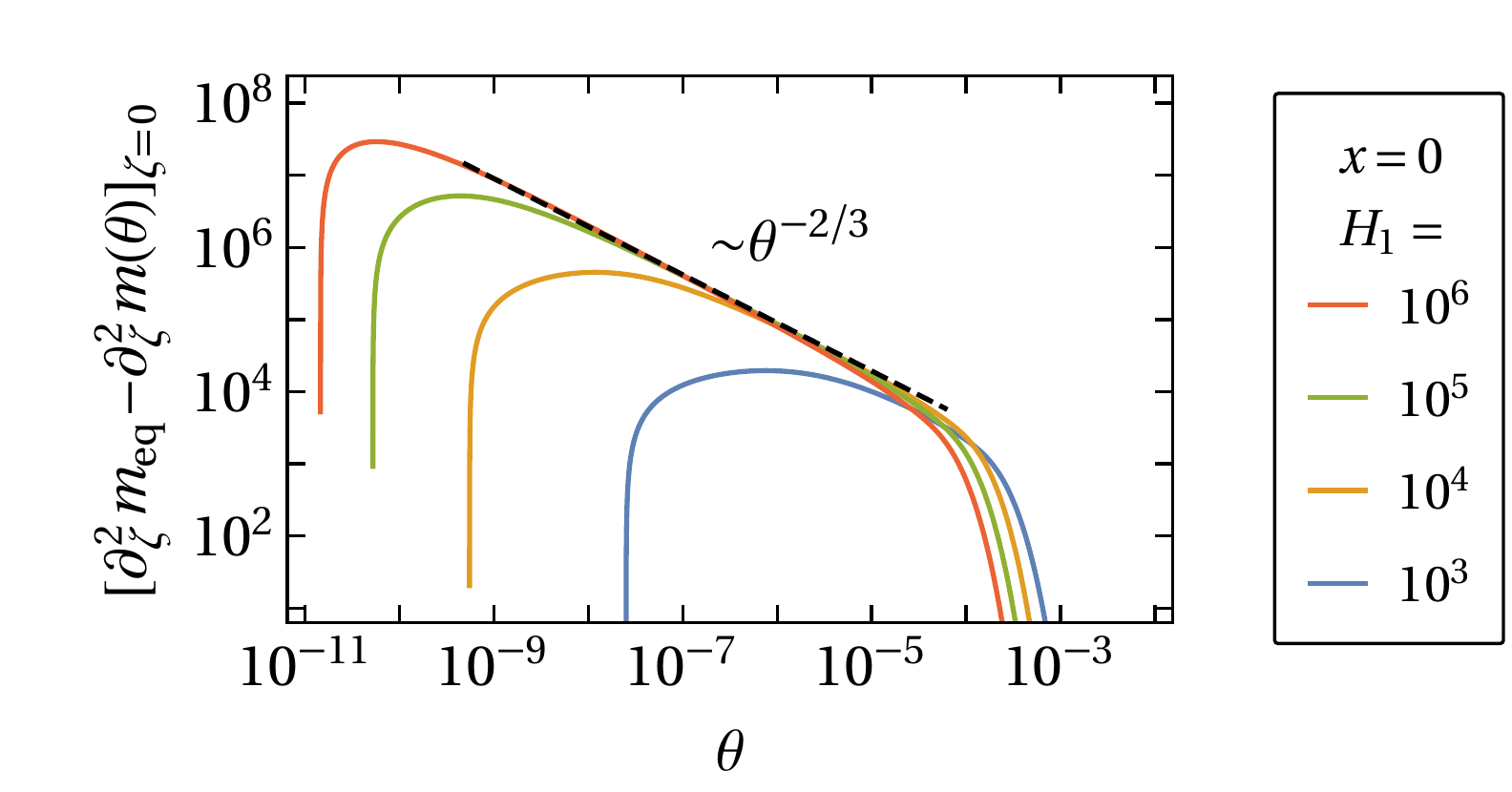} } 
    \caption{Behavior of $\pd^2_\zeta \op$ as a function of time within the nonlinear model B for various strengths of the surface field $H_1$ and at criticality $\tscal=0$. For small $H_1$, $\pd_\zeta^2 \op$ varies in accordance with \cref{eq_linB_wall_Dz2_early} [dashed line in panel (a)], which follows from the linear model B. For large $H_1$, panel (b) shows that an intermediate asymptotic regime emerges, where $\pd_\zeta^2\op$ approaches its equilibrium value algebraically, i.e., $[\pd_\zeta^2\op]_{\zeta=0} \simeq [\pd_\zeta^2\op\st{\,eq}]_{\zeta=0} - C \time^{-2/3}$, with $C\simeq 9$ as determined from a fit [see \cref{eq_nlinB_wall_Dz2_interm}].}
    \label{fig_nlin_OPwall_deriv2}
\end{figure}

In this section, within MFT, we determine the dynamics of the fully nonlinear model B. This is carried out numerically by discretizing \cref{eq_modelB_resc,eq_bcs_red} in terms of finite differences on a spatial grid and by integrating the resulting system of first-order ordinary differential equations in time \cite{moin_fundamentals_2010}.
As before, a flat, vanishing profile [\cref{eq_prof_init}] is used as initial configuration \footnote{We have checked that by using as initial condition an equilibrium profile, which corresponds to a sufficiently high temperature $\tscal\gg 1$, one obtains essentially the same results after a short transient period, which we do not consider here.}.

Within the linearized model B, $H_1$ merely appears as an overall scaling factor of the profile $\op(\time,\zeta)$ [see \cref{eq_linB_lapl_sol}]. The characteristic time scales are independent of $H_1$ [see \cref{eq_linB_wall_evol,eq_linB_wall_evol_0x}].
In contrast, the dynamics of the fully nonlinear model B [\cref{eq_modelB_resc,eq_bcs_red}] is expected to exhibit an interplay between $H_1$ and the scaling variable $\tscal$. 
As shown in Ref.\ \cite{gross_critical_2016}, this applies already to the properties of the equilibrium profile. Specifically, for $\tscal\geq 0 $,  MFT predicts the following equilibrium behavior of the OP value at the wall \cite{gross_critical_2016}:
\begin{subequations}
\label{eq_MFT_OPwall_eq_x}
  \begin{empheq}[left ={ \op\eq(\zeta=0) \simeq  \empheqlbrace }]{align}
    & H_1 \left(\frac{1}{6} - \frac{\tscal}{360}\right), & H_1 \lesssim 1,\, \tscal\lesssim 1, \label{eq_MFT_OPwall_eq_smallH1_x1}\\
    & \frac{H_1}{\tscal^{1/2}}, & H_1 \lesssim \tscal,\, \tscal\gg 1, \label{eq_MFT_OPwall_eq_smallH1_x}\\
    & 2^{1/4} H_1^{1/2} - \frac{\tscal}{3\times 2^{1/4} H_1^{1/2}}, & H_1^{1/2}\gg \tscal,\, H_1\gg 1. \label{eq_MFT_OPwall_eq_largeH1_x}
   \end{empheq}
\end{subequations}
\Cref{eq_MFT_OPwall_eq_smallH1_x1,eq_MFT_OPwall_eq_smallH1_x} in fact coincide with the results of linear MFT [see \cref{eq_eqprof_lin_Tc,eq_eqprof_lin_wall}], while \cref{eq_MFT_OPwall_eq_largeH1_x} can be obtained from a short-distance expansion within nonlinear MFT (see Ref.\ \cite{gross_critical_2016}). 
The ranges of validity of the asymptotic behaviors reported above result from a comparison between linear and nonlinear MFT.
Notably, in the presence of a mass constraint, linear MFT remains a valid approximation for $H_1\lesssim 1$, even as $\tscal\to 0$ \cite{gross_critical_2016}.
For $H_1\gg 1$ and $\tscal\ll H_1$, instead, linear MFT fails to accurately capture the equilibrium profile \cite{gross_critical_2016}. Accordingly, we expect the nonlinear term $\op^3$ [see \cref{eq_modelB_resc}] to play also a role in the dynamics.

\Cref{fig_quench_Tc_nlin,fig_quench_abvTc_nlin} summarize the dynamics of the OP profile governed by the nonlinear model B for $(++)$ \bcs after a quench to criticality ($\tscal=0$) and to a  supercritical temperature ($\tscal\gg 1$), respectively. 
As in the linear case (see \cref{fig_quench_Tc,fig_quench_abvTc}), the increase of the OP at the wall is supported by transport of mass from the interior of the film, giving rise to a pronounced minimum of the profile moving towards the center.
A numerical analysis, illustrated in \cref{fig_nlin_minPos_Tc}, shows that this minimum follows a subdiffusive law: 
\beq \zeta\st{min}\simeq 2.4\times \time^{1/4}.
\label{eq_nlinB_OPmin}\eeq 
Except for the slightly larger value of the prefactor \footnote{For large $H_1$, the value of this prefactor becomes independent of $H_1$.}, this relation is identical to the one obtained within the linear model B [see \cref{eq_linB_wall_zmin}].
Panels (c)--(e) of \cref{fig_quench_Tc_nlin} illustrate the increase of the OP at the wall, $\op(\time,\zeta=0)$. 
For large $H_1$ one can identify three asymptotic regimes: in addition to the initial algebraic growth $\op(\time,0)\propto \time^{1/4}$ [\cref{fig_quench_Tc_nlin}(c)] and the late-time exponential saturation [\cref{fig_quench_Tc_nlin}(e)], an intermediate asymptotic regime emerges for $H_1\gg 1$, in which the OP saturates algebraically.
For $\tscal=0$, the latter regime is not present in linear MFT [see \cref{eq_linB_wall_evol_0x,fig_quench_Tc}].
Below, these various regimes are analyzed further.

The effect of a nonzero reduced temperature $\tscal>0$ on the OP dynamics is illustrated in \cref{fig_quench_abvTc_nlin}. As shown in panel (a), upon increasing $\tscal$, the intermediate asymptotic law identified in \cref{fig_quench_Tc_nlin}(d) crosses over from a behavior dominated by the nonlinearity towards that of the linear model B reported in \cref{eq_linB_wall_sum_interm}. At late times, the OP at the wall $\op(\time,\zeta=0)$ always saturates exponentially.
\Cref{fig_quench_abvTc_nlin}(b) shows that, upon increasing $\tscal$ at fixed $H_1\gg 1$, the dynamics of the OP minimum significantly deviates from \cref{eq_nlinB_OPmin}, which holds at criticality $\tscal=0$.
Due to limited numerical accuracy, for large $\tscal$ the evolution of $\zeta\st{min}$ cannot be followed up to values of $\zeta\st{min}\sim \Ocal(1)$. However, an inspection of the actual profile $\op(\time,\zeta)$ reveals that, at a time around $\time\sim \tscal^{-1}$, the two minima in the film profile have merged at the center of the system and the profile has essentially reached equilibrium.
Accordingly, we conclude that, also for the nonlinear model B, the characteristic time scale for the onset of the interaction between the two walls is given by \cref{eq_min_merge_time}.

We now return to a quantitative discussion of the OP dynamics at the wall, i.e., of $\op(\time,\zeta=0)$, within the nonlinear model B. The asymptotic behavior can be analyzed based on the relative weight of the individual terms on the r.h.s.\ of \cref{eq_modelB_resc}, i.e., $ -\pd_\zeta^2 \op + \tscal \op + \op^3$.
We emphasize that the term $\pd_\zeta^2 \op$ is always relevant, because it is responsible for the emergence of the nontrivial equilibrium profile $\op\eq$ (see also Ref.\ \cite{gross_critical_2016}). In fact, it provides the driving force for the evolution of the profile away from the flat initial configuration.
The analysis of the asymptotic dynamics is facilitated by the knowledge of the linear mean-field solutions reported in \cref{sec_lin_dyn}. 

At sufficiently early times $\time\ll 1$, $\op$ is small owing to the initial condition in \cref{eq_prof_init}.
Consequently, $\tscal\op$ and $\op^3$ are negligible and the behavior reported in \cref{eq_linBTc_earlytime_wall_D2z,eq_linB_wall_sum_early} applies, which is solely driven by the term $\pd_\zeta^2\op$ on the r.h.s.\ of \cref{eq_modelB_resc}.
In the course of the early-time dynamics given in \cref{eq_linB_wall_sum_early,eq_linBTc_earlytime_wall_D2z}, the term $\tscal\op$ increases and it becomes comparable to $\pd_\zeta^2\op$ around the time 
\beq \time^*_\tscal \sim \tscal^{-2},
\eeq 
while $\pd_\zeta^2\op$ becomes comparable to $\op^3$ around the time 
\beq \time^*_3 \sim H_1^{-2}.
\label{eq_timescale_nlin}\eeq 
The relative magnitude of $\time^*_\tscal$ and $\time^*_3$ plays a central role in characterizing the deviation of the time evolution from the linear mean-field behavior, as it is analyzed in the following.

\underline{Case $H_1\ll \tscal$, $\tscal\gtrsim 100$:}
Since in this case $\time^*_\tscal \ll \time^*_3$, the evolution of $\op(\time,0)$ crosses over at $\time\sim \time^*_\tscal$ from the behavior described in \cref{eq_linB_wall_sum_early} to the intermediate asymptotic regime of linear MFT reported in \cref{eq_linB_wall_sum_interm}, which requires $\tscal\gtrsim 100$. 
According to \cref{eq_linB_wall_sum_early,eq_MFT_OPwall_eq_smallH1_x}, one has $\op(\time^*_x,0)\simeq 0.78 \times \op\eq(0)$ at the time $\time^*_x$.
Since, in addition, $\op\eq^3/(\tscal\op\eq) \sim H_1^2/\tscal^2\ll 1$, we conclude that in the present case, the term $\op^3$ never exceeds $\tscal\op$. Accordingly, the time evolution for $\time\gtrsim \time^*_x$ closely follows the behavior of the linear model B given in \cref{eq_linB_wall_evol}. 
A numerical analysis confirms that, as expected from \cref{eq_linB_wall_sum_late}, the crossover to the final exponential saturation regime occurs at a time $\time^*_e\sim \tscal^{-1}$. This crossover corresponds to the point of maximum curvature of the curves plotted in \cref{fig_nlin_OPwall_interm_t_abvTc}.

\underline{Case $H_1\ll \tscal,\tscal \lesssim 100$:}
For $\tscal\lesssim 100$, the early-time regime, which is described by the linear model B behavior in \cref{eq_linB_wall_sum_early}, proceeds up to the time $\time^*_e \simeq 10^{-4}$ [\cref{eq_time_early_crit}], at which $\op(\time^*_e,0)\simeq 0.48 \times \op\eq(0)$ [see \cref{eq_MFT_OPwall_eq_smallH1_x1}].
Since $\op\eq^3/(\tscal \op\eq) \sim H_1^2/(36 \tscal)\ll 1$, the dynamics is governed by the linear model B also for $\time>\time^*_e$, where $\op(\time,0)$ follows the exponential saturation law reported in \cref{eq_linB_wall_sum_late}.

\underline{Case $H_1\gg \tscal$, $H_1\gtrsim 100$:}
In this case one has $\time^*_3 \ll \time^*_\tscal$ (and $\time^*_3\lesssim 10^{-4}$), such that the initial algebraic growth law in \cref{eq_linB_wall_sum_early} is expected to cross over at $\time\sim \time^*_3$ to a different behavior characteristic of nonlinear dynamics.
A numerical analysis, which is illustrated in \cref{fig_nlin_OPwall_interm_t_Tc} for $\tscal=0$, reveals an effective intermediate asymptotic behavior of the form
\beq \op(\time\gtrsim \time^*_3, 0) \simeq \op\eq(0) \left(1- C \time^{-2/3}\right),\qquad \text{with}\qquad  C\simeq 0.2 \times H_1^{-1.2},
\label{eq_nlinB_wall_interm}\eeq 
where the dependence of $C$ on $H_1$ as well as the value of the exponent of $\theta$ have been determined from a fit of the numerical data. 
Using \cref{eq_MFT_OPwall_eq_largeH1_x}, one obtains $\op(\time^*_3,0)\simeq 0.66\times \op\eq(0)$ and furthermore $\op^3\eq/(\tscal\op\eq)\sim 2^{1/2} H_1/\tscal\gg 1$, indicating that the linear term $\tscal\op$ remains negligible compared to $\op^3$ for times $\time \gtrsim \time^*_3$.
Accordingly, no further intermediate asymptotic law is expected in this case.
Instead, we find that, at a time $\time^*_e \gg \time^*_3$, the dynamics crosses over to a final exponential saturation regime,
\beq \op(\time\gtrsim \time^*_e, 0) \simeq \op\eq(0)\left(1 - \tilde C e^{-\tilde A \time}\right), 
\label{eq_nlinB_wall_late}\eeq 
which is illustrated in \cref{fig_nlin_OPwall_large_t_Tc} for $\tscal=0$.
The parameters $\tilde A$ and $\tilde C$ are numerically found to depend on $H_1$ and $\tscal$ in a way which does not lend itself to a meaningful fit based on the presently available data. 
As one infers from \cref{fig_nlin_OPwall_interm_t_Tc}, for $\tscal=0$ and for the considered values of $H_1$, the crossover time $\time^*_e$ depends only weakly on $H_1$ and it can be estimated as 
\beq \time^*_e\simeq 10^{-4},
\label{eq_timescale_nlin_exp}
\eeq 
which agrees with \cref{eq_time_early_crit}. 
For $\tscal\gg 1$, instead, a numerical analysis (data not shown) indicates that the crossover time is approximately given by $\time^*_e \sim \tscal^{-1}$, which is the same scaling law as in \cref{eq_linB_wall_sum_late}.
\Cref{eq_nlinB_wall_interm,eq_nlinB_wall_late} replace the intermediate and late-time asymptotics in \cref{eq_linB_wall_evol} in the nonlinear case.

\underline{Case $H_1\gg \tscal$, $H_1\lesssim 100$:} In this case one has $\time^*_3\gtrsim \time^*_e$, such that, according to \cref{eq_linB_wall_evol_0x}, before the nonlinearity becomes relevant, the dynamics crosses over from the early-time growth to an exponential saturation regime at a time $\time\simeq \time^*_e$.
A numerical analysis reveals that this exponential saturation in fact persists until equilibrium is reached. 
The irrelevance of the nonlinear term is consistent with the fact that, in this case, static linear MFT accurately approximates $\op\eq$ [see \cref{eq_MFT_OPwall_eq_smallH1_x1}].

In \cref{fig_nlin_OPwall_deriv2}, the temporal evolution of the second derivative of the profile at the wall, $\pd^2_\zeta \op(\time,\zeta)|_{\zeta=0}$, is illustrated for $\tscal=0$ and for various values of the surface field $H_1$. For times $\time\lesssim \time^*_3$ [\cref{eq_timescale_nlin}], $\pd_\zeta^2\op$ generally evolves following the predictions of the linear model B [see \cref{eq_linB_wall_Dz2_early}]. 
For large values of $H_1\gg 1$ and for $\tscal\ll H_1$, the nonlinear term becomes relevant and causes the appearance of an intermediate asymptotic saturation regime of the form [see \cref{fig_nlin_OPwall_deriv2}(b)]
\beq \pd_\zeta^2\op|_{\zeta=0} \simeq \pd_\zeta^2\op\eq|_{\zeta=0} - C\time^{-2/3},\qquad \text{for}\qquad \time^*_3\ll \time\ll \time^*_e,
\label{eq_nlinB_wall_Dz2_interm}\eeq 
where $C\simeq 9$ follows together with the value of the exponent of $\time$ from a fit to the numerical data, while $\time_e^*$ coincides with \cref{eq_timescale_nlin_exp}.

In conclusion, except for the case $H_1\gg \tscal$, $H_1\gtrsim 100$, the expressions in \cref{eq_linB_wall_evol,eq_linB_wall_evol_0x}, based on the linearized dynamics, generally provide an accurate description of the asymptotic dynamics of the nonlinear model B.

\section{Critical Casimir force}
\label{sec_CCF}

\begin{figure}[t!]\centering
    \subfigure[]{\includegraphics[width=0.15\linewidth]{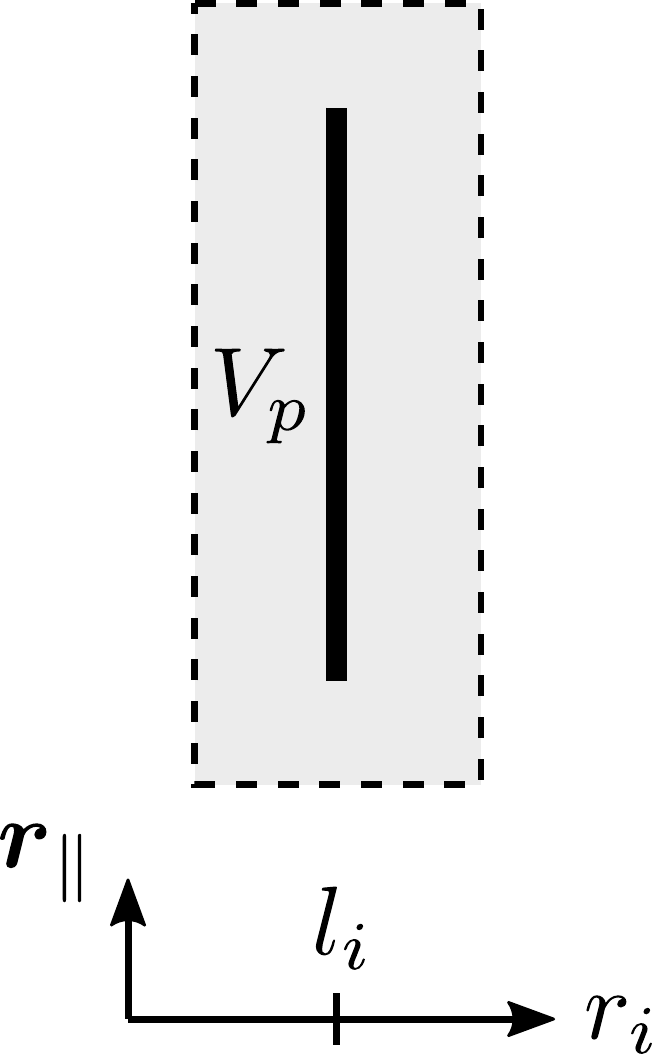} }\hspace{2cm}
    \subfigure[]{\includegraphics[width=0.15\linewidth]{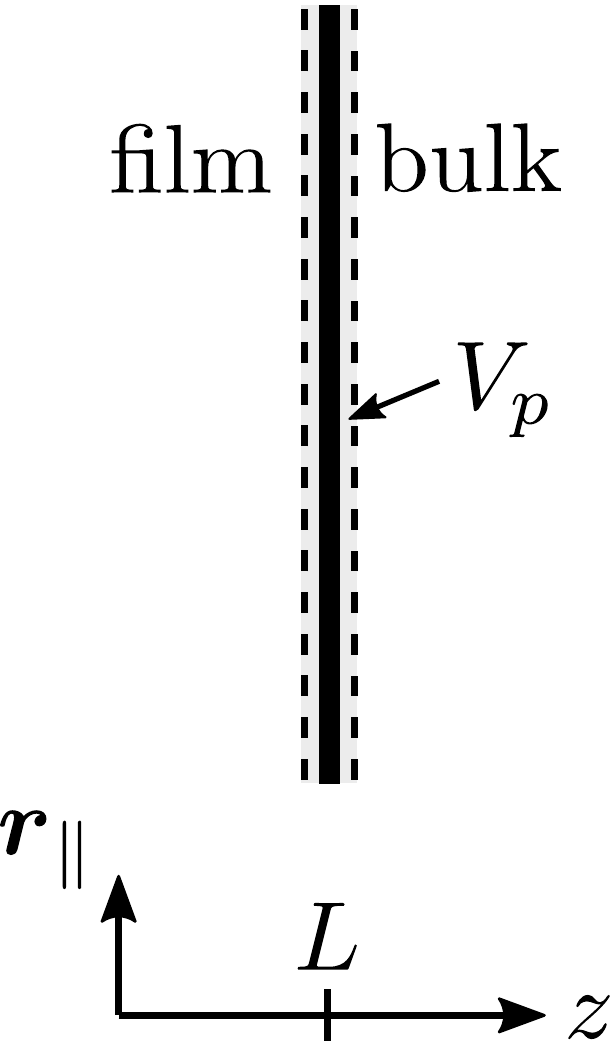} }
    \caption{(a) The generalized force $K_i$, which is acting along the coordinate direction $r_i$ ($i\in\{x,y,z\}$) onto a plate of finite extent (thin black rectangle), is obtained via an integration of the derivative of the Hamiltonian density over an arbitrary enclosing volume $V_p$ [shaded rectangle; see \cref{eq_inst_force}]. This plate of transverse area $A$ is located at $r_i=l_i$ and lies in the plane spanned by the lateral coordinates $\rv_\parallel = \sum_{j\in \{x,y,z\}\backslash i} r_j \bv{e}_j$. (b) Here, the plate represents one of the two boundaries between the film and the surrounding bulk system. Along the lateral directions, both the plate and the volume $V_p$ are infinitely extended. The force acting on the plate can be interpreted as the CCF [see \cref{eq_CCF}], which is determined by shifting the surfaces of $V_p$ (dashed lines) directly next to the corresponding surfaces of the plate.}
    \label{fig_wall}
\end{figure}

Following Ref.\ \cite{dean_out--equilibrium_2010}, in this section we obtain the dynamic CCF based on the notion of a generalized force associated with the interaction between the OP field $\phi$ and the boundary.
In order to recall the corresponding formalism, we first consider an arbitrary volume $V$ containing a single plate with finite transverse area [see \cref{fig_wall}(a)] 
and subsequently show how this approach renders the CCF in a film.
For simplicity, we assume the plate to be perpendicular to the coordinate axis $i\in \{x,y,z\}$ and to be located at the position $r_i=l_i$. 
The OP $\phi$ gives rise to the free energy functional
\beq \Fcal = \int_V \d^d r\left\{ \Hcal_0(\phi(\rv),\nabla\phi(\rv)) + \Ucal_{p}(l_i,\phi(\rv)) \right\} \equiv \int_V \d^d r\, \Hcal(l_i, \phi(\rv),\nabla\phi(\rv)) ,
\eeq 
where $\Hcal = \Hcal_0 + \Ucal_{p}$, $\Hcal_0$ is an arbitrary bulk Hamiltonian density (energy per volume), and the boundary potential
\beq \Ucal_{p}(l_i,\phi) = \delta(r_i-l_i) U_{p}(\phi)
\label{eq_boundpot}\eeq 
accounts for the interaction with strength $U_{p}(\phi)$ between the plate and the OP field.
(Note that, in contrast to \cref{eq_freeEn}, here $\Fcal$ is defined not per unit area and per $k_B T$.)
According to the principle of virtual work, for any configuration of $\phi$ the generalized force $K_i$ acting on the plate is given by \cite{dean_out--equilibrium_2010}
\beq K_i \equiv -\frac{\pd \Fcal}{\pd l_i} = \int_{V_p} \d^d r \frac{\pd\, \Ucal_{p}(l_i,\phi)}{\pd l_i} = \int_{V_p} \d^d r \frac{\pd\, \Hcal(l_i, \phi,\nabla\phi) }{\pd l_i},
\label{eq_inst_force}\eeq 
where $V_p$ is an arbitrary volume enclosing the \emph{p}late as sketched in \cref{fig_wall}(a). The last two expressions in \cref{eq_inst_force} follow from the spatially localized nature of the interaction $\Ucal_{p}$, i.e., from the finite extent of the surface. 
Note that $K_i$ depends only on the \emph{static} free energy functional and is therefore independent of the actual dynamics or conservation laws.
Introducing the bulk chemical potential associated with $\Fcal$ [see \cref{eq_chempot}],
\beq \mu \equiv \frac{\delta\Fcal}{\delta \phi} = \frac{\pd\Hcal}{\pd\phi} - \nabla_j\left(\frac{\pd\Hcal}{\pd\nabla_j\phi}\right), 
\label{eq_chempot_K}\eeq 
$K_i$ can, after some algebra, be expressed as
\beq\begin{split}
K_i &= -\int_{V_p}\d^d r\, \nabla_j \Tcal_{ij} - \int_{V_p} \d^d r \mu\nabla_i\phi \\
&= -\int_{\pd V_p} \d^{d-1} s_j\, \left( \Tcal_{ij} + \mu\phi \delta_{ij}\right) + \int_{V_p}\d^d r\, (\nabla_i \mu) \phi, 
\end{split}\label{eq_force_stress_gen}
\eeq
in terms of the standard stress tensor
\beq \Tcal_{ij} = \frac{\pd\Hcal}{\pd\nabla_i\phi} \nabla_j \phi - \delta_{ij}\Hcal,
\eeq 
where, in the previous expressions, summing over repeated indices is understood and $\pd V_p$ denotes the boundaries of $V_p$.

We now apply this formalism to a film, i.e., a volume bounded by two infinitely extended plates located at $l_z=0$ and $l_z=L$.
At each plate, the volume $V_p$ is chosen such that its surfaces are located directly next to the corresponding surfaces of the plate, as sketched in \cref{fig_wall}(b).
We define the CCF $\Kcal$ as the generalized force [\cref{eq_inst_force}] per area $A$ acting on the boundary, 
\beq \Kcal \equiv K_{z}/A,
\label{eq_dyn_CCF_def}\eeq 
in the limit $A\to\infty$.
We describe the location of the right and the left boundary as $l_z=L/2\pm \tilde l_z$, respectively, and compute the derivative in \cref{eq_inst_force} with respect to $\tilde l_z$, which ensures that at each boundary a variation $\d \tilde l_z>0$ increases the film thickness.
Accordingly, \cref{eq_inst_force} turns into $K_z = \mp \pd \Fcal / \pd \tilde l_z$ at the right and the left boundary, respectively.
Consistently with \cref{eq_CCF_eq_def}, the resulting CCF is therefore repulsive if the film free energy decreases upon increasing the film thickness. 
Owing to the no-flux \bcs [\cref{eq_bcs_noflx}], the last term in the second equation of \cref{eq_force_stress_gen} vanishes for this choice of $V_p$. 
Upon taking into account the direction of the surface normals, the CCF finally follows from \cref{eq_force_stress_gen} as 
\beq \Kcal  = \bar \Tcal_{zz}\big|_{z=0,L} - \bar \Tcal_{zz,b}\: ,
\label{eq_CCF}\eeq  
where 
\beq \bar \Tcal_{ij} \equiv \Tcal_{ij} + \mu \phi \delta_{ij} = \frac{\pd\Hcal}{\pd\nabla_i\phi} \nabla_j \phi - \delta_{ij}(\Hcal - \mu \phi),
\label{eq_dyn_stressten}\eeq 
and $\bar \Tcal_{ij,b}$ denotes the corresponding expression of $\bar \Tcal_{ij}$ in the bulk (where, within MFT, gradient terms are absent).
Note that \cref{eq_CCF} has to be evaluated for the actual time-dependent solution $\phi(t,z)$ of the model B equations [\cref{eq_modelB}].

Henceforth we shall call $\bar \Tcal_{ij}$ the \emph{dynamical} stress tensor \footnote{In the case of \emph{model A} dynamics, i.e., for $\pd_t \phi = -\mu(\phi)$, the dynamical stress tensor in \cref{eq_dyn_stressten} can be written as ${\mathcal{\bar T}}_{ij} = \Tcal_{ij} - \onehalf \pd_t(\phi^2)$ and thus reduces to the alternative formulation given in Eq.~(56) in Ref.\ \cite{dean_out--equilibrium_2010} (apart from an overall minus sign in its definition).}. 
Remarkably, \cref{eq_CCF,eq_dyn_stressten} coincide with the formal expressions for the equilibrium CCF and the equilibrium stress tensor in the \emph{canonical} ensemble, respectively, as derived in Ref.\ \cite{gross_critical_2016}.
The dynamical stress tensor $\bar\Tcal_{ij}$ differs from the standard equilibrium stress tensor $\Tcal_{ij}$ used in the grand canonical ensemble \cite{krech_casimir_1994} by the term $\mu\phi\delta_{ij}$ involving the chemical potential [\cref{eq_chempot,eq_chempot_K}].
In fact, in an unconstrained (grand canonical) equilibrium, $\mu=\delta\Fcal/\delta\phi=0$, such that in this case $\bar\Tcal_{ij}= \Tcal_{ij}$.
However, in nonequilibrium, one generally has $\mu\phi\neq 0$, independently of the presence of a global or a local OP conservation law (see also Ref.\ \cite{dean_out--equilibrium_2010}).
For the Landau-Ginzburg free energy functional considered in \cref{eq_freeEn,eq_freeEn_pot}, one has 
\beq \bar\Tcal_{zz} =  \frac{1}{2}(\pd_z\phi)^2 + \frac{1}{2}\tLG \phi^2 + \frac{1}{8}g \phi^4 - \phi \pd_z^2 \phi .
\label{eq_CCF_LG}\eeq 
In Ref.\ \cite{kruger_stresses_2018}, a related nonequilibrium stress formulation has been analyzed for fluids far from criticality .

In order to proceed, we recall that the film is taken to have a vanishing mass [see \cref{eq_mass}].
In accordance with Ref.\ \cite{gross_critical_2016}, we therefore also assume that the bulk medium surrounding the film has a vanishing mean OP:
\beq \phi_b=0.
\label{eq_bulkOP_zero}\eeq 
Consequently, the associated bulk pressure $p_b = \bar \Tcal_{zz,b}=0$ vanishes, too. 
Upon introducing the MFT scaling variables defined in \cref{eq_FSS_vars}, \cref{eq_CCF} [taking into account \cref{eq_bulkOP_zero}] can be brought into the scaling form given in \cref{eq_FSS_CCF}:
\beq \Kcal  = L^{-4} \Delta_0 \left[ \onehalf (\pd_\zeta \op)^2 + \onehalf \tscal \op^2 + \frac{3}{4} \op^4 - \op \pd_\zeta^2 \op \right]_{\zeta=0,1},
\label{eq_CCF_scal}\eeq 
where 
\beq \Delta_0 = \frac{6}{g} = (\amplPhit \amplXip)^2
\label{eq_MFT_CCF_ampl}\eeq 
represents a mean-field amplitude, which can be expressed in terms of the non-universal critical amplitudes $\amplPhit$ and $\amplXip$ defined in \cref{eq_FSS_op,eq_FSS_vars}.
Consequently, the expression multiplying $L^{-4}$ on the r.h.s.\ of \cref{eq_CCF_scal} represents the suitably normalized scaling function $\Xi(\time,\tscal,H_1)$ of the CCF.
A ratio of observables independent of $\Delta_0$ is provided by $\Kcal/\Kcal\eq$, where $\Kcal\eq\equiv \Kcal|_{t\to\infty}$ denotes the equilibrium CCF.

\subsection{Critical Casimir forces within linear mean field theory}

Here, we analyze the dynamics of the CCF [\cref{eq_CCF}] emerging within the linear model B, using a flat profile [\cref{eq_prof_init}] as initial condition. Accordingly, the CCF is completely determined by the expression of the profile given in \cref{eq_linB_lapl_sol}.
Due to the \bcs in \cref{eq_bcs_CA_red}, the first term in the square brackets in \cref{eq_CCF} is constant and equal to $H_1^2/2$.
Therefore the scaling function $\Xi$ of the CCF [see \cref{eq_FSS_CCF,eq_CCF,eq_CCF_scal,eq_MFT_CCF_ampl} reduces to
\beq \Xi/\Delta_0 = \onehalf H_1^2 + \onehalf x\op^2 - \op \pd_\zeta^2\op.
\label{eq_CCF_lin}\eeq 
Note that, within the linear model B, $\Xi\propto H_1^2$, because $H_1$ enters the profile $\op$ as an overall prefactor [see \cref{eq_linB_lapl_sol}].

\subsubsection{Critical quench ($\tscal=0$)}

According to \cref{eq_linB_wall_evol_0x,eq_linBTc_earlytime_wall_D2z}, at early times one has 
\beq \op \pd_\zeta^2\op \simeq \frac{\sqrt{2}}{\pi} H_1^2, \qquad \time\lesssim \time^*_e.
\eeq 
This implies that, correspondingly, the CCF in \cref{eq_CCF_lin} attains a non-vanishing value:
\beq \Xi(\time\lesssim \time^*_e, \tscal=0, H_1)/\Delta_0 \simeq \left(\onehalf-\frac{\sqrt{2}}{\pi}\right) H_1^2 >0.
\label{eq_CCF_lin_crit_early}\eeq 
At late times ($\time\to\infty$), instead, by using \cref{eq_eqprof_lin_Tc}, we recover the equilibrium CCF at criticality in the canonical ensemble (see Ref.\ \cite{gross_critical_2016}):
\beq \Xi\eq(\tscal=0,H_1) \equiv \Xi(\time\to\infty, \tscal=0, H_1) = \frac{1}{6}H_1^2 \Delta_0.
\label{eq_CCF_lin_crit_late}\eeq 
The CCF scaling function $\Xi(\time,\tscal=0,H_1)$ at criticality is illustrated as a function of time in \cref{fig_CCF_lin_typ}. 
Interestingly, the CCF shows a non-monotonic transient behavior which interpolates between the early- and the late-time behaviors indicated above.

\subsubsection{Non-critical quench ($\tscal\neq 0$)}

\begin{figure}[t]\centering
    \subfigure[]{\includegraphics[width=0.44\linewidth]{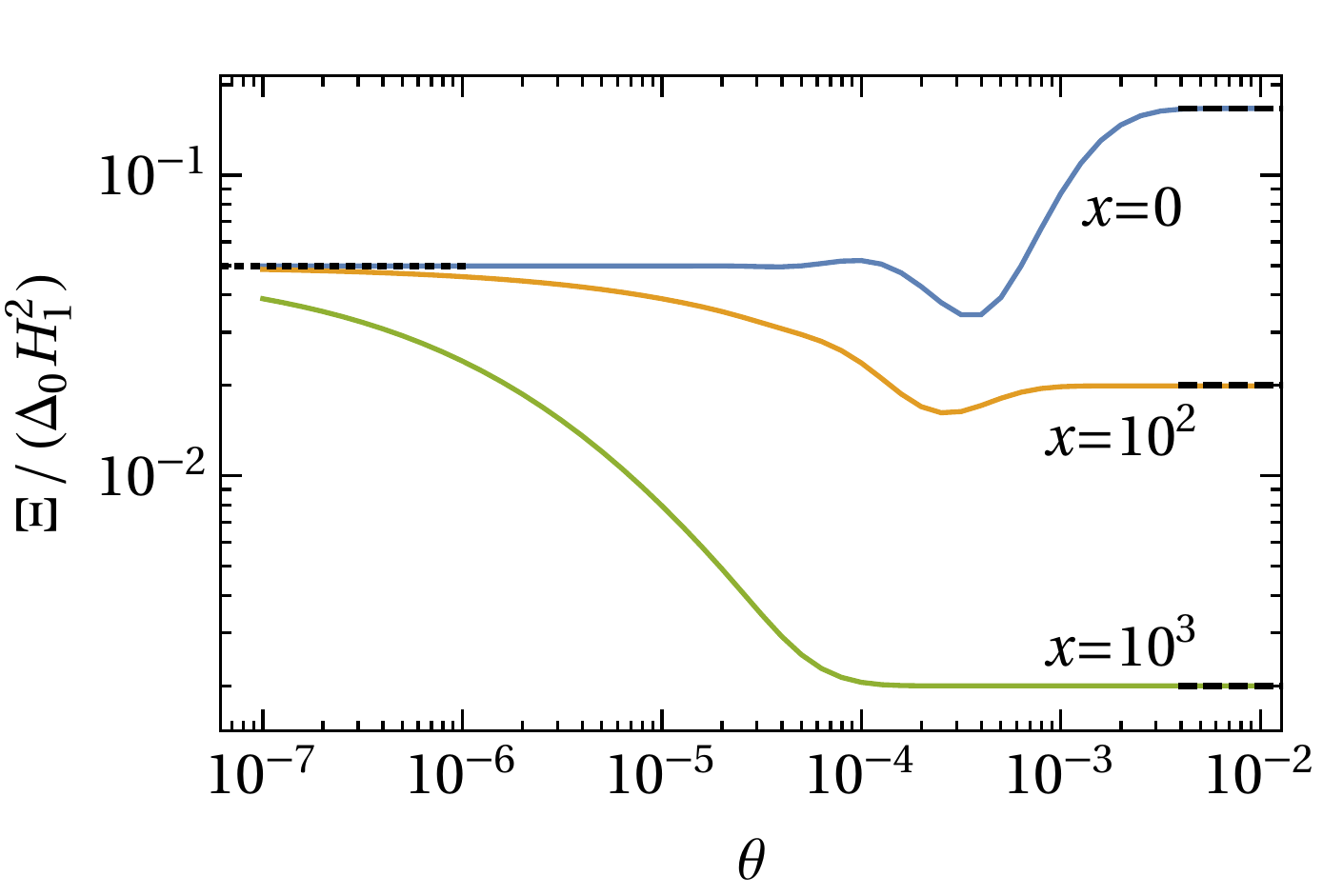} \label{fig_CCF_lin_typ}} \qquad
    \subfigure[]{\includegraphics[width=0.44\linewidth]{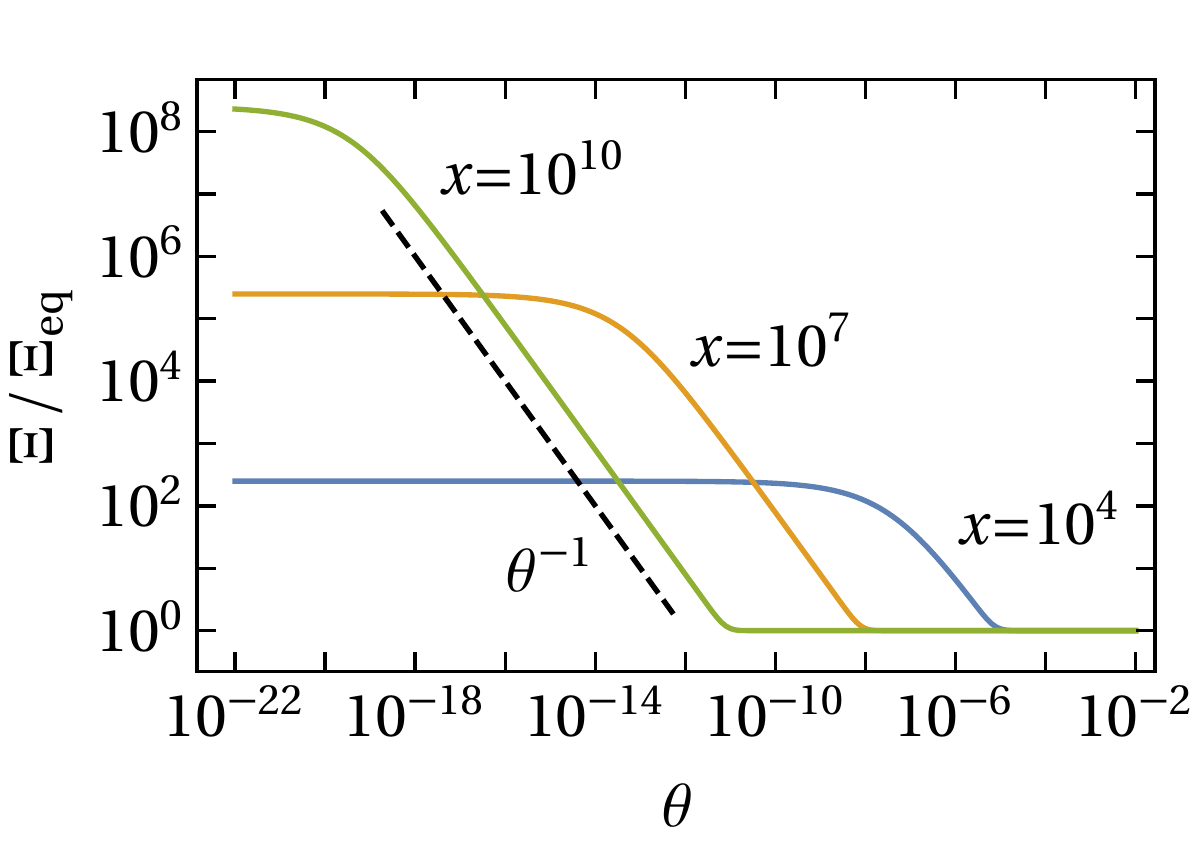} } 
    \caption{Scaling function $\Xi$ of the CCF as a function of time $\time$ within the linear model B and for various values of the rescaled temperature $\tscal$. In (a), the scaling function $\Xi$ is normalized by $\Delta_0 H_1^2$, which is a common prefactor arising within linear MFT and which can be expressed in terms of physically accessible quantities [see \cref{eq_MFT_CCF_ampl,eq_FSS_H1} and the related discussions]. The dotted and dashed lines represent the early- and late-time (equilibrium) values of the CCF reported in \cref{eq_CCF_lin_crit_early,eq_CCF_lin_crit_late,eq_CCF_lin_crit_eq_largeT}. In (b), instead, $\Xi$ is normalized by its late-time, equilibrium value. For large $\tscal\gg 1$, the CCF follows an intermediate asymptotic behavior $\Xi\propto \time^{-1}$ [\cref{eq_linB_CCF_sum_interm}; dashed line in (b)]. At late times the scaling functions smoothly attain the value 1.}
    \label{fig_CCF_lin}
\end{figure}

From \cref{eq_linB_wall_evol,eq_linB_wall_evol_Dz2} one concludes that, for $\tscal\gg 1$, the early-time behavior in \cref{eq_CCF_lin_crit_early} applies to times $\time\ll \tscal^{-2}$.
For $\time\to\infty$, instead, the equilibrium value of the CCF \cite{gross_critical_2016},
\beq \Xi\eq(\tscal\gg 1,H_1) \equiv \Xi(\time\to\infty, \tscal\gg 1, H_1) \simeq \frac{2 H_1^2}{\tscal} \Delta_0,
\label{eq_CCF_lin_crit_eq_largeT}\eeq 
is recovered.
The non-equilibrium CCF for $\tscal>0$ is generally weaker than that at criticality [see \cref{fig_CCF_lin}(a)], while one observes that, according to \cref{eq_CCF_lin_crit_early,eq_CCF_lin_crit_eq_largeT}, for $\tscal\gg 1$ the magnitude of the CCF generally decreases with time.
In the same limit but at intermediate times $\time$ with $\tscal^{-2}\lesssim \time \lesssim \tscal^{-1}$, the CCF decays as $\Xi\simeq (2\pi \tscal^2 \time)^{-1}$ [see \cref{fig_CCF_lin}(b)]. This asymptotic expression follows upon inserting the corresponding expressions for $\op$ [\cref{eq_linB_wall_sum_interm}] and $\pd_\zeta^2\op$ [\cref{eq_linB_wall_Dz2_interm}] into \cref{eq_CCF_lin} and by identifying the dominant term in the result.
For times $\time\gtrsim \tscal^{-1}$ the CCF relaxes exponentially towards its equilibrium value (not shown).

In summary, within the linear model B and far from criticality (i.e., $\tscal\gg 1$) the scaling function of the CCF exhibits the following asymptotic behavior:
\begin{subequations}
\label{eq_linB_CCF_evol}
  \begin{empheq}[left ={ \Xi(\time,\tscal\gg 1,H_1)/(\Delta_0 H_1^2) \simeq \empheqlbrace }]{align}
    & \onehalf-\frac{\sqrt{2}}{\pi} ,\qquad & \time\ll \tscal^{-2}, \label{eq_linB_CCF_sum_early}\\
    & \frac{1}{2\pi \tscal^2} \time^{-1}, & \tscal^{-2}\ll \time \ll \tscal^{-1}, \label{eq_linB_CCF_sum_interm}\\
    & \frac{2}{\tscal}, & \time\gg \tscal^{-1}. \label{eq_linB_CCF_sum_late}
   \end{empheq}
\end{subequations}
For $0\leq  \tscal \,\lesssim \Ocal(1)$, instead, the critical early- and late-time expressions given in \cref{eq_CCF_lin_crit_early,eq_CCF_lin_crit_late} apply, which occur for times $\time \lesssim 10^{-4}$ and $\time\gtrsim 10^{-2}$, respectively.

\subsection{Critical Casimir force within nonlinear mean field theory}

\begin{figure}[t]\centering
    \includegraphics[width=0.2\linewidth]{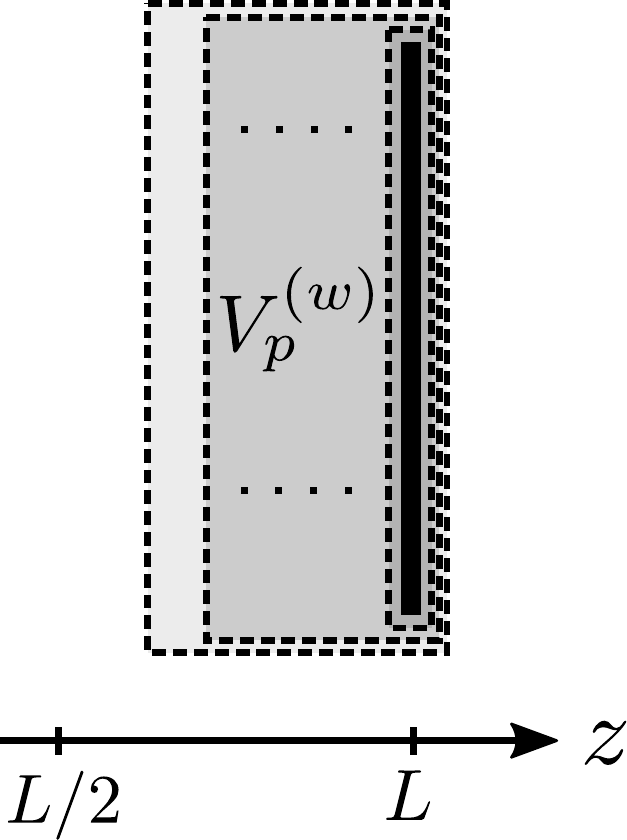}  \hfill
    \caption{Illustration of the method used for determining numerically the CCF within the nonlinear model B. In order to minimize numerical errors, the CCF is calculated as an average over various integration volumes $V_p^{(w)}$ [see \cref{eq_CCF_num_avg}], which, starting from the outer one, progressively shrink towards the confining surface, as indicated by the dots in the figure. The left surface of $V_p^{(w)}$ is located at $z=w$ with $L/2<w<L$.}
    \label{fig_int_avg_CCF}
\end{figure}

For the nonlinear model B, the time-dependent OP profile required to evaluate the CCF is determined numerically based on a finite-difference approximation of the dynamical equations as described in \cref{sec_nonlin_dyn}.
However, this approach can lead to numerical inaccuracies whenever the gradient of the OP profile is large, which is typically the case near a wall.
In order to reduce the influence of this source of error for the CCF, we evaluate the latter by using the freedom in the choice of the integration volume $V_p$ [see Ref.\ \cite{dean_out--equilibrium_2010} and \cref{eq_inst_force}]. 
Specifically, we consider a volume $V_p^{(w)}$ enclosing the right plate which has a surface within the film at position $z=w$ with $L/2<w<L$ (see \cref{fig_int_avg_CCF}).
Since this surface is \emph{not} infinitesimally close to the plate, the second term in in the second equation of \cref{eq_force_stress_gen} does not vanish and the generalized force (per area) is given by
\beq \Kcal(w) = \bar\Tcal_{zz}\big|_{z=w} - \bar\Tcal_{zz,b} + \int_{w}^{L-\epsilon} \d z'\, \phi(z')\pd_{z'} \mu(z') .
\label{eq_CCF_arbvol}\eeq 
Here, $\epsilon$ is an infinitesimal quantity which ensures that the integration ends just next to the inner surface of the boundary, noting that next to the outer boundary (i.e., in the bulk) the corresponding contribution would vanish owing to no-flux \bcs.
The CCF (which is taken per area), being actually independent of $w$, can be conveniently obtained from \cref{eq_CCF_arbvol} as an average over all possible locations $w$ between $L/2$ and $L$ \footnote{The last expression in \cref{eq_CCF_num_avg} follows by noting that, for an arbitrary function $f(z')$, one has $\int_{L/2}^{L-\epsilon}\d w \int_w^{L-\epsilon} \d z' f(z') = \int_{L/2}^{L-\epsilon}\d z' \int_{L/2}^{z'} \d w f(z') = \int_{L/2}^{L-\epsilon} \d z' (z'-(L/2)) f(z')$.}:
\beq\begin{split} \Kcal =  \frac{1}{L/2} \int_{L/2}^{L-\epsilon} \d w\, \Kcal(w) &= \frac{1}{L/2} \int_{L/2}^{L-\epsilon} \d w\, \left[ \bar\Tcal_{zz}\big|_{z=w}  + \int_{w}^{L-\epsilon} \d z'\, \phi(z') \pd_{z'} \mu(z') \right] - \bar \Tcal_{zz,b} \\
&=  \frac{2}{L} \int_{L/2}^{L-\epsilon} \d w\, \left[ \bar\Tcal_{zz}\big|_{z=w}  + \left(w-\frac{L}{2}\right) \phi(w) \pd_{w} \mu(w) \right] - \bar \Tcal_{zz,b} .
\end{split}\label{eq_CCF_num_avg}\eeq 
We have checked numerically that this procedure renders the equilibrium value of the CCF reported in \cref{eq_CCF_lin_crit_late,eq_CCF_lin_crit_eq_largeT} as well as in Ref.\ \cite{gross_critical_2016}.

\begin{figure}[t]\centering
    \includegraphics[width=0.44\linewidth]{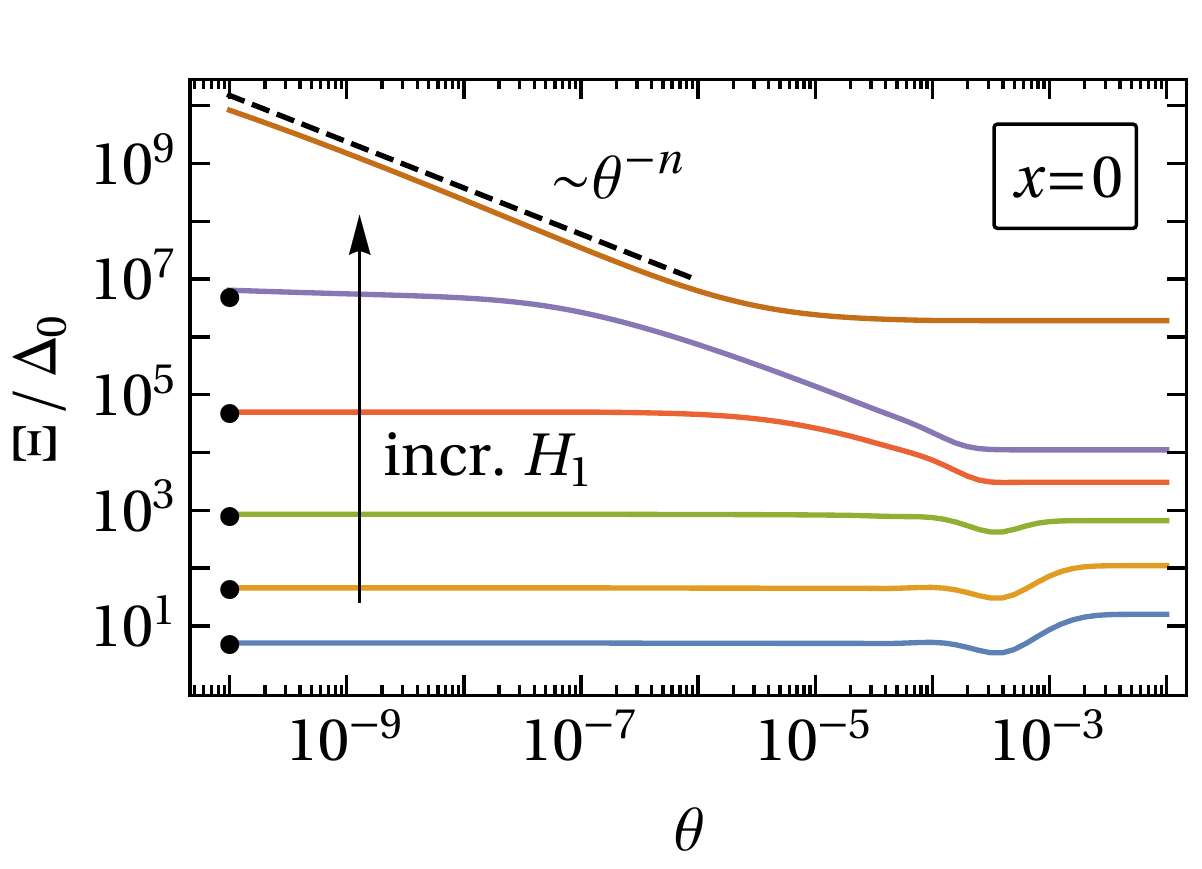} 
    \caption{Dynamical CCF [\cref{eq_CCF}] within the nonlinear model B, computed according to \cref{eq_CCF_num_avg}. The values of the surface field $H_1$ are (from bottom to top) $H_1=10, 30, 130, 10^3, 10^4$, and $10^6$. The black dots correspond to the values of $\Xi$ at early times [\cref{eq_CCF_lin_crit_early}], which are determined by the linear MFT. The dashed line indicates an algebraic behavior with an effective exponent $n\simeq 0.8$. }
    \label{fig_CCF_nlin_Tc}
\end{figure}

In \cref{fig_CCF_nlin_Tc}, the CCF obtained from \cref{eq_CCF_num_avg} is shown as a function of time at criticality ($\tscal=0$) for various values of the surface field $H_1$.
According to the analysis in \cref{sec_nonlin_dyn}, for $\tscal=0$ the dynamics of the profile is governed by the linear model B for times $\time \lesssim \time^*_3\sim H_1^{-2}$ [\cref{eq_timescale_nlin}].
This applies also to the CCF, which, in this regime, follows the prediction given in \cref{eq_CCF_lin_crit_early} (black dots in \cref{fig_CCF_nlin_Tc}).
Conversely, at times $\time\gtrsim 100\times \time^*_e\simeq 10^{-2}$, the CCF has essentially reached its equilibrium value, given by (see \cref{eq_CCF_lin_crit_late} and Ref.\ \cite{gross_critical_2016})
\begin{subequations}
\label{eq_nlinB_CCF}
  \begin{empheq}[left ={ \Xi\eq(\tscal=0,H_1)/\Delta_0 \equiv \Xi(\time\gg \time^*_e,\tscal = 0, H_1)/\Delta_0 \simeq \empheqlbrace }]{align}
    & \frac{1}{6}H_1^2,\qquad & H_1 \ll 100, \label{eq_linB_CCF_eq}\\
    & 1.5\times (\ln H_1)^4, &  H_1 \gg 100. \label{eq_nlinB_CCF_eq}
   \end{empheq}
\end{subequations}
We recall that the monotonic increase of $\Xi\eq$ upon increasing $H_1$ in \cref{eq_nlinB_CCF_eq} is a consequence of the conserved mass in the film [\cref{eq_mass}] and of the fact that $\op\eq(\zeta\to 0)\sim \zeta^{-1}$ within the nonlinear MFT (see Ref.\ \cite{gross_critical_2016}).
For large $H_1\gg 1$, the dynamics of the profile is affected by the nonlinear term (see \cref{sec_nonlin_dyn}), which is reflected in the CCF by the emergence of a characteristic intermediate asymptotic regime occurring for times $\time^*_3\lesssim \time \lesssim \time^*_e$ (see \cref{fig_CCF_nlin_Tc}). 
A numerical analysis reveals that, within this regime, the CCF follows an algebraic decay, $\Xi\propto \time^{-n}$, with an effective exponent $n\simeq 0.8$ (dashed line in \cref{fig_CCF_nlin_Tc}).

It was shown in Ref.\ \cite{gross_critical_2016} that, as a consequence of the mass constraint, the equilibrium CCF in the canonical ensemble is repulsive for $(++)$ \bcs.
Here, we find that this repulsive character applies also to the non-equilibrium CCF over the whole time evolution.

\section{Summary and Outlook}

In the present study, we have investigated the dynamics of the order parameter (OP) and of the critical Casimir force (CCF) in a fluid film after an instantaneous quench from a homogeneous high-temperature phase to a (rescaled) near-critical temperature $\tscal$ [\cref{eq_FSS_temp}]. 
The dynamics of the medium is taken to be described by model B within mean field theory, i.e., by a diffusive transport equation without noise. 
Initially, the OP profile $\op(\time=0,\zeta)=0$ vanishes across the film [\cref{eq_FSS_op,eq_FSS_time,eq_FSS_z}].
The driving force for the dynamics stems from the presence of symmetric surface fields, which, in the long-time limit $\time\to\infty$, give rise to an inhomogeneous OP profile across the film characteristic of critical adsorption with $(++)$ \bcs.
The model B dynamics is supplemented by no-flux \bcs acting at the boundaries of the film, such that the total integrated OP within the film is constant at all times: $\int_0^1 \d\zeta\, \op(\time,\zeta)=0$. 
Accordingly, the model used here realizes the canonical ensemble---the equilibrium, time-independent properties of which have been previously studied in Ref.\ \cite{gross_critical_2016} and have been shown to lead to pronounced differences in the behavior of the CCF compared to the usual grand canonical ensemble.
The analytical solution of the linearized model B is supplemented by a numerical solution of the full nonlinear model B equations.
Our main findings are summarized as follows:
\begin{enumerate}

 \item For small values of the surface field $H_1\lesssim 100$ [\cref{eq_FSS_H1}] as well as for large rescaled temperatures $\tscal\gg H_1$, the linear model B provides an accurate description of the full mean field dynamics of the OP in the film.  

 \item Both at criticality ($\tscal=0$) and away from it ($\tscal\gg 1$), the OP $\op(\time,\zeta \in \{0,1\})$ at the walls increases algebraically $\propto \time^{1/\zdyn}$ at early times, where $\zdyn=4$ is the dynamic critical exponent of model B within MFT [see \cref{fig_OPwall_small_t_Tc,fig_OPwall_small_t_abvTc,fig_nlin_OPwall_small_t_Tc}]. At late times, the OP at the wall saturates exponentially towards its nonzero equilibrium value [see \cref{fig_OPwall_large_t_Tc,fig_OPwall_large_t_abvTc,fig_nlin_OPwall_large_t_Tc}]. 
 These two characteristic behaviors occur both within linear and nonlinear model B dynamics. 
 
 \item For quench temperatures far from criticality ($\tscal\gg 1$) as well as for large values of the surface field ($H_1\gg 1$), an intermediate asymptotic regime emerges between the early- and late-time regime of the OP. Within this intermediate asymptotic regime the OP at the wall saturates algebraically [see \cref{fig_OPwall_interm_t_abvTc,fig_nlin_OPwall_interm_t_Tc}].
 
 \item As a consequence of the diffusive nature of the OP transport, in the film two symmetric minima of the OP profile emerge immediately after the quench, moving subdiffusively (with a temporal exponent $1/\zdyn$) from each boundary towards the center of the film [see \cref{fig_minpos_evol_Tc,fig_minpos_evol_abvTc,fig_nlin_minPos_Tc}]. 
 
 \item We have introduced a dynamical nonequilibrium CCF $\Kcal$ [see \cref{eq_CCF_scal}] based on the notion of a generalized force generated by an OP field interacting with an inclusion \cite{dean_out--equilibrium_2010}. In the presence of no-flux \bcs, the dynamical CCF can be expressed in terms of a dynamic stress tensor [see \cref{eq_dyn_stressten}], which, in equilibrium, reduces to the expressions well-known for the canonical \cite{gross_critical_2016} and for the grand canonical ensemble \cite{krech_casimir_1994}, respectively.
 
 \item For model B in the film geometry with $(++)$ \bcs, we find that the nonequilibrium CCF $\Kcal$ is typically repulsive at all times. At late times, $\Kcal$ approaches the equilibrium value of the CCF in the canonical ensemble, which has been analyzed previously in Ref.\ \cite{gross_critical_2016}. At short times, the nonequilibrium CCF is non-vanishing and its value is reliably predicted by the linear model B (see \cref{fig_CCF_nlin_Tc}). Depending on the values of the various parameters, the time-dependence of $\Kcal$ may be non-monotonic.
 
\end{enumerate}

As far as future studies are concerned, it would be interesting to assess to which extent the actual critical dynamics of a fluid film (model H) differs from that of model B. This could be addressed, e.g., via Molecular Dynamics or Lattice Boltzmann simulations \cite{roy_structure_2016, puosi_direct_2016, belardinelli_fluctuating_2015}. 
Extending the present study to $(+-)$ \bcs appears to be a natural and rewarding step.
Furthermore, the quench dynamics of the CCF for \bcs, which differ from the ones describing critical adsorption, deserves to be studied.
In particular, for non-symmetry breaking \bcs, such as Dirichlet \bcs, the CCF stems solely from thermal fluctuations and the resulting quench dynamics in such films is yet unexplored. 
Finally, the non-equilibrium dynamics of colloids immersed in a near-critical solvent and driven by CCFs \cite{furukawa_nonequilibrium_2013} promises to be a fruitful topic for future studies.

\appendix

\section{Model B in a half-space}
\label{app_halfspace}

We consider \cref{eq_linB_lapl_flat} in Laplace space:
\beq s \hat\op(s,\zeta) = - \pd_\zeta^4 \hat \op(s,\zeta) + \tscal\pd_\zeta^2\hat\op(s,\zeta),
\label{eqHS_linB_lapl}\eeq
subject to a flat initial configuration [see \cref{eq_prof_init}] and to the appropriate \bcs in the half-space: 
\begin{subequations}\begin{align}
\pd_\zeta \op(\time,\zeta=0) &= -H_1, \label{eqHS_bcs_CA}\\
\pd_\zeta^3 \op(\time,\zeta=0) - \tscal \pd_\zeta \op(\time,\zeta=0)  &= 0,\\
\pd_\zeta^{(n)} \op(\time,\zeta\to\infty)&=0, \qquad \text{for all $n\geq 0$,}
\label{eqHS_bcs_noflx}
\end{align}\label{eqHS_bcs}\end{subequations}
which represent the critical adsorption and no-flux conditions of the OP at the wall, and the flatness of the OP profile far from the wall, respectively.
Following Refs.\ \cite{ball_spinodal_1990, lee_filler-induced_1999}, we introduce the Fourier cosine transform,
\beq \tilde \op(s,k) = \int_0^\infty \d \zeta\, \hat \op(s,\zeta) \cos(k\zeta)
\label{eq_FourCosTransf}\eeq 
and its inverse,
\beq \hat \op(s,\zeta) = \frac{2}{\pi} \int_0^\infty \d k\, \tilde \op(s,k) \cos(k\zeta).
\eeq 
Applying these transforms and using the boundary conditions in \cref{eqHS_bcs}, yields the solution of \cref{eqHS_linB_lapl} in Laplace-Fourier space:
\beq \tilde \op(s, k) = - \frac{k^2\, \pd_\zeta \hat\op(s,\zeta=0)}{s+k^2(k^2+\tscal)} = \frac{H_1 k^2}{s [s+k^2(k^2+\tscal)]} .
\label{eqHS_linB_solLF}\eeq 
In the second equation, we have used the fact that in Laplace space \cref{eqHS_bcs_CA} turns into $\pd_\zeta\hat\op(s,\zeta=0) = -H_1/s$ .
Performing the Laplace inversion of \cref{eqHS_linB_solLF} yields
\beq \tilde \op(\time, k) = H_1\frac{1-\exp(-k^2 (k^2+\tscal) \time)}{k^2+\tscal}.
\label{eqHS_linB_solF}\eeq 

At criticality ($\tscal=0$), the inverse Fourier transform of \cref{eqHS_linB_solF} yields 
\beq \op(\time, \zeta, \tscal, H_1)\big|_{\tscal=0} = H_1\time^{1/4} \Mcal(\zeta/\time^{1/4}), 
\label{eqHS_linB_sol}\eeq 
with the scaling function 
\beq \Mcal(\Theta) =  \frac{2}{\pi} \Gamma \pfrac{3}{4} {_1 F_3} \left(-\frac{1}{4}; \frac{1}{4}, \frac{1}{2}, \frac{3}{4}; \pfrac{\Theta}{4}^4\right) + \frac{1}{\pi} \Theta^2  \Gamma\pfrac{5}{4} {_1 F_3}\left(\frac{1}{4}; \frac{3}{4}, \frac{5}{4}, \frac{3}{2}; \pfrac{\Theta}{4}^4\right) - \Theta,
\label{eqHS_linB_solTc_scalf}\eeq 
where ${_1 F_3}$ is a hypergeometric function \cite{olver_nist_2010}.
This expression also provides the asymptotic short-time scaling function for model B in a film [see \cref{eq_linB_profTc_scalf}].
Accordingly, for $\op(\time,\zeta=0)$ we recover the same expression as that given in \cref{eq_linB_wall_sum_early_0x} and obtain the same scaling behavior of the global minimum of $\op(\time,\zeta)$ as in \cref{eq_linB_wall_zmin}. 
However, in contrast to the film, the \emph{critical} profile in a half-space does not saturate but instead increases at all times according to \cref{eqHS_linB_sol}.

\begin{figure}[t]\centering
    \includegraphics[width=0.41\linewidth]{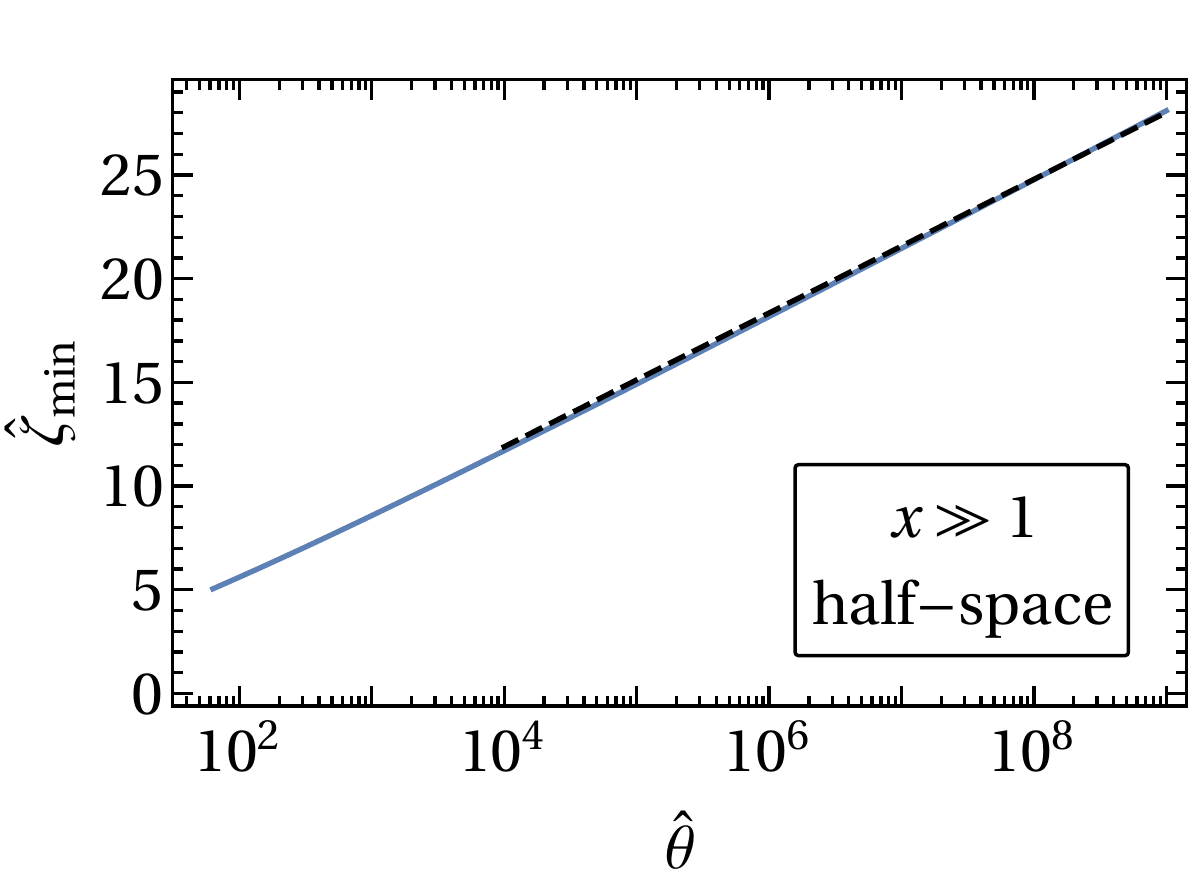} 
    \caption{Rescaled position $\hat\zeta\st{min} = \sqrt{\tscal}\zeta\st{min}$ of the minimum of the profile $\op(\hat\time,\hat\zeta)$ in the half-space $\zeta>0$ for $\tscal\gg 1$ [\cref{eqHS_linB_solSupc_resc}] as a function of the rescaled time $\hat\time = \tscal^2\time$. For $\hat\zeta\st{min}\gg 1$ it turns out that the numerical data for $\hat\zeta\st{min}$ are captured accurately by a logarithmic dependence on $\time$: $\hat\zeta\st{min}\simeq 1.4\times \ln(\hat\time)-1$ (dashed line). The numerical prefactor and the constant involved follow from a fit.}
    \label{fig_HS_minpos}
\end{figure}

Asymptotically for large $\tscal\gg 1$, in \cref{eqHS_linB_solF} one can use the approximation $k^2+\tscal\simeq \tscal$, which renders the inverse Fourier transform 
\beq \op(\time,\zeta\gg \tscal^{-1/2})\big|_{\tscal\gg 1} \simeq \frac{H_1}{\sqrt{\tscal}} \exp(-\zeta\tscal) - \frac{H_1}{\sqrt{\pi} \tscal^{3/2} \sqrt{\time}}\exp\left(-\frac{\zeta^2}{4 \time\tscal}\right).
\label{eqHS_linB_solSupc}\eeq 
This expression applies to distances $\zeta\gg \tscal^{-1/2}$ from the wall.
The first term on the r.h.s.\ in \cref{eqHS_linB_solSupc} represents the asymptotic equilibrium profile [see \cref{eq_eqprof_largeT}]. 
In fact, for $\tscal\neq 0$ the profile saturates at late times, in contrast to the situation at criticality [see \cref{eqHS_linB_sol}].
In order to proceed, we introduce the rescaled variables $\hat\zeta=\sqrt{\tscal}\zeta$ and $\hat\time = \tscal^2 \time$, in terms of which \cref{eqHS_linB_solSupc} turns into
\beq \op(\hat\time,\hat\zeta\gg 1)\big|_{\tscal\gg 1} \simeq \frac{H_1}{\sqrt{\tscal}}\left[ \exp(-\hat\zeta) - \frac{1}{\sqrt{\pi}\hat\time}\exp\left(-\frac{\hat\zeta^2}{4\hat\time}\right) \right] .
\label{eqHS_linB_solSupc_resc}\eeq 
The position $\hat\zeta\st{min}$ of the global minimum of these profiles is found to increase approximately logarithmically in time,
\beq \hat\zeta\st{min}(\hat\time) \simeq 1.4\times \ln(\hat\time) -1,
\label{eqHS_linB_minScal}\eeq 
as illustrated in \cref{fig_HS_minpos}. 
Accordingly, in terms of the original scaling variables, one has $\zeta\st{min}\simeq \tscal^{-1/2}[1.4\times \ln(\tscal^2\time)-1]$.

\begin{thebibliography}{62}%
\makeatletter
\providecommand \@ifxundefined [1]{%
 \@ifx{#1\undefined}
}%
\providecommand \@ifnum [1]{%
 \ifnum #1\expandafter \@firstoftwo
 \else \expandafter \@secondoftwo
 \fi
}%
\providecommand \@ifx [1]{%
 \ifx #1\expandafter \@firstoftwo
 \else \expandafter \@secondoftwo
 \fi
}%
\providecommand \natexlab [1]{#1}%
\providecommand \enquote  [1]{``#1''}%
\providecommand \bibnamefont  [1]{#1}%
\providecommand \bibfnamefont [1]{#1}%
\providecommand \citenamefont [1]{#1}%
\providecommand \href@noop [0]{\@secondoftwo}%
\providecommand \href [0]{\begingroup \@sanitize@url \@href}%
\providecommand \@href[1]{\@@startlink{#1}\@@href}%
\providecommand \@@href[1]{\endgroup#1\@@endlink}%
\providecommand \@sanitize@url [0]{\catcode `\\12\catcode `\$12\catcode
  `\&12\catcode `\#12\catcode `\^12\catcode `\_12\catcode `\%12\relax}%
\providecommand \@@startlink[1]{}%
\providecommand \@@endlink[0]{}%
\providecommand \url  [0]{\begingroup\@sanitize@url \@url }%
\providecommand \@url [1]{\endgroup\@href {#1}{\urlprefix }}%
\providecommand \urlprefix  [0]{URL }%
\providecommand \Eprint [0]{\href }%
\providecommand \doibase [0]{http://dx.doi.org/}%
\providecommand \selectlanguage [0]{\@gobble}%
\providecommand \bibinfo  [0]{\@secondoftwo}%
\providecommand \bibfield  [0]{\@secondoftwo}%
\providecommand \translation [1]{[#1]}%
\providecommand \BibitemOpen [0]{}%
\providecommand \bibitemStop [0]{}%
\providecommand \bibitemNoStop [0]{.\EOS\space}%
\providecommand \EOS [0]{\spacefactor3000\relax}%
\providecommand \BibitemShut  [1]{\csname bibitem#1\endcsname}%
\let\auto@bib@innerbib\@empty
\bibitem [{\citenamefont {Hohenberg}\ and\ \citenamefont
  {Halperin}(1977)}]{hohenberg_theory_1977}%
  \BibitemOpen
  \bibfield  {author} {\bibinfo {author} {\bibfnamefont {P.~C.}\ \bibnamefont
  {Hohenberg}}\ and\ \bibinfo {author} {\bibfnamefont {B.~I.}\ \bibnamefont
  {Halperin}},\ }\bibfield  {title} {\enquote {\bibinfo {title} {Theory of
  dynamic critical phenomena},}\ }\href@noop {} {\bibfield  {journal} {\bibinfo
   {journal} {Rev. Mod. Phys.}\ }\textbf {\bibinfo {volume} {49}},\ \bibinfo
  {pages} {435} (\bibinfo {year} {1977})}\BibitemShut {NoStop}%
\bibitem [{\citenamefont {Folk}\ and\ \citenamefont
  {Moser}(2006)}]{folk_critical_2006}%
  \BibitemOpen
  \bibfield  {author} {\bibinfo {author} {\bibfnamefont {R.}~\bibnamefont
  {Folk}}\ and\ \bibinfo {author} {\bibfnamefont {G.}~\bibnamefont {Moser}},\
  }\bibfield  {title} {\enquote {\bibinfo {title} {Critical dynamics: a
  field-theoretical approach},}\ }\href@noop {} {\bibfield  {journal} {\bibinfo
   {journal} {J. Phys. A}\ }\textbf {\bibinfo {volume} {39}},\ \bibinfo {pages}
  {R207} (\bibinfo {year} {2006})}\BibitemShut {NoStop}%
\bibitem [{\citenamefont {T\"auber}(2014)}]{tauber_critical_2014}%
  \BibitemOpen
  \bibfield  {author} {\bibinfo {author} {\bibfnamefont {U.~C.}\ \bibnamefont
  {T\"auber}},\ }\href@noop {} {\emph {\bibinfo {title} {Critical {Dynamics}:
  {A} {Field} {Theory} {Approach} to {Equilibrium} and {Non}-{Equilibrium}
  {Scaling} {Behavior}}}}\ (\bibinfo  {publisher} {Cambridge University
  Press},\ \bibinfo {year} {2014})\BibitemShut {NoStop}%
\bibitem [{\citenamefont {Dietrich}\ and\ \citenamefont
  {Diehl}(1983)}]{dietrich_effects_1983}%
  \BibitemOpen
  \bibfield  {author} {\bibinfo {author} {\bibfnamefont {S.}~\bibnamefont
  {Dietrich}}\ and\ \bibinfo {author} {\bibfnamefont {H.~W.}\ \bibnamefont
  {Diehl}},\ }\bibfield  {title} {\enquote {\bibinfo {title} {The effects of
  surfaces on dynamic critical behavior},}\ }\href {\doibase
  10.1007/BF01319217} {\bibfield  {journal} {\bibinfo  {journal} {Z. Phys. B}\
  }\textbf {\bibinfo {volume} {51}},\ \bibinfo {pages} {343} (\bibinfo {year}
  {1983})}\BibitemShut {NoStop}%
\bibitem [{\citenamefont {Diehl}\ and\ \citenamefont
  {Janssen}(1992)}]{diehl_boundary_1992}%
  \BibitemOpen
  \bibfield  {author} {\bibinfo {author} {\bibfnamefont {H.~W.}\ \bibnamefont
  {Diehl}}\ and\ \bibinfo {author} {\bibfnamefont {H.~K.}\ \bibnamefont
  {Janssen}},\ }\bibfield  {title} {\enquote {\bibinfo {title} {Boundary
  conditions for the field theory of dynamic critical behavior in semi-infinite
  systems with conserved order parameter},}\ }\href {\doibase
  10.1103/PhysRevA.45.7145} {\bibfield  {journal} {\bibinfo  {journal} {Phys.
  Rev. A}\ }\textbf {\bibinfo {volume} {45}},\ \bibinfo {pages} {7145}
  (\bibinfo {year} {1992})}\BibitemShut {NoStop}%
\bibitem [{\citenamefont {Diehl}(1994)}]{diehl_universality_1994}%
  \BibitemOpen
  \bibfield  {author} {\bibinfo {author} {\bibfnamefont {H.~W.}\ \bibnamefont
  {Diehl}},\ }\bibfield  {title} {\enquote {\bibinfo {title} {Universality
  classes for the dynamic surface critical behavior of systems with
  relaxational dynamics},}\ }\href@noop {} {\bibfield  {journal} {\bibinfo
  {journal} {Phys. Rev. B}\ }\textbf {\bibinfo {volume} {49}},\ \bibinfo
  {pages} {2846} (\bibinfo {year} {1994})}\BibitemShut {NoStop}%
\bibitem [{\citenamefont {Wichmann}\ and\ \citenamefont
  {Diehl}(1995)}]{wichmann_dynamic_1995}%
  \BibitemOpen
  \bibfield  {author} {\bibinfo {author} {\bibfnamefont {F.}~\bibnamefont
  {Wichmann}}\ and\ \bibinfo {author} {\bibfnamefont {H.~W.}\ \bibnamefont
  {Diehl}},\ }\bibfield  {title} {\enquote {\bibinfo {title} {Dynamic surface
  critical behavior of systems with conserved bulk order parameter: {Detailed}
  {RG} analysis of the semi-infinite extensions of model {B} with and without
  nonconservative surface terms},}\ }\href {\doibase 10.1007/BF01307476}
  {\bibfield  {journal} {\bibinfo  {journal} {Z. Phys. B}\ }\textbf {\bibinfo
  {volume} {97}},\ \bibinfo {pages} {251} (\bibinfo {year} {1995})}\BibitemShut
  {NoStop}%
\bibitem [{\citenamefont {Ritschel}\ and\ \citenamefont
  {Czerner}(1995)}]{ritschel_universal_1995}%
  \BibitemOpen
  \bibfield  {author} {\bibinfo {author} {\bibfnamefont {U.}~\bibnamefont
  {Ritschel}}\ and\ \bibinfo {author} {\bibfnamefont {P.}~\bibnamefont
  {Czerner}},\ }\bibfield  {title} {\enquote {\bibinfo {title} {Universal
  {Short}-{Time} {Behavior} in {Critical} {Dynamics} near {Surfaces}},}\ }\href
  {\doibase 10.1103/PhysRevLett.75.3882} {\bibfield  {journal} {\bibinfo
  {journal} {Phys. Rev. Lett.}\ }\textbf {\bibinfo {volume} {75}},\ \bibinfo
  {pages} {3882} (\bibinfo {year} {1995})}\BibitemShut {NoStop}%
\bibitem [{\citenamefont {Majumdar}\ and\ \citenamefont
  {Sengupta}(1996)}]{majumdar_nonequilibrium_1996}%
  \BibitemOpen
  \bibfield  {author} {\bibinfo {author} {\bibfnamefont {S.~N.}\ \bibnamefont
  {Majumdar}}\ and\ \bibinfo {author} {\bibfnamefont {A.~M.}\ \bibnamefont
  {Sengupta}},\ }\bibfield  {title} {\enquote {\bibinfo {title} {Nonequilibrium
  {Dynamics} following a {Quench} to the {Critical} {Point} in a
  {Semi}-infinite {System}},}\ }\href {\doibase 10.1103/PhysRevLett.76.2394}
  {\bibfield  {journal} {\bibinfo  {journal} {Phys. Rev. Lett.}\ }\textbf
  {\bibinfo {volume} {76}},\ \bibinfo {pages} {2394} (\bibinfo {year}
  {1996})}\BibitemShut {NoStop}%
\bibitem [{\citenamefont {Pleimling}(2004)}]{pleimling_aging_2004}%
  \BibitemOpen
  \bibfield  {author} {\bibinfo {author} {\bibfnamefont {M.}~\bibnamefont
  {Pleimling}},\ }\bibfield  {title} {\enquote {\bibinfo {title} {Aging
  phenomena in critical semi-infinite systems},}\ }\href {\doibase
  10.1103/PhysRevB.70.104401} {\bibfield  {journal} {\bibinfo  {journal} {Phys.
  Rev. B}\ }\textbf {\bibinfo {volume} {70}},\ \bibinfo {pages} {104401}
  (\bibinfo {year} {2004})}\BibitemShut {NoStop}%
\bibitem [{\citenamefont {Diehl}\ and\ \citenamefont
  {Ritschel}(1993)}]{diehl_dynamical_1993}%
  \BibitemOpen
  \bibfield  {author} {\bibinfo {author} {\bibfnamefont {H.~W.}\ \bibnamefont
  {Diehl}}\ and\ \bibinfo {author} {\bibfnamefont {U.}~\bibnamefont
  {Ritschel}},\ }\bibfield  {title} {\enquote {\bibinfo {title} {Dynamical
  relaxation and universal short-time behavior of finite systems},}\ }\href
  {\doibase 10.1007/BF01052748} {\bibfield  {journal} {\bibinfo  {journal} {J.
  Stat. Phys.}\ }\textbf {\bibinfo {volume} {73}},\ \bibinfo {pages} {1}
  (\bibinfo {year} {1993})}\BibitemShut {NoStop}%
\bibitem [{\citenamefont {Ritschel}\ and\ \citenamefont
  {Diehl}(1995)}]{ritschel_long-time_1995}%
  \BibitemOpen
  \bibfield  {author} {\bibinfo {author} {\bibfnamefont {U.}~\bibnamefont
  {Ritschel}}\ and\ \bibinfo {author} {\bibfnamefont {H.~W.}\ \bibnamefont
  {Diehl}},\ }\bibfield  {title} {\enquote {\bibinfo {title} {Long-time traces
  of the initial condition in relaxation phenomena near criticality},}\ }\href
  {\doibase 10.1103/PhysRevE.51.5392} {\bibfield  {journal} {\bibinfo
  {journal} {Phys. Rev. E}\ }\textbf {\bibinfo {volume} {51}},\ \bibinfo
  {pages} {5392} (\bibinfo {year} {1995})}\BibitemShut {NoStop}%
\bibitem [{\citenamefont {Ritschel}\ and\ \citenamefont
  {Diehl}(1996)}]{ritschel_dynamical_1996}%
  \BibitemOpen
  \bibfield  {author} {\bibinfo {author} {\bibfnamefont {U.}~\bibnamefont
  {Ritschel}}\ and\ \bibinfo {author} {\bibfnamefont {H.~W.}\ \bibnamefont
  {Diehl}},\ }\bibfield  {title} {\enquote {\bibinfo {title} {Dynamical
  relaxation and universal short-time behavior in finite systems. {The}
  renormalization-group approach},}\ }\href {\doibase
  10.1016/0550-3213(96)00012-0} {\bibfield  {journal} {\bibinfo  {journal}
  {Nucl. Phys. B}\ }\textbf {\bibinfo {volume} {464}},\ \bibinfo {pages} {512}
  (\bibinfo {year} {1996})}\BibitemShut {NoStop}%
\bibitem [{\citenamefont {Gambassi}\ and\ \citenamefont
  {Dietrich}(2006)}]{gambassi_critical_2006}%
  \BibitemOpen
  \bibfield  {author} {\bibinfo {author} {\bibfnamefont {A.}~\bibnamefont
  {Gambassi}}\ and\ \bibinfo {author} {\bibfnamefont {S.}~\bibnamefont
  {Dietrich}},\ }\bibfield  {title} {\enquote {\bibinfo {title} {Critical
  {Dynamics} in {Thin} {Films}},}\ }\href@noop {} {\bibfield  {journal}
  {\bibinfo  {journal} {J. Stat. Phys.}\ }\textbf {\bibinfo {volume} {123}},\
  \bibinfo {pages} {929} (\bibinfo {year} {2006})}\BibitemShut {NoStop}%
\bibitem [{\citenamefont {Diehl}\ and\ \citenamefont
  {Chamati}(2009)}]{diehl_dynamic_2009}%
  \BibitemOpen
  \bibfield  {author} {\bibinfo {author} {\bibfnamefont {H.~W.}\ \bibnamefont
  {Diehl}}\ and\ \bibinfo {author} {\bibfnamefont {H.}~\bibnamefont
  {Chamati}},\ }\bibfield  {title} {\enquote {\bibinfo {title} {Dynamic
  critical behavior of model {A} in films: {Zero}-mode boundary conditions and
  expansion near four dimensions},}\ }\href@noop {} {\bibfield  {journal}
  {\bibinfo  {journal} {Phys. Rev. B}\ }\textbf {\bibinfo {volume} {79}},\
  \bibinfo {pages} {104301} (\bibinfo {year} {2009})}\BibitemShut {NoStop}%
\bibitem [{\citenamefont {Gambassi}(2008)}]{gambassi_relaxation_2008}%
  \BibitemOpen
  \bibfield  {author} {\bibinfo {author} {\bibfnamefont {A.}~\bibnamefont
  {Gambassi}},\ }\bibfield  {title} {\enquote {\bibinfo {title} {Relaxation
  phenomena at criticality},}\ }\href {\doibase 10.1140/epjb/e2008-00043-y}
  {\bibfield  {journal} {\bibinfo  {journal} {Eur. Phys. J. B}\ }\textbf
  {\bibinfo {volume} {64}},\ \bibinfo {pages} {379} (\bibinfo {year}
  {2008})}\BibitemShut {NoStop}%
\bibitem [{\citenamefont {Ball}\ and\ \citenamefont
  {Essery}(1990)}]{ball_spinodal_1990}%
  \BibitemOpen
  \bibfield  {author} {\bibinfo {author} {\bibfnamefont {R.~C.}\ \bibnamefont
  {Ball}}\ and\ \bibinfo {author} {\bibfnamefont {R.~L.~H.}\ \bibnamefont
  {Essery}},\ }\bibfield  {title} {\enquote {\bibinfo {title} {Spinodal
  decomposition and pattern formation near surfaces},}\ }\href {\doibase
  10.1088/0953-8984/2/51/006} {\bibfield  {journal} {\bibinfo  {journal} {J.
  Phys.: Condens. Matter}\ }\textbf {\bibinfo {volume} {2}},\ \bibinfo {pages}
  {10303} (\bibinfo {year} {1990})}\BibitemShut {NoStop}%
\bibitem [{\citenamefont {Binder}\ and\ \citenamefont
  {Frisch}(1991)}]{binder_dynamics_1991}%
  \BibitemOpen
  \bibfield  {author} {\bibinfo {author} {\bibfnamefont {K.}~\bibnamefont
  {Binder}}\ and\ \bibinfo {author} {\bibfnamefont {H.~L.}\ \bibnamefont
  {Frisch}},\ }\bibfield  {title} {\enquote {\bibinfo {title} {Dynamics of
  surface enrichment: {A} theory based on the {Kawasaki} spin-exchange model in
  the presence of a wall},}\ }\href {\doibase 10.1007/BF01314015} {\bibfield
  {journal} {\bibinfo  {journal} {Z. Phys. B}\ }\textbf {\bibinfo {volume}
  {84}},\ \bibinfo {pages} {403} (\bibinfo {year} {1991})}\BibitemShut
  {NoStop}%
\bibitem [{\citenamefont {Puri}\ and\ \citenamefont
  {Frisch}(1993)}]{puri_dynamics_1993}%
  \BibitemOpen
  \bibfield  {author} {\bibinfo {author} {\bibfnamefont {S.}~\bibnamefont
  {Puri}}\ and\ \bibinfo {author} {\bibfnamefont {H.~L.}\ \bibnamefont
  {Frisch}},\ }\bibfield  {title} {\enquote {\bibinfo {title} {Dynamics of
  surface enrichment. {Phenomenology} and numerical results above the bulk
  critical temperature},}\ }\href {\doibase 10.1063/1.465948} {\bibfield
  {journal} {\bibinfo  {journal} {J. Chem. Phys.}\ }\textbf {\bibinfo {volume}
  {99}},\ \bibinfo {pages} {5560} (\bibinfo {year} {1993})}\BibitemShut
  {NoStop}%
\bibitem [{\citenamefont {Lee}\ \emph {et~al.}(1999)\citenamefont {Lee},
  \citenamefont {Douglas},\ and\ \citenamefont
  {Glotzer}}]{lee_filler-induced_1999}%
  \BibitemOpen
  \bibfield  {author} {\bibinfo {author} {\bibfnamefont {B.~P.}\ \bibnamefont
  {Lee}}, \bibinfo {author} {\bibfnamefont {J.~F.}\ \bibnamefont {Douglas}}, \
  and\ \bibinfo {author} {\bibfnamefont {S.~C.}\ \bibnamefont {Glotzer}},\
  }\bibfield  {title} {\enquote {\bibinfo {title} {Filler-induced composition
  waves in phase-separating polymer blends},}\ }\href {\doibase
  10.1103/PhysRevE.60.5812} {\bibfield  {journal} {\bibinfo  {journal} {Phys.
  Rev. E}\ }\textbf {\bibinfo {volume} {60}},\ \bibinfo {pages} {5812}
  (\bibinfo {year} {1999})}\BibitemShut {NoStop}%
\bibitem [{\citenamefont {Onuki}(2002)}]{onuki_phase_2002}%
  \BibitemOpen
  \bibfield  {author} {\bibinfo {author} {\bibfnamefont {A.}~\bibnamefont
  {Onuki}},\ }\href@noop {} {\emph {\bibinfo {title} {Phase {Transition}
  {Dynamics}}}}\ (\bibinfo  {publisher} {Cambridge University Press},\ \bibinfo
  {year} {2002})\BibitemShut {NoStop}%
\bibitem [{\citenamefont {Fisher}\ and\ \citenamefont
  {de~Gennes}(1978)}]{fisher_wall_1978}%
  \BibitemOpen
  \bibfield  {author} {\bibinfo {author} {\bibfnamefont {M.~E.}\ \bibnamefont
  {Fisher}}\ and\ \bibinfo {author} {\bibfnamefont {P.~G.}\ \bibnamefont
  {de~Gennes}},\ }\bibfield  {title} {\enquote {\bibinfo {title} {Wall
  {Phenomena} in a {Critical} {Binary} {Mixture}},}\ }\href@noop {} {\bibfield
  {journal} {\bibinfo  {journal} {C. R. Acad. Sci. Paris B}\ }\textbf {\bibinfo
  {volume} {287}},\ \bibinfo {pages} {207} (\bibinfo {year}
  {1978})}\BibitemShut {NoStop}%
\bibitem [{\citenamefont {Krech}\ and\ \citenamefont
  {Dietrich}(1992)}]{krech_free_1992}%
  \BibitemOpen
  \bibfield  {author} {\bibinfo {author} {\bibfnamefont {M.}~\bibnamefont
  {Krech}}\ and\ \bibinfo {author} {\bibfnamefont {S.}~\bibnamefont
  {Dietrich}},\ }\bibfield  {title} {\enquote {\bibinfo {title} {Free energy
  and specific heat of critical films and surfaces},}\ }\href {\doibase
  10.1103/PhysRevA.46.1886} {\bibfield  {journal} {\bibinfo  {journal} {Phys.
  Rev. A}\ }\textbf {\bibinfo {volume} {46}},\ \bibinfo {pages} {1886}
  (\bibinfo {year} {1992})}\BibitemShut {NoStop}%
\bibitem [{\citenamefont {Krech}(1994)}]{krech_casimir_1994}%
  \BibitemOpen
  \bibfield  {author} {\bibinfo {author} {\bibfnamefont {M.}~\bibnamefont
  {Krech}},\ }\href@noop {} {\emph {\bibinfo {title} {The {Casimir} effect in
  critical systems}}}\ (\bibinfo  {publisher} {World Scientific},\ \bibinfo
  {address} {Singapore},\ \bibinfo {year} {1994})\BibitemShut {NoStop}%
\bibitem [{\citenamefont {Brankov}\ \emph {et~al.}(2000)\citenamefont
  {Brankov}, \citenamefont {Dantchev},\ and\ \citenamefont
  {Tonchev}}]{brankov_theory_2000}%
  \BibitemOpen
  \bibfield  {author} {\bibinfo {author} {\bibfnamefont {J.~G.}\ \bibnamefont
  {Brankov}}, \bibinfo {author} {\bibfnamefont {D.~M.}\ \bibnamefont
  {Dantchev}}, \ and\ \bibinfo {author} {\bibfnamefont {N.~S.}\ \bibnamefont
  {Tonchev}},\ }\href@noop {} {\emph {\bibinfo {title} {The {Theory} of
  {Critical} {Phenomena} in {Finite}-{Size} {Systems}}}}\ (\bibinfo
  {publisher} {World Scientific},\ \bibinfo {address} {Singapore},\ \bibinfo
  {year} {2000})\BibitemShut {NoStop}%
\bibitem [{\citenamefont {Gambassi}(2009)}]{gambassi_casimir_2009}%
  \BibitemOpen
  \bibfield  {author} {\bibinfo {author} {\bibfnamefont {A.}~\bibnamefont
  {Gambassi}},\ }\bibfield  {title} {\enquote {\bibinfo {title} {The {Casimir}
  effect: {From} quantum to critical fluctuations},}\ }\href {\doibase
  10.1088/1742-6596/161/1/012037} {\bibfield  {journal} {\bibinfo  {journal}
  {J. Phys.: Conf. Ser.}\ }\textbf {\bibinfo {volume} {161}},\ \bibinfo {pages}
  {012037} (\bibinfo {year} {2009})}\BibitemShut {NoStop}%
\bibitem [{\citenamefont {Brito}\ \emph {et~al.}(2007)\citenamefont {Brito},
  \citenamefont {Marini Bettolo~Marconi},\ and\ \citenamefont
  {Soto}}]{brito_generalized_2007}%
  \BibitemOpen
  \bibfield  {author} {\bibinfo {author} {\bibfnamefont {R.}~\bibnamefont
  {Brito}}, \bibinfo {author} {\bibfnamefont {U.}~\bibnamefont {Marini
  Bettolo~Marconi}}, \ and\ \bibinfo {author} {\bibfnamefont {R.}~\bibnamefont
  {Soto}},\ }\bibfield  {title} {\enquote {\bibinfo {title} {Generalized
  {Casimir} forces in nonequilibrium systems},}\ }\href {\doibase
  10.1103/PhysRevE.76.011113} {\bibfield  {journal} {\bibinfo  {journal} {Phys.
  Rev. E}\ }\textbf {\bibinfo {volume} {76}},\ \bibinfo {pages} {011113}
  (\bibinfo {year} {2007})}\BibitemShut {NoStop}%
\bibitem [{\citenamefont {Rodriguez-Lopez}\ \emph {et~al.}(2011)\citenamefont
  {Rodriguez-Lopez}, \citenamefont {Brito},\ and\ \citenamefont
  {Soto}}]{rodriguez-lopez_dynamical_2011}%
  \BibitemOpen
  \bibfield  {author} {\bibinfo {author} {\bibfnamefont {P.}~\bibnamefont
  {Rodriguez-Lopez}}, \bibinfo {author} {\bibfnamefont {R.}~\bibnamefont
  {Brito}}, \ and\ \bibinfo {author} {\bibfnamefont {R.}~\bibnamefont {Soto}},\
  }\bibfield  {title} {\enquote {\bibinfo {title} {Dynamical approach to the
  {Casimir} effect},}\ }\href {\doibase 10.1103/PhysRevE.83.031102} {\bibfield
  {journal} {\bibinfo  {journal} {Phys. Rev. E}\ }\textbf {\bibinfo {volume}
  {83}},\ \bibinfo {pages} {031102} (\bibinfo {year} {2011})}\BibitemShut
  {NoStop}%
\bibitem [{\citenamefont {Dean}\ and\ \citenamefont
  {Gopinathan}(2010)}]{dean_out--equilibrium_2010}%
  \BibitemOpen
  \bibfield  {author} {\bibinfo {author} {\bibfnamefont {D.~S.}\ \bibnamefont
  {Dean}}\ and\ \bibinfo {author} {\bibfnamefont {A.}~\bibnamefont
  {Gopinathan}},\ }\bibfield  {title} {\enquote {\bibinfo {title}
  {Out-of-equilibrium behavior of {Casimir}-type fluctuation-induced forces for
  free classical fields},}\ }\href {\doibase 10.1103/PhysRevE.81.041126}
  {\bibfield  {journal} {\bibinfo  {journal} {Phys. Rev. E}\ }\textbf {\bibinfo
  {volume} {81}},\ \bibinfo {pages} {041126} (\bibinfo {year}
  {2010})}\BibitemShut {NoStop}%
\bibitem [{\citenamefont {Dean}\ \emph {et~al.}(2012)\citenamefont {Dean},
  \citenamefont {Demery}, \citenamefont {Parsegian},\ and\ \citenamefont
  {Podgornik}}]{dean_out--equilibrium_2012}%
  \BibitemOpen
  \bibfield  {author} {\bibinfo {author} {\bibfnamefont {D.~S.}\ \bibnamefont
  {Dean}}, \bibinfo {author} {\bibfnamefont {V.}~\bibnamefont {Demery}},
  \bibinfo {author} {\bibfnamefont {V.~A.}\ \bibnamefont {Parsegian}}, \ and\
  \bibinfo {author} {\bibfnamefont {R.}~\bibnamefont {Podgornik}},\ }\bibfield
  {title} {\enquote {\bibinfo {title} {Out-of-equilibrium relaxation of the
  thermal {Casimir} effect in a model polarizable material},}\ }\href {\doibase
  10.1103/PhysRevE.85.031108} {\bibfield  {journal} {\bibinfo  {journal} {Phys.
  Rev. E}\ }\textbf {\bibinfo {volume} {85}},\ \bibinfo {pages} {031108}
  (\bibinfo {year} {2012})}\BibitemShut {NoStop}%
\bibitem [{\citenamefont {Dean}\ and\ \citenamefont
  {Podgornik}(2014)}]{dean_relaxation_2014}%
  \BibitemOpen
  \bibfield  {author} {\bibinfo {author} {\bibfnamefont {D.~S.}\ \bibnamefont
  {Dean}}\ and\ \bibinfo {author} {\bibfnamefont {R.}~\bibnamefont
  {Podgornik}},\ }\bibfield  {title} {\enquote {\bibinfo {title} {Relaxation of
  the thermal {Casimir} force between net neutral plates containing {Brownian}
  charges},}\ }\href {\doibase 10.1103/PhysRevE.89.032117} {\bibfield
  {journal} {\bibinfo  {journal} {Phys. Rev. E}\ }\textbf {\bibinfo {volume}
  {89}},\ \bibinfo {pages} {032117} (\bibinfo {year} {2014})}\BibitemShut
  {NoStop}%
\bibitem [{\citenamefont {Hanke}(2013)}]{hanke_non-equilibrium_2013}%
  \BibitemOpen
  \bibfield  {author} {\bibinfo {author} {\bibfnamefont {A.}~\bibnamefont
  {Hanke}},\ }\bibfield  {title} {\enquote {\bibinfo {title} {Non-{Equilibrium}
  {Casimir} {Force} between {Vibrating} {Plates}},}\ }\href {\doibase
  10.1371/journal.pone.0053228} {\bibfield  {journal} {\bibinfo  {journal}
  {PLOS ONE}\ }\textbf {\bibinfo {volume} {8}},\ \bibinfo {pages} {e53228}
  (\bibinfo {year} {2013})}\BibitemShut {NoStop}%
\bibitem [{\citenamefont {Rohwer}\ \emph {et~al.}(2017)\citenamefont {Rohwer},
  \citenamefont {Kardar},\ and\ \citenamefont
  {Kr\"uger}}]{rohwer_transient_2017}%
  \BibitemOpen
  \bibfield  {author} {\bibinfo {author} {\bibfnamefont {C.~M.}\ \bibnamefont
  {Rohwer}}, \bibinfo {author} {\bibfnamefont {M.}~\bibnamefont {Kardar}}, \
  and\ \bibinfo {author} {\bibfnamefont {M.}~\bibnamefont {Kr\"uger}},\
  }\bibfield  {title} {\enquote {\bibinfo {title} {Transient {Casimir} {Forces}
  from {Quenches} in {Thermal} and {Active} {Matter}},}\ }\href {\doibase
  10.1103/PhysRevLett.118.015702} {\bibfield  {journal} {\bibinfo  {journal}
  {Phys. Rev. Lett.}\ }\textbf {\bibinfo {volume} {118}},\ \bibinfo {pages}
  {015702} (\bibinfo {year} {2017})}\BibitemShut {NoStop}%
\bibitem [{\citenamefont {Gross}\ \emph {et~al.}(2016)\citenamefont {Gross},
  \citenamefont {Vasilyev}, \citenamefont {Gambassi},\ and\ \citenamefont
  {Dietrich}}]{gross_critical_2016}%
  \BibitemOpen
  \bibfield  {author} {\bibinfo {author} {\bibfnamefont {M.}~\bibnamefont
  {Gross}}, \bibinfo {author} {\bibfnamefont {O.}~\bibnamefont {Vasilyev}},
  \bibinfo {author} {\bibfnamefont {A.}~\bibnamefont {Gambassi}}, \ and\
  \bibinfo {author} {\bibfnamefont {S.}~\bibnamefont {Dietrich}},\ }\bibfield
  {title} {\enquote {\bibinfo {title} {Critical adsorption and critical
  {Casimir} forces in the canonical ensemble},}\ }\href {\doibase
  10.1103/PhysRevE.94.022103} {\bibfield  {journal} {\bibinfo  {journal} {Phys.
  Rev. E}\ }\textbf {\bibinfo {volume} {94}},\ \bibinfo {pages} {022103}
  (\bibinfo {year} {2016})}\BibitemShut {NoStop}%
\bibitem [{\citenamefont {Gross}\ \emph {et~al.}(2017)\citenamefont {Gross},
  \citenamefont {Gambassi},\ and\ \citenamefont
  {Dietrich}}]{gross_statistical_2017}%
  \BibitemOpen
  \bibfield  {author} {\bibinfo {author} {\bibfnamefont {M.}~\bibnamefont
  {Gross}}, \bibinfo {author} {\bibfnamefont {A.}~\bibnamefont {Gambassi}}, \
  and\ \bibinfo {author} {\bibfnamefont {S.}~\bibnamefont {Dietrich}},\
  }\bibfield  {title} {\enquote {\bibinfo {title} {Statistical field theory
  with constraints: {Application} to critical {Casimir} forces in the canonical
  ensemble},}\ }\href {\doibase 10.1103/PhysRevE.96.022135} {\bibfield
  {journal} {\bibinfo  {journal} {Phys. Rev. E}\ }\textbf {\bibinfo {volume}
  {96}},\ \bibinfo {pages} {022135} (\bibinfo {year} {2017})}\BibitemShut
  {NoStop}%
\bibitem [{\citenamefont {Diehl}(1986)}]{diehl_field-theoretical_1986}%
  \BibitemOpen
  \bibfield  {author} {\bibinfo {author} {\bibfnamefont {H.~W.}\ \bibnamefont
  {Diehl}},\ }\bibfield  {title} {\enquote {\bibinfo {title} {Field-theoretical
  {Approach} to {Critical} {Behavior} at {Surfaces}},}\ }in\ \href@noop {}
  {\emph {\bibinfo {booktitle} {Phase {Transitions} and {Critical}
  {Phenomena}}}},\ Vol.~\bibinfo {volume} {10},\ \bibinfo {editor} {edited by\
  \bibinfo {editor} {\bibfnamefont {C.}~\bibnamefont {Domb}}\ and\ \bibinfo
  {editor} {\bibfnamefont {J.~L.}\ \bibnamefont {Lebowitz}}}\ (\bibinfo
  {publisher} {Academic},\ \bibinfo {address} {London},\ \bibinfo {year}
  {1986})\ p.~\bibinfo {pages} {76}\BibitemShut {NoStop}%
\bibitem [{\citenamefont {Pelissetto}\ and\ \citenamefont
  {Vicari}(2002)}]{pelissetto_critical_2002}%
  \BibitemOpen
  \bibfield  {author} {\bibinfo {author} {\bibfnamefont {A.}~\bibnamefont
  {Pelissetto}}\ and\ \bibinfo {author} {\bibfnamefont {E.}~\bibnamefont
  {Vicari}},\ }\bibfield  {title} {\enquote {\bibinfo {title} {Critical
  phenomena and renormalization-group theory},}\ }\href {\doibase
  10.1016/S0370-1573(02)00219-3} {\bibfield  {journal} {\bibinfo  {journal}
  {Phys. Rep.}\ }\textbf {\bibinfo {volume} {368}},\ \bibinfo {pages} {549}
  (\bibinfo {year} {2002})}\BibitemShut {NoStop}%
\bibitem [{\citenamefont {Gambassi}\ \emph {et~al.}(2009)\citenamefont
  {Gambassi}, \citenamefont {Maciolek}, \citenamefont {Hertlein}, \citenamefont
  {Nellen}, \citenamefont {Helden}, \citenamefont {Bechinger},\ and\
  \citenamefont {Dietrich}}]{gambassi_critical_2009}%
  \BibitemOpen
  \bibfield  {author} {\bibinfo {author} {\bibfnamefont {A.}~\bibnamefont
  {Gambassi}}, \bibinfo {author} {\bibfnamefont {A.}~\bibnamefont {Maciolek}},
  \bibinfo {author} {\bibfnamefont {C.}~\bibnamefont {Hertlein}}, \bibinfo
  {author} {\bibfnamefont {U.}~\bibnamefont {Nellen}}, \bibinfo {author}
  {\bibfnamefont {L.}~\bibnamefont {Helden}}, \bibinfo {author} {\bibfnamefont
  {C.}~\bibnamefont {Bechinger}}, \ and\ \bibinfo {author} {\bibfnamefont
  {S.}~\bibnamefont {Dietrich}},\ }\bibfield  {title} {\enquote {\bibinfo
  {title} {Critical {Casimir} effect in classical binary liquid mixtures},}\
  }\href {\doibase 10.1103/PhysRevE.80.061143} {\bibfield  {journal} {\bibinfo
  {journal} {Phys. Rev. E}\ }\textbf {\bibinfo {volume} {80}},\ \bibinfo
  {pages} {061143} (\bibinfo {year} {2009})}\BibitemShut {NoStop}%
\bibitem [{Note1()}]{Note1}%
  \BibitemOpen
  \bibinfo {note} {Note that in Laplace space \protect \cref {eq_bcs_CA_red},
  being time-independent, reduces to $\partial _\zeta \protect \mathaccentV
  {hat}05E{\protect \mathpzc {m}}(s,\zeta \in \protect \{ 0,1 \protect \})=\mp
  H_1/s$. These boundary conditions\protect \xspace also fix the prefactor of
  the solution in \protect \cref {eq_linB_lapl_sol}.}\BibitemShut {Stop}%
\bibitem [{\citenamefont {Poularikas}(2010)}]{poularikas_transforms_2010}%
  \BibitemOpen
  \bibfield  {author} {\bibinfo {author} {\bibfnamefont {A.~D.}\ \bibnamefont
  {Poularikas}},\ }\href@noop {} {\emph {\bibinfo {title} {Transforms and
  {Applications} {Handbook}, {Third} {Edition}}}}\ (\bibinfo  {publisher} {CRC
  Press},\ \bibinfo {address} {Boca Raton, FL},\ \bibinfo {year}
  {2010})\BibitemShut {NoStop}%
\bibitem [{\citenamefont {Abate}\ and\ \citenamefont
  {Valkó}(2004)}]{abate_multi-precision_2004}%
  \BibitemOpen
  \bibfield  {author} {\bibinfo {author} {\bibfnamefont {J.}~\bibnamefont
  {Abate}}\ and\ \bibinfo {author} {\bibfnamefont {P.~P.}\ \bibnamefont
  {Valkó}},\ }\bibfield  {title} {\enquote {\bibinfo {title} {Multi-precision
  {Laplace} transform inversion},}\ }\href {\doibase 10.1002/nme.995}
  {\bibfield  {journal} {\bibinfo  {journal} {Int. J. Num. Meth. Eng.}\
  }\textbf {\bibinfo {volume} {60}},\ \bibinfo {pages} {979} (\bibinfo {year}
  {2004})}\BibitemShut {NoStop}%
\bibitem [{\citenamefont {Gradshteyn}\ and\ \citenamefont
  {Ryzhik}(2014)}]{gradshteyn_table_2014}%
  \BibitemOpen
  \bibfield  {author} {\bibinfo {author} {\bibfnamefont {I.~S.}\ \bibnamefont
  {Gradshteyn}}\ and\ \bibinfo {author} {\bibfnamefont {I.~M.}\ \bibnamefont
  {Ryzhik}},\ }\href@noop {} {\emph {\bibinfo {title} {Table of {Integrals},
  {Series}, and {Products}}}}\ (\bibinfo  {publisher} {Academic},\ \bibinfo
  {address} {London},\ \bibinfo {year} {2014})\BibitemShut {NoStop}%
\bibitem [{\citenamefont {Gross}(2018)}]{gross_first-passage_2018}%
  \BibitemOpen
  \bibfield  {author} {\bibinfo {author} {\bibfnamefont {M.}~\bibnamefont
  {Gross}},\ }\bibfield  {title} {\enquote {\bibinfo {title} {First-passage
  dynamics of linear stochastic interface models: weak-noise theory and
  influence of boundary conditions},}\ }\href {\doibase
  10.1088/1742-5468/aaa386} {\bibfield  {journal} {\bibinfo  {journal} {J.
  Stat. Mech. Theor. Exp.}\ }\textbf {\bibinfo {volume} {2018}},\ \bibinfo
  {pages} {033213} (\bibinfo {year} {2018})}\BibitemShut {NoStop};
    \bibfield  {author} {\bibinfo {author} {\bibfnamefont {M.}~\bibnamefont
  {Gross}},\ }\bibfield  {title} {\enquote {\bibinfo {title} {First-passage
  dynamics of linear stochastic interface models: numerical simulations and entropic repulsion effect},}\ }\href {\doibase
  10.1088/1742-5468/aaa792} {\bibfield  {journal} {\bibinfo  {journal} {J.
  Stat. Mech. Theor. Exp.}\ }\textbf {\bibinfo {volume} {2018}},\ \bibinfo
  {pages} {033212} (\bibinfo {year} {2018})}\BibitemShut {NoStop}%
\bibitem [{Note2()}]{Note2}%
  \BibitemOpen
  \bibinfo {note} {Upon performing the asymptotic expansion of the expression
  for $\chi _0$ given below \protect \cref {eq_laplsol_wall_inv0}, one has to
  take into account that $\kappa \zeta \ll \kappa $ because we assume $\zeta
  \ll 1/2$.}\BibitemShut {Stop}%
\bibitem [{Note3()}]{Note3}%
  \BibitemOpen
  \bibinfo {note} {This is formally a consequence of the lemma of
  Riemann-Lebesgue \cite {bender_advanced_1999}}\BibitemShut {NoStop}%
\bibitem [{Note4()}]{Note4}%
  \BibitemOpen
  \bibinfo {note} {The calculation is facilitated by using known Laplace
  transforms of hypergeometric functions (see \protect \S 3.38.1 in Ref.\ \cite
  {prudnikov_direct_1992})}\BibitemShut {NoStop}%
\bibitem [{\citenamefont {Olver}\ \emph {et~al.}(2010)\citenamefont {Olver},
  \citenamefont {Lozier}, \citenamefont {Boisvert},\ and\ \citenamefont
  {Clark}}]{olver_nist_2010}%
  \BibitemOpen
  \bibfield  {author} {\bibinfo {author} {\bibfnamefont {F.~W.~J.}\
  \bibnamefont {Olver}}, \bibinfo {author} {\bibfnamefont {D.~W.}\ \bibnamefont
  {Lozier}}, \bibinfo {author} {\bibfnamefont {R.~F.}\ \bibnamefont
  {Boisvert}}, \ and\ \bibinfo {author} {\bibfnamefont {C.~W.}\ \bibnamefont
  {Clark}},\ }\href {http://dlmf.nist.gov/} {\emph {\bibinfo {title} {{NIST}
  {Handbook} of {Mathematical} {Functions}}}},\ \bibinfo {edition} {1st}\ ed.\
  (\bibinfo  {publisher} {Cambridge University Press},\ \bibinfo {year}
  {2010})\BibitemShut {NoStop}%
\bibitem [{Note5()}]{Note5}%
  \BibitemOpen
  \bibinfo {note} {The asymptotic behaviors of a function in real space and in
  Laplace space are inter-related by means of so-called Tauberian theorems (see
  \protect \S XIII.5 in Ref.\ \cite {feller_introduction_1971})}\BibitemShut
  {NoStop}%
\bibitem [{Note6()}]{Note6}%
  \BibitemOpen
  \bibinfo {note} {The limit $\theta \to 0$ is singular because the initial
  condition [\protect \cref {eq_prof_init}] is incompatible with the boundary
  conditions\protect \xspace [\protect \cref {eq_bcs_CA_red}].}\BibitemShut
  {Stop}%
\bibitem [{\citenamefont {Moin}(2010)}]{moin_fundamentals_2010}%
  \BibitemOpen
  \bibfield  {author} {\bibinfo {author} {\bibfnamefont {P.}~\bibnamefont
  {Moin}},\ }\href@noop {} {\emph {\bibinfo {title} {Fundamentals of
  {Engineering} {Numerical} {Analysis}}}},\ \bibinfo {edition} {2nd}\ ed.\
  (\bibinfo  {publisher} {Cambridge University Press},\ \bibinfo {year}
  {2010})\BibitemShut {NoStop}%
\bibitem [{Note7()}]{Note7}%
  \BibitemOpen
  \bibinfo {note} {We have checked that by using as initial condition an
  equilibrium profile, which corresponds to a sufficiently high temperature
  $x\gg 1$, one obtains essentially the same results after a short transient
  period, which we do not consider here.}\BibitemShut {Stop}%
\bibitem [{Note8()}]{Note8}%
  \BibitemOpen
  \bibinfo {note} {For large $H_1$, the value of this prefactor becomes
  independent of $H_1$.}\BibitemShut {Stop}%
\bibitem [{Note9()}]{Note9}%
  \BibitemOpen
  \bibinfo {note} {In the case of \protect \emph {model A} dynamics, i.e., for
  $\partial _t \phi = -\mu (\phi )$, the dynamical stress tensor in \protect
  \cref {eq_dyn_stressten} can be written as ${\protect \mathcal {\protect
  \mathaccentV {bar}016T}}_{ij} = \protect \mathcal {T}_{ij} - \protect \frac
  {1}{2}\partial _t(\phi ^2)$ and thus reduces to the alternative formulation
  given in Eq.~(56) in Ref.\ \cite {dean_out--equilibrium_2010} (apart from an
  overall minus sign in its definition).}\BibitemShut {Stop}%
\bibitem [{\citenamefont {Kr\"uger}\ \emph {et~al.}(2018)\citenamefont
  {Kr\"uger}, \citenamefont {Solon}, \citenamefont {Demery}, \citenamefont
  {Rohwer},\ and\ \citenamefont {Dean}}]{kruger_stresses_2018}%
  \BibitemOpen
  \bibfield  {author} {\bibinfo {author} {\bibfnamefont {M.}~\bibnamefont
  {Kr\"uger}}, \bibinfo {author} {\bibfnamefont {A.}~\bibnamefont {Solon}},
  \bibinfo {author} {\bibfnamefont {V.}~\bibnamefont {Demery}}, \bibinfo
  {author} {\bibfnamefont {C.~M.}\ \bibnamefont {Rohwer}}, \ and\ \bibinfo
  {author} {\bibfnamefont {D.~S.}\ \bibnamefont {Dean}},\ }\bibfield  {title}
  {\enquote {\bibinfo {title} {Stresses in non-equilibrium fluids: {Exact}
  formulation and coarse-grained theory},}\ }\href {\doibase 10.1063/1.5019424}
  {\bibfield  {journal} {\bibinfo  {journal} {J. Chem. Phys.}\ }\textbf
  {\bibinfo {volume} {148}},\ \bibinfo {pages} {084503} (\bibinfo {year}
  {2018})}\BibitemShut {NoStop}%
\bibitem [{Note10()}]{Note10}%
  \BibitemOpen
  \bibinfo {note} {The last expression in \protect \cref {eq_CCF_num_avg}
  follows by noting that, for an arbitrary function $f(z')$, one has $\DOTSI
  \intop \ilimits@ _{L/2}^{L-\epsilon }\protect \mathrm {d}w \DOTSI \intop
  \ilimits@ _w^{L-\epsilon } \protect \mathrm {d}z' f(z') = \DOTSI \intop
  \ilimits@ _{L/2}^{L-\epsilon }\protect \mathrm {d}z' \DOTSI \intop \ilimits@
  _{L/2}^{z'} \protect \mathrm {d}w f(z') = \DOTSI \intop \ilimits@
  _{L/2}^{L-\epsilon } \protect \mathrm {d}z' (z'-(L/2)) f(z')$.}\BibitemShut
  {Stop}%
\bibitem [{\citenamefont {Roy}\ \emph {et~al.}(2016)\citenamefont {Roy},
  \citenamefont {Dietrich},\ and\ \citenamefont
  {H\"ofling}}]{roy_structure_2016}%
  \BibitemOpen
  \bibfield  {author} {\bibinfo {author} {\bibfnamefont {S.}~\bibnamefont
  {Roy}}, \bibinfo {author} {\bibfnamefont {S.}~\bibnamefont {Dietrich}}, \
  and\ \bibinfo {author} {\bibfnamefont {F.}~\bibnamefont {H\"ofling}},\
  }\bibfield  {title} {\enquote {\bibinfo {title} {Structure and dynamics of
  binary liquid mixtures near their continuous demixing transitions},}\ }\href
  {\doibase 10.1063/1.4963771} {\bibfield  {journal} {\bibinfo  {journal} {J.
  Chem. Phys.}\ }\textbf {\bibinfo {volume} {145}},\ \bibinfo {pages} {134505}
  (\bibinfo {year} {2016})}\BibitemShut {NoStop}%
\bibitem [{\citenamefont {Puosi}\ \emph {et~al.}(2016)\citenamefont {Puosi},
  \citenamefont {Cardozo}, \citenamefont {Ciliberto},\ and\ \citenamefont
  {Holdsworth}}]{puosi_direct_2016}%
  \BibitemOpen
  \bibfield  {author} {\bibinfo {author} {\bibfnamefont {F.}~\bibnamefont
  {Puosi}}, \bibinfo {author} {\bibfnamefont {D.~L.}\ \bibnamefont {Cardozo}},
  \bibinfo {author} {\bibfnamefont {S.}~\bibnamefont {Ciliberto}}, \ and\
  \bibinfo {author} {\bibfnamefont {P.~C.~W.}\ \bibnamefont {Holdsworth}},\
  }\bibfield  {title} {\enquote {\bibinfo {title} {Direct calculation of the
  critical {Casimir} force in a binary fluid},}\ }\href {\doibase
  10.1103/PhysRevE.94.040102} {\bibfield  {journal} {\bibinfo  {journal} {Phys.
  Rev. E}\ }\textbf {\bibinfo {volume} {94}},\ \bibinfo {pages} {040102}
  (\bibinfo {year} {2016})}\BibitemShut {NoStop}%
\bibitem [{\citenamefont {Belardinelli}\ \emph {et~al.}(2015)\citenamefont
  {Belardinelli}, \citenamefont {Sbragaglia}, \citenamefont {Biferale},
  \citenamefont {Gross},\ and\ \citenamefont
  {Varnik}}]{belardinelli_fluctuating_2015}%
  \BibitemOpen
  \bibfield  {author} {\bibinfo {author} {\bibfnamefont {D.}~\bibnamefont
  {Belardinelli}}, \bibinfo {author} {\bibfnamefont {M.}~\bibnamefont
  {Sbragaglia}}, \bibinfo {author} {\bibfnamefont {L.}~\bibnamefont
  {Biferale}}, \bibinfo {author} {\bibfnamefont {M.}~\bibnamefont {Gross}}, \
  and\ \bibinfo {author} {\bibfnamefont {F.}~\bibnamefont {Varnik}},\
  }\bibfield  {title} {\enquote {\bibinfo {title} {Fluctuating multicomponent
  lattice {Boltzmann} model},}\ }\href {\doibase 10.1103/PhysRevE.91.023313}
  {\bibfield  {journal} {\bibinfo  {journal} {Phys. Rev. E}\ }\textbf {\bibinfo
  {volume} {91}},\ \bibinfo {pages} {023313} (\bibinfo {year}
  {2015})}\BibitemShut {NoStop}%
\bibitem [{\citenamefont {Furukawa}\ \emph {et~al.}(2013)\citenamefont
  {Furukawa}, \citenamefont {Gambassi}, \citenamefont {Dietrich},\ and\
  \citenamefont {Tanaka}}]{furukawa_nonequilibrium_2013}%
  \BibitemOpen
  \bibfield  {author} {\bibinfo {author} {\bibfnamefont {A.}~\bibnamefont
  {Furukawa}}, \bibinfo {author} {\bibfnamefont {A.}~\bibnamefont {Gambassi}},
  \bibinfo {author} {\bibfnamefont {S.}~\bibnamefont {Dietrich}}, \ and\
  \bibinfo {author} {\bibfnamefont {H.}~\bibnamefont {Tanaka}},\ }\bibfield
  {title} {\enquote {\bibinfo {title} {Nonequilibrium {Critical} {Casimir}
  {Effect} in {Binary} {Fluids}},}\ }\href {\doibase
  10.1103/PhysRevLett.111.055701} {\bibfield  {journal} {\bibinfo  {journal}
  {Phys. Rev. Lett.}\ }\textbf {\bibinfo {volume} {111}},\ \bibinfo {pages}
  {055701} (\bibinfo {year} {2013})}\BibitemShut {NoStop}%
\bibitem [{\citenamefont {Bender}\ and\ \citenamefont
  {Orszag}(1999)}]{bender_advanced_1999}%
  \BibitemOpen
  \bibfield  {author} {\bibinfo {author} {\bibfnamefont {C.~M.}\ \bibnamefont
  {Bender}}\ and\ \bibinfo {author} {\bibfnamefont {S.~A.}\ \bibnamefont
  {Orszag}},\ }\href@noop {} {\emph {\bibinfo {title} {Advanced {Mathematical}
  {Methods} for {Scientists} and {Engineers} {I}: {Asymptotic} {Methods} and
  {Perturbation} {Theory}}}}\ (\bibinfo  {publisher} {Springer Science \&
  Business Media},\ \bibinfo {address} {New York, NY},\ \bibinfo {year}
  {1999})\BibitemShut {NoStop}%
\bibitem [{\citenamefont {Prudnikov}(1992)}]{prudnikov_direct_1992}%
  \BibitemOpen
  \bibfield  {author} {\bibinfo {author} {\bibfnamefont {A.~P.}\ \bibnamefont
  {Prudnikov}},\ }\href@noop {} {\emph {\bibinfo {title} {Integrals and series, Vol.\ 4: Direct {Laplace}
  transforms}}},\ (\bibinfo  {publisher} {Gordon and
  Breach},\ \bibinfo {address} {New York, NY},\ \bibinfo {year}
  {1992})\BibitemShut {NoStop}%
\bibitem [{\citenamefont {Feller}(1971)}]{feller_introduction_1971}%
  \BibitemOpen
  \bibfield  {author} {\bibinfo {author} {\bibfnamefont {W.}~\bibnamefont
  {Feller}},\ }\href@noop {} {\emph {\bibinfo {title} {An {Introduction} to
  {Probability} {Theory} and {Its} {Applications}}}},\ \bibinfo {edition}
  {2nd}\ ed.,\ Vol.~\bibinfo {volume} {2}\ (\bibinfo  {publisher} {John Wiley
  \& Sons, Inc.},\ \bibinfo {address} {New York, NY},\ \bibinfo {year}
  {1971})\BibitemShut {NoStop}%
\end{thebibliography}


%

\end{document}